\documentclass[11pt]{article}
\usepackage{appendix}
\usepackage{amsmath,amssymb,,amscd,amsthm,ifthen,color,framed,bm}
\usepackage{graphicx}
\usepackage{rays_defs_18}
\usepackage[OT1]{fontenc}
\usepackage{bbm}
\usepackage{comment}
\usepackage{tikz}
\usetikzlibrary{pgfplots.groupplots, positioning, matrix, arrows,decorations.pathmorphing}
\usepackage{pgfplots,pgfplotstable}
\usepgfplotslibrary{groupplots,fillbetween} 
\pgfplotsset{compat=1.10}
\usepackage{hyperref}
\hypersetup{
    unicode=false,          
    pdftoolbar=true,        
    pdfmenubar=true,        
    pdffitwindow=false,     
    pdfstartview={FitH},    
    pdfauthor={Reinhard Heckel},     
    pdfsubject={Subject},   
    pdfcreator={Reinhard Heckel},   
    pdfproducer={Producer}, 
    pdfnewwindow=true,      
    colorlinks=true,        
    linkcolor=red,          
    citecolor=green,        
    filecolor=magenta,      
    urlcolor=cyan           
}

\usepackage[
backend=bibtex,
url=false,isbn=false,doi=false,
date=year,
hyperref=auto,
style=alphabetic,
natbib=true,
sorting=nty,
giveninits=true,
maxnames=2, maxbibnames=10,
]{biblatex}
\bibliography{./bib.bib}

\theoremstyle{definition}

\long\def\comment#1{}

\pgfplotsset{select coords between index/.style 2 args={
    x filter/.code={
        \ifnum\coordindex<#1\fi
        \ifnum\coordindex>#2\fi
    }
}}

\usepackage{makecell}
\usepackage{caption}
\usepackage{adjustbox}
\usepackage{diagbox}
\usepackage{booktabs}
\definecolor{DarkBlue}{rgb}{0,0,0.7} 
\definecolor{BrickRed}{RGB}{203,65,84}
\definecolor{skyblue}{rgb}{0.53, 0.81, 0.92}
\newcommand{\vsparagraph}[1]{\vspace{-0.1in}\paragraph{#1}}

\begin{document}

\begin{center}

{\bf{\LARGE{
Robustness of Deep Learning for Accelerated MRI:\\ \vspace{2mm} Benefits of Diverse Training Data
}}}

\vspace*{.2in}

{\large{
\begin{tabular}{cccc}
Kang Lin and Reinhard Heckel
\end{tabular}
}}

\vspace*{.05in}

\begin{tabular}{c}
Department of Computer Engineering, Technical University of Munich \\
\href{mailto:ka.lin@tum.de}{ka.lin@tum.de},
\href{mailto:reinhard.heckel@tum.de}{reinhard.heckel@tum.de}
\end{tabular}

\vspace*{.1in}

\today

\vspace*{.1in}

\end{center}

\begin{abstract}
Deep learning based methods for image reconstruction are state-of-the-art for a variety of imaging tasks. However, neural networks often perform worse if the training data differs significantly from the data they are applied to. For example, a model trained for accelerated magnetic resonance imaging (MRI) on one scanner performs worse on another scanner. In this work, we investigate the impact of the training data on a model's performance and robustness for accelerated MRI. We find that models trained on the combination of various data distributions, such as those obtained from different MRI scanners and anatomies, exhibit robustness equal or superior to models trained on the best single distribution for a specific target distribution. Thus training on such diverse data tends to improve robustness. Furthermore, training on such a diverse dataset does not compromise in-distribution performance, i.e., a model trained on diverse data yields in-distribution performance at least as good as models trained on the more narrow individual distributions. Our results suggest that training a model for imaging on a variety of distributions tends to yield a more effective and robust model than maintaining separate models for individual distributions.
\end{abstract}

\section{Introduction}
Deep learning models trained end-to-end for image reconstruction are fast and accurate and outperform traditional methods for a variety of imaging tasks ranging from denoising over super-resolution to accelerated magnetic resonance imaging (MRI)~\citep{7949028, dongLearningDeepConvolutional2014, muckleyResults2020FastMRI2021}. 

Imaging accuracy is typically measured as in-distribution performance: A model trained on data from one source is applied to data from the same source. However, in practice a neural network for imaging is typically applied to slightly different data than it is trained on. For example, a neural network for accelerated magnetic resonance imaging trained on data from one hospital is applied in a different hospital. 

Neural networks for imaging often perform significantly worse under such distribution-shifts. For accelerated MRI, a model trained on knees performs worse on brains when compared to the same model trained on brains. Similar performance loss occurs for other natural distribution-shifts~\citep{knollAssessmentGeneralizationLearned2019,johnsonEvaluationRobustnessLearned2021a,darestaniMeasuringRobustnessDeep2021}.  

To date, much of research in deep learning for imaging has focused on developing better models and algorithms to improve in-distribution performance. Nevertheless, recent literature on computer vision models, in particular multi-modal models, suggest that a model's robustness is largely impacted by the training data, and a key ingredient for robust models are large and diverse training sets~\citep{fangDataDeterminesDistributional2022a,nguyenQualityNotQuantity2022a,gadreDataCompSearchNext2023a}. 

In this work, we take a step towards a better understanding of the training data for learning robust deep networks for imaging, in particular for accelerated MRI. 

\begin{figure}
    \begin{minipage}{1\linewidth}
    \centering
        \begin{tikzpicture}
            \node at (0.0\linewidth,0.0\linewidth) (a2)
            {\includegraphics[width=0.3\linewidth]{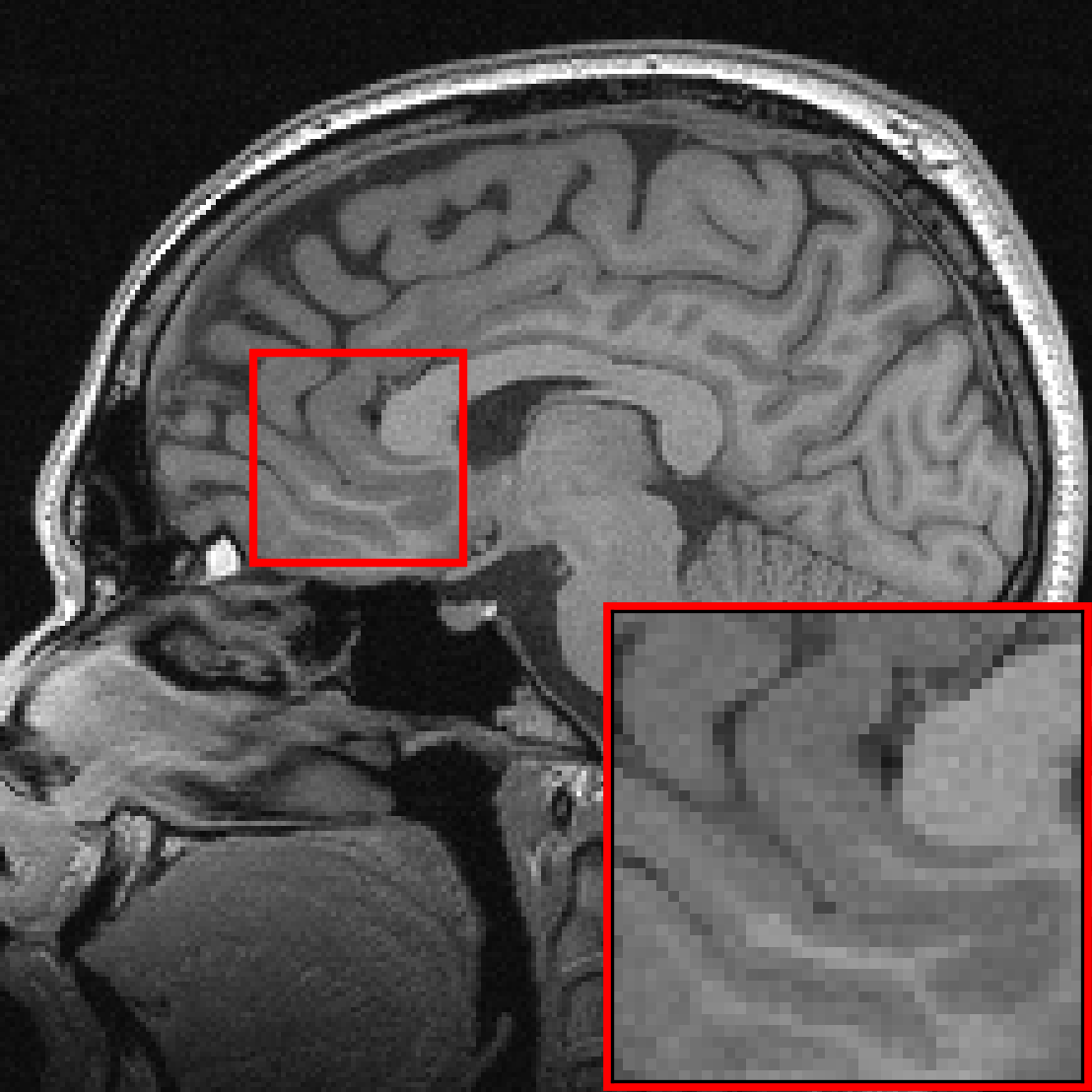}};
            \node[below = -0.0\linewidth of a2]  {Ground truth};
            \node[right = -0.0\linewidth of a2] (a3)
            {\includegraphics[width=0.3\linewidth]{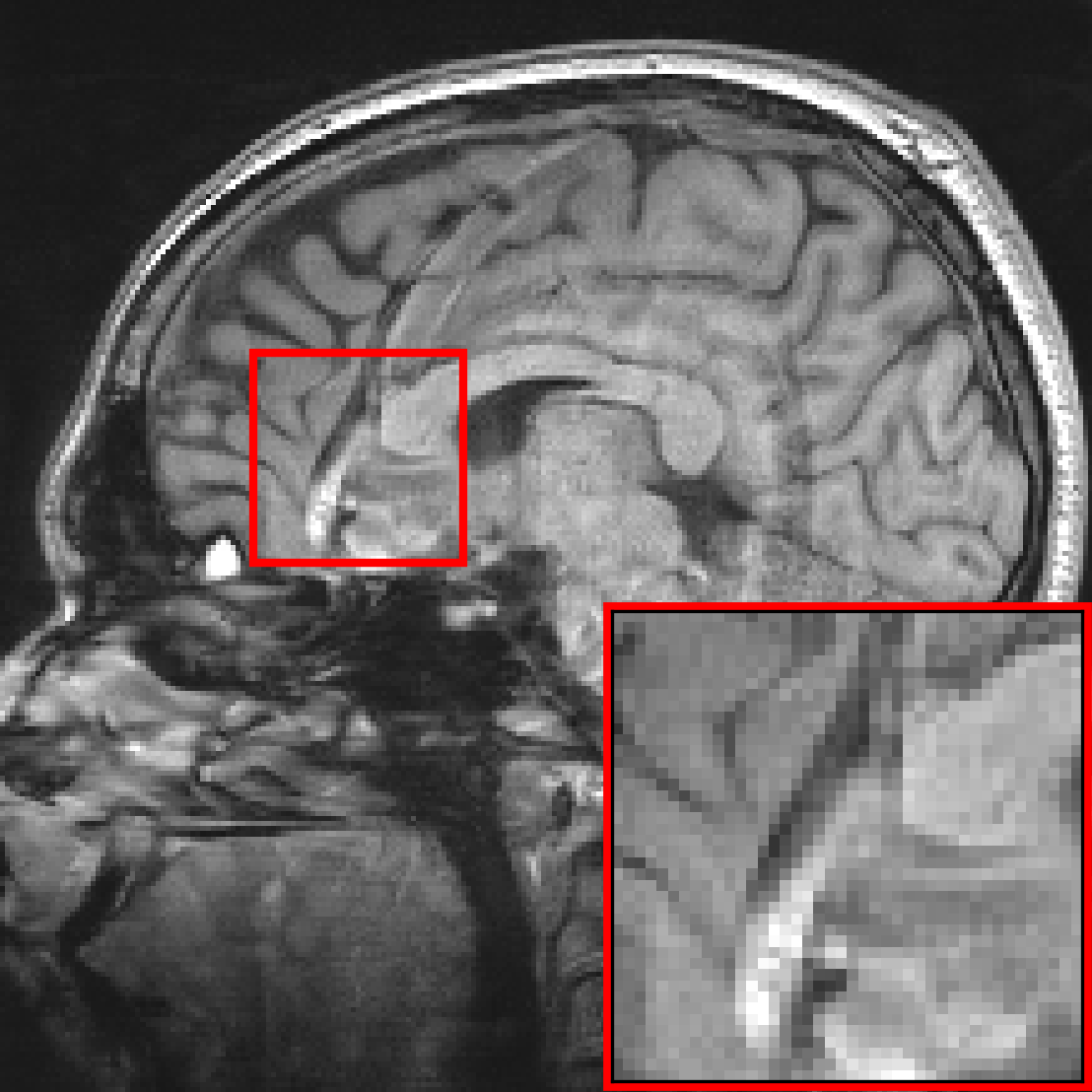}};
            \node[below = -0.0\linewidth of a3]  {Trained on fastMRI brain};
            \node[right = -0.0\linewidth of a3] (a4)
            {\includegraphics[width=0.3\linewidth]{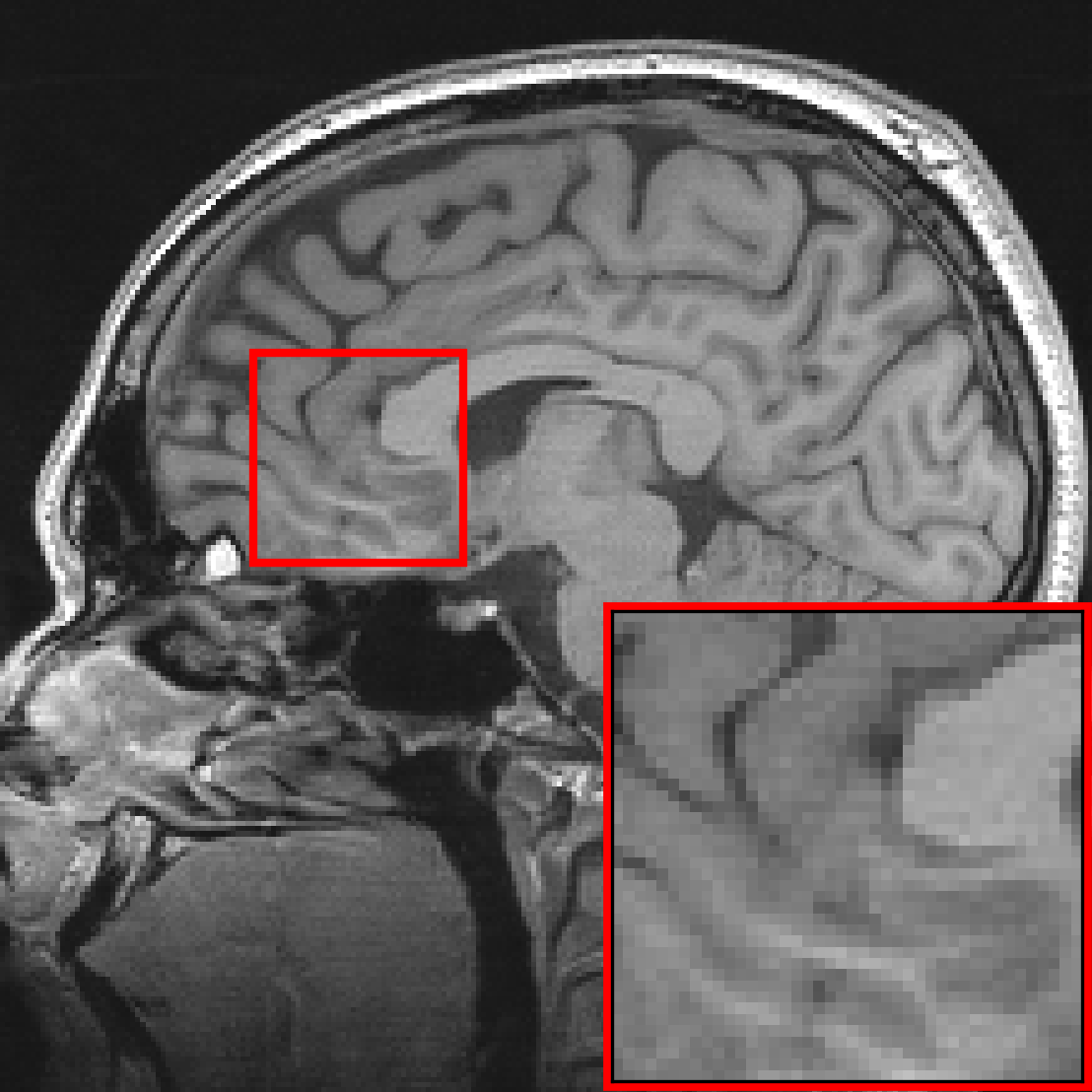}};
            \node[below = -0.0\linewidth of a4]  {Trained on $\mathcal{D}_P$};
        \end{tikzpicture}
    \captionof{figure}{
    An illustrative (randomly chosen) example to demonstrate benefits of training on a large and diverse dataset: Shown are reconstructions from two VarNets~\citep{sriramEndtoEndVariationalNetworks2020b}, one trained on fastMRI brain, the largest single dataset of brain images for accelerated MRI, and one trained on a diverse collection of datasets $\mathcal{D}_P$. Both models are evaluated out-of-distribution on an image from the CC-359-sagittal \citep{souzaOpenMultivendorMultifieldstrength2018} dataset. The model trained on fastMRI brain shows severe artifact whereas the model trained $\mathcal{D}_P$ provides better details and fewer artifacts.
    }
    \label{fig:P_recon}
    \end{minipage}
\end{figure}
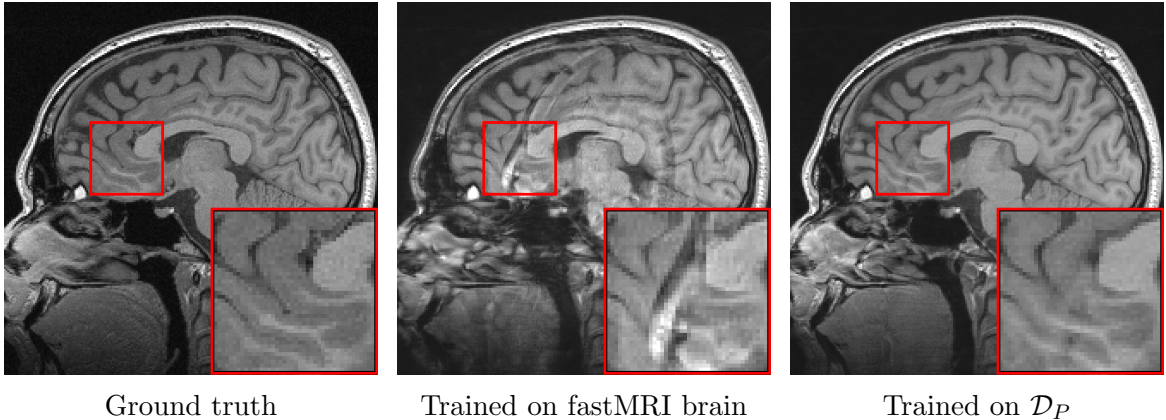

\begin{itemize}
\item First, we investigate whether deep networks for accelerated MRI compromise performance on individual distributions when trained on more than one distribution. We find for various pairs of distributions (different anatomies, image contrasts, and magnetic fields), that training a single model on two distributions yields the same performance as training two individual models.

\item Second, we demonstrate for a variety of distribution-shifts (anatomy, image contrast, and magnetic field shift) that the robustness of models, regardless of its architecture, is largely determined by the training set. A diverse set enhances robustness towards distribution-shifts.  We further show that robustness improvements highly correlate with similarity between train and test set.

\item Third, we consider a distribution-shift from healthy to non-healthy subjects and find that models trained on a diverse set of healthy subjects can reconstruct images with pathologies as accurately as models trained on images containing pathologies.

\item Fourth, we empirically find for several distribution-shifts that what we call `distributional overfitting' occurs: When training for long, in-distribution performance continues to improve slightly while out-of-distribution performance sharply drops. A related observation was made by~\citet{wortsmanRobustFinetuningZeroshot2022} for fine-tuning of CLIP models.  Therefore, early stopping can be helpful for training a robust model as it can yield a model with almost optimal in-distribution performance without losing robustness.
\end{itemize}

Taken together, those four findings suggest that training a single model on a diverse set of data distributions and incorporating early stopping yields a robust model. We test this hypothesis by training a model on a \textbf{large and diverse pool of data} significantly larger than the fastMRI dataset~\citep{zbontarFastMRIOpenDataset2019a}. The resulting model, as shown in Figure~\ref{fig:P_recon}, is significantly more robust than a model trained on the fastMRI dataset, the single largest dataset for accelerated MRI, without compromising performance on fastMRI data.

\vsparagraph{Related Work.}  
Several works have shown that deep learning models for accelerated MRI are sensitive to distribution-shifts. \citet{johnsonEvaluationRobustnessLearned2021a} found the models submitted to the 2019 fastMRI challenge~\citep{knollAdvancingMachineLearning2020} to be sensitive to distribution-shifts. Furthermore, \citet{darestaniMeasuringRobustnessDeep2021} show that reconstruction methods for MRI, regardless of whether they are trained or only tuned on data, all exhibit similar performance loss under distribution-shifts. 
Contrary to our work, both works do not propose robustness enhancing strategies, such as training on a diverse dataset. 

Moreover, there are several works that characterise the severity of specific distribution-shifts and propose transfer learning as a mitigation strategy~\citep{knollAssessmentGeneralizationLearned2019, huangEvaluationGeneralizationLearned2022, darTransferLearningApproachAccelerated2020}. Those works fine-tune on data from the test distribution, whereas we study a setup without access to data from the test distribution.  

A potential solution to enhance robustness in accelerated MRI is test-time training to narrow the performance gap on out-of-distribution data~\cite{darestaniTestTimeTrainingCan2022}, albeit at high computational costs. \citet{liuUniversalUndersampledMRI2021} propose a special model architecture for improving performance of training on multiple anatomies. \citet{ouyangGeneralizingDeepLearning2023} proposes an approach that modifies natural images for training MRI reconstruction models. In ultrasound imaging, \citet{khunjushDeepLearningUltrasound2023} demonstrate that diversifying simulated training data can improve robustness on real-world data. 

More broadly, several influential papers have shown that machine learning methods for problems ranging from image classification to natural language processing perform worse under distribution-shifts~\citep{rechtImageNetClassifiersGeneralize2019,millerEffectNaturalDistribution2020,taoriMeasuringRobustnessNatural2020a,hendrycksManyFacesRobustness2021}.

Shifting to computer vision, OpenAI's CLIP model~\citep{radfordLearningTransferableVisual2021a} is robust under distribution-shifts. \citet{fangDataDeterminesDistributional2022a} finds that the key contributor to CLIP's robustness is the diversity of the training set. However, \citet{nguyenQualityNotQuantity2022a} show that blindly combining data sources can weaken robustness compared to training on the best individual data source. 

These studies underscore the pivotal role of dataset design, particularly data diversity, for a model's performance and robustness. In light of concerns regarding the robustness of deep learning in medical imaging, we explore the impact of data diversity on models trained for accelerated MRI. 

While increasing the training set size generally improves performance, often following a power law~\citep{kaplanScalingLawsNeural2020a, zhaiScalingVisionTransformers2022,klugScalingLawsDeep2023}, this work focuses on out-of-distribution improvements through diversity, rather than in-distribution improvements through dataset size. It's worth noting that we specifically address out-of-distribution robustness, while other notions exist, such as worst-case robustness~\cite{antunInstabilitiesDeepLearning2020,ducotterdImprovingLipschitzConstrainedNeural2022,krainovicLearningProvablyRobust2023}.

\begin{table*}[t]
\centering
\scriptsize
    \caption{Fully-sampled k-space datasets used here. The percentages are the proportions of the data within a dataset. Scans containing multiple echoes or averages are separated as such and counted as separate volumes.}
    \label{tbl:datasets}
    \begin{adjustbox}{max width=\linewidth}
    \begin{tabular}{l r r r r r r r r}
    \toprule
    Dataset & Anatomy & View & Image contrast & Vendor & Magnet & Coils & Vol./Subj. & Slices\\
    \midrule
    fastMRI knee~\citep{zbontarFastMRIOpenDataset2019a} & knee & coronal & PD (50\%), PDFS (50\%) & Siemens &\makecell[tr]{1.5T (45\%),\\ 3T (55\%)} & 15 & 1.2k/1.2k & 42k\\
    fastMRI brain~\citep{zbontarFastMRIOpenDataset2019a} & brain & axial & \makecell[tr]{T1 (11\%), T1POST (21\%),\\ T2 (60\%), FLAIR (8\%)} & Siemens & \makecell[tr]{1.5T (43\%), \\ 3T (67\%)} & 4-20 & 6.4k/6.4k & 100k\\
    fastMRI prostate~\citep{tibrewalaFastMRIProstatePublicly2023} & prostate & axial & T2 & Siemens & 3T & 10-30 & 312/312 & 9.5k\\
    M4Raw~\citep{lyuM4RawMulticontrastMultirepetition2023a} & brain & axial & \makecell[tr]{T1 (37\%), T2 (37\%), \\ FLAIR (26\%)} & XGY & 0.3T & 4 & 1.4k/183 & 25k\\
    SKM-TEA, 3D~\citep{desaiSKMTEADatasetAccelerated2021} & knee & sagittal & qDESS & GE & 3T & 8, 16 & 310/155 & 50k\\
    Stanford 3D~\citep{eppersonCreationFullySampled2013} & knee & axial & PDFS & GE & 3T & 8 & 19/19 & 6k\\
    Stanford 3D~\citep{eppersonCreationFullySampled2013} & knee & coronal & PDFS & GE & 3T & 8 & 19/19 & 6k\\
    Stanford 3D~\citep{eppersonCreationFullySampled2013} & knee & sagittal & PDFS & GE & 3T & 8 & 19/19 & 4.8k\\
    7T database, 3D~\citep{caanmatthanQuantitativeMotioncorrected7T2022} & brain & axial & MP2RAGE-ME & Philips & 7T & 32 & 385/77 & 112k\\
    7T database, 3D~\citep{caanmatthanQuantitativeMotioncorrected7T2022} & brain & coronal & MP2RAGE-ME & Philips & 7T & 32 & 385/77 & 112k\\
    7T database, 3D~\citep{caanmatthanQuantitativeMotioncorrected7T2022} & brain & sagittal & MP2RAGE-ME & Philips & 7T & 32 & 385/77 & 91k\\
    CC-359, 3D~\citep{souzaOpenMultivendorMultifieldstrength2018} & brain & axial & GRE & GE & 3T & 12 & 67/67 & 17k\\
    CC-359, 3D~\citep{souzaOpenMultivendorMultifieldstrength2018} & brain & coronal & GRE & GE & 3T & 12 & 67/67 & 14k\\
    \midrule
    CC-359, 3D~\citep{souzaOpenMultivendorMultifieldstrength2018} & brain & sagittal & GRE & GE & 3T & 12 & 67/67 & 11k\\
    Stanford 2D~\citep{chengStanford2DFSE2018} & various &various & various & GE & 3T & 3-32 & 89/89 & 2k\\
    NYU data~\citep{hammernikLearningVariationalNetwork2018a}  & knee & various & \makecell[tr]{PD (40\%), PDFS (20\%),\\ T2FS(40\%)} & Siemens & 3T & 15 & 100/20 & 3.5k\\
    M4Raw GRE~\citep{lyuM4RawMulticontrastMultirepetition2023a} & brain & axial & GRE & XGY & 0.3T & 4 & 366/183 & 6.6k\\
    \bottomrule
    \end{tabular}
    \end{adjustbox}
\end{table*}
\begin{figure*}[t]
    \centering
    \includegraphics[width=0.99\linewidth]{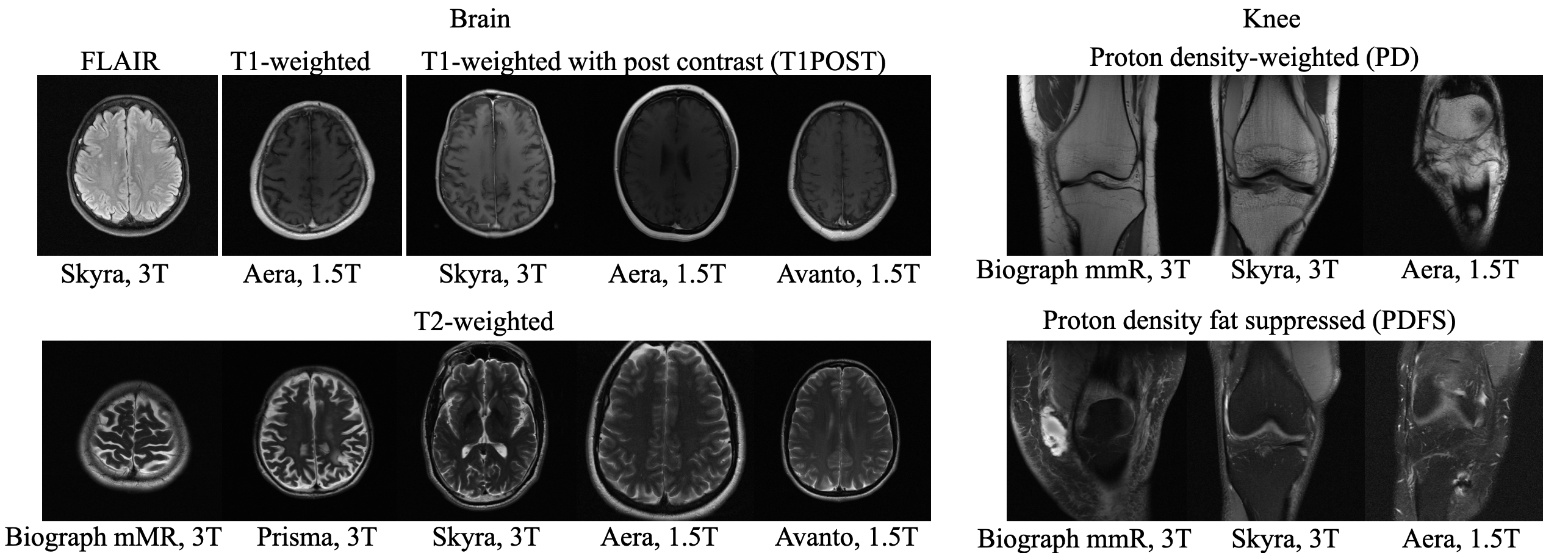}
    \caption{Example images for a selection of distributions from the fastMRI dataset~\citep{zbontarFastMRIOpenDataset2019a} we consider here. Axial view brain images are on the left, coronal view knee images are on the right. The caption above an image describes the image contrast, and the caption below is the name of the MRI scanner used.}
    \label{fig:sources}
\end{figure*}

\section{Setup and Background} \label{sec:setup}
We consider multi-coil accelerated MRI, where the goal is to reconstruct a complex-valued image $\vx \in \complexset^N$ from measurements of electromagnetic signals obtained through $C$ receiver coils according to
\begin{align}
\label{eq:fwd_map}
    \vy_i = \mM\mF\mS_i\vx + \vz_i \in \complexset^m, \quad i=1,\ldots,C.
\end{align}
Here, $\mS_i$ is the sensitivity map of the $i$-th coil, $\mF$ is the 2D discrete Fourier transform, $\mM$ is an undersampling mask, and $\vz_i$ models additive white Gaussian noise. The measurements $\vy_i$ are often called k-space measurements. 

In this work, we consider 4-fold accelerated (i.e., $m = N/4$) multi-coil 2D MRI reconstruction with Cartesian undersampling. The central k-space region is fully sampled including 8\% of all k-space lines, and the remaining lines are sampled equidistantly with a random offset from the start. We choose 4-fold acceleration as going beyond 4-fold acceleration, radiologists tend to reject the reconstructions by neural networks and other methods as not sufficiently good~\citep{muckleyResults2020FastMRI2021, radmaneshExploringAccelerationLimits2022}. Equidistant sampling is chosen due to the ease of implementation on existing machines~\citep{zbontarFastMRIOpenDataset2019a}. 

\vsparagraph{Class of reconstruction methods.} We focus on deep learning models trained end-to-end for accelerated MRI because they give state-of-the-art performance in accuracy and speed~\citep{hammernikLearningVariationalNetwork2018a, aggarwalMoDLModelBasedDeep2019, sriramEndtoEndVariationalNetworks2020b, fabianHUMUSNetHybridUnrolled2022}. 
There are different methods to image reconstruction with neural networks including un-trained neural networks~\cite{ulyanovDeepImagePrior2020a,heckelDeepDecoderConcise2019,darestaniAcceleratedMRIUnTrained2021} and methods based on generative neural networks~\cite{boraCompressedSensingUsing2017a,jalalRobustCompressedSensing2021,zachStableDeepMRI2023}.  

A neural network $f_\vth$ with parameters $\vth$ mapping measurements $\vy = \{\vy_1,\ldots, \vy_C\}$ to an image is commonly trained to reconstruct an image from the measurements $\vy$ by minimizing the supervised loss $\mathcal L(\vth) = \sum_{i=1}^n \text{loss}(f_\vth(\vy_i),\vx_i)$
over a training set consisting of target images and corresponding measurements $\{(\vx_1,\vy_1),\ldots, (\vx_n,\vy_n)\}$. 
This dataset is typically constructed from fully-sampled k-space data (i.e., where the undersampling mask $M$ is identity). From the fully-sampled data, a target image $\vx$ is estimated, and retrospectively undersampled measurements $\vy$ are generated by applying the undersampling mask to the fully-sampled data.

Several choices of network architectures work well. A standard baseline is a U-net~\citep{ronnebergerUNetConvolutionalNetworks2015a} trained to reconstruct the image from a coarse least-squares reconstruction of the measurements \citep{zbontarFastMRIOpenDataset2019a}. A vision transformer~\cite{dosovitskiyImageWorth16x162021b} for image reconstruction applied in the same fashion as the U-net also works well~\citep{linVisionTransformersEnable2022}. 
The best-performing models are unrolled networks such as the variational network~\citep{hammernikLearningVariationalNetwork2018a} and a deep cascade of convolutional neural networks \citep{schlemperDeepCascadeConvolutional2018a}. The unrolled networks often use either the U-net as backbone, like the end-to-end VarNet~\citep{sriramEndtoEndVariationalNetworks2020b}, or a transformer based architecture~\citep{fabianHUMUSNetHybridUnrolled2022}. 

We expect our results in this paper to be model agnostic, and show that this is indeed the case for the U-net, ViT, and end-to-end VarNet.

\vsparagraph{Datasets.}
We consider the fully-sampled MRI dataset with varying attributes listed in Table~\ref{tbl:datasets}.
The datasets include the largest publicly available fully-sampled MRI datasets, and contain altogether around \textbf{500k slices}. 

Many of our experiments are based on splits of the fastMRI dataset~\citep{zbontarFastMRIOpenDataset2019a}, the most commonly used dataset for MRI reconstruction research. 
Figure~\ref{fig:sources} depicts samples from the fastMRI dataset and shows that MRI data vary in appearance across different anatomies and image contrasts (T1, T2, etc.). The image distribution also varies across vendors and magnetic field strengths of scanners, as the strength of the magnet impacts the signal-to-noise ratio (SNR), with stronger magnets leading to higher SNRs. 

The fastMRI dataset stands out for its diversity and size, making it particularly well-suited for exploring how different data distributions can affect the performance of deep learning models for accelerated MRI. In our experiments in  Section~\ref{sec:separate},~\ref{sec:robustness},~\ref{sec:pathology}, and~\ref{sec:dist_overfit} we split the fastMRI dataset according to different attributes of the data. In Section~\ref{sec:robust_models}, we showcase the generalizability of our findings on a diverse collection of 17 different datasets.  

\begin{figure}[t!]
    \centering
    \begin{minipage}[c]{1\textwidth}
    \begin{minipage}[c]{0.32\linewidth}
        \begin{minipage}[c]{1\linewidth}
            \begin{tikzpicture}
                \node (x) at (0.0\linewidth,0.0\linewidth) {\includegraphics[width=0.6\linewidth]{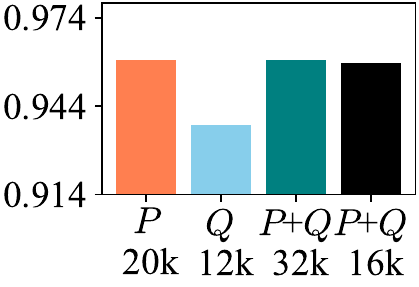}};
                \node[above = -0.03\linewidth of x, xshift=0.2\linewidth] {\footnotesize Anatomy};
                \node (a) [right = -0.04\linewidth of x, yshift=0.07\linewidth] {\includegraphics[width=0.26\linewidth, ]{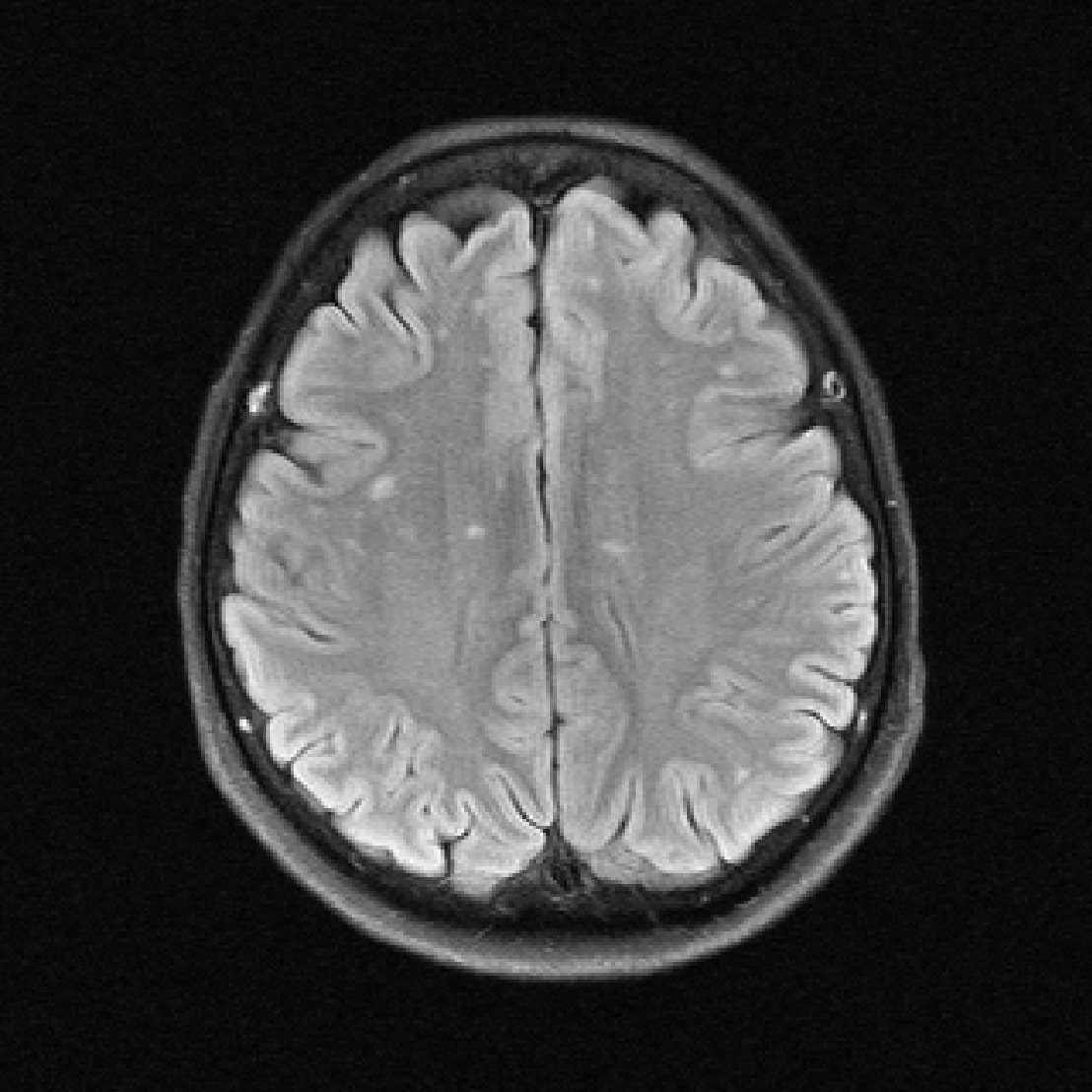}};
                \node[below = -0.03\linewidth of a, align=center, font=\scriptsize] {\textbf{$P$: Brain}};
                \node[rotate=90] [left = 0.01\linewidth of x, xshift=0.22\linewidth] {\scriptsize SSIM on $P$};
                
                \node (y) [below = -0.05\linewidth of x]  {\includegraphics[width=0.6\linewidth]{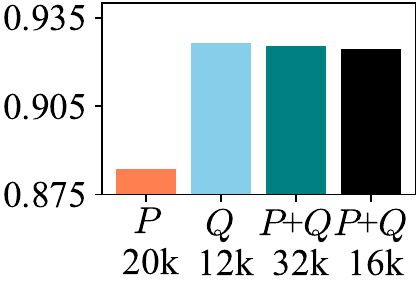}};
                \node (b) [right = -0.04\linewidth of y, yshift=0.07\linewidth]  {\includegraphics[width=0.26\linewidth, ]{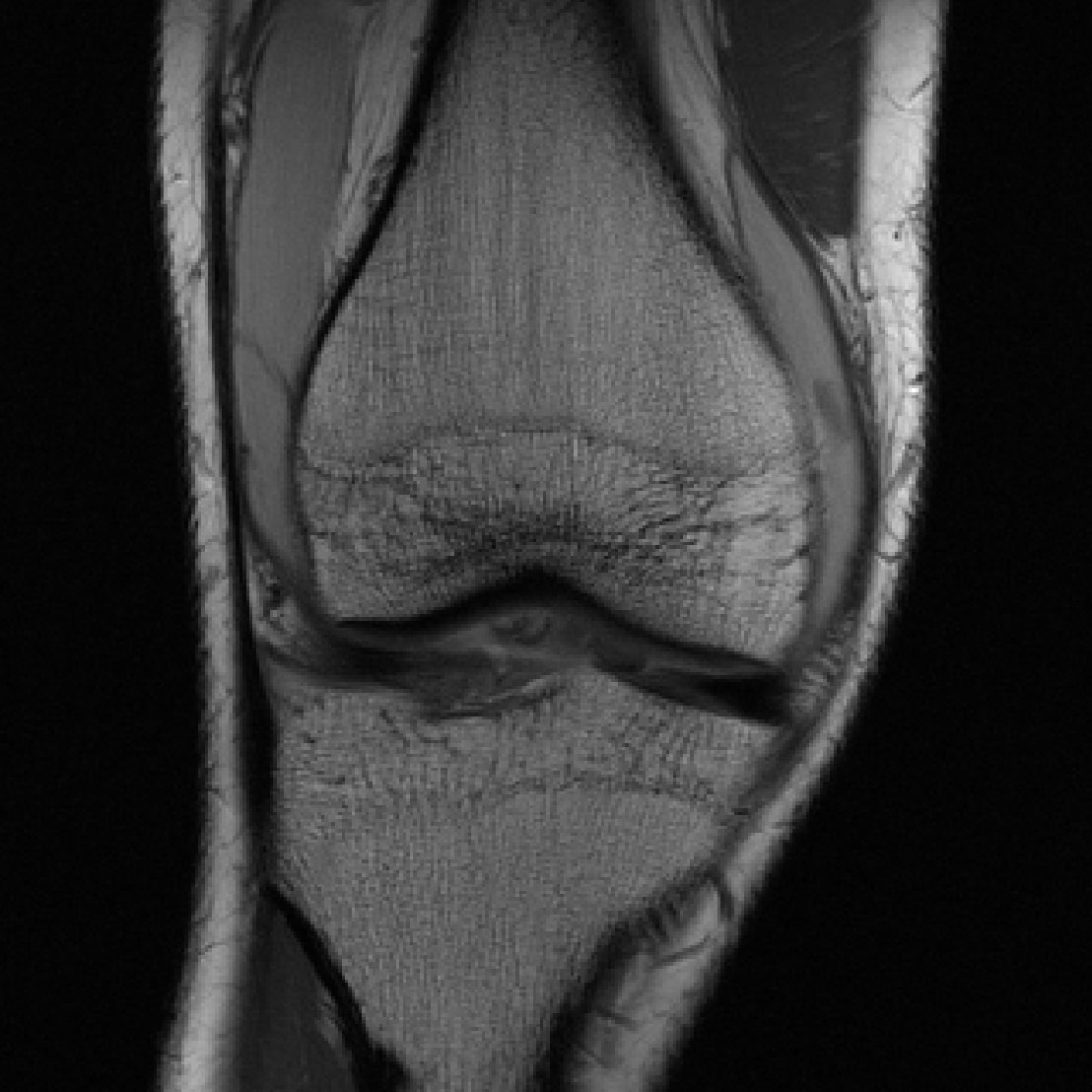}};
                \node[below = -0.03\linewidth of b, align=center, font=\scriptsize] {\textbf{$Q$: Knee}};
                \node[rotate=90] [left = 0.01\linewidth of y, xshift=0.22\linewidth] {\scriptsize SSIM on $Q$};
            \end{tikzpicture}
        \end{minipage}
    \end{minipage}\hfill
    \begin{minipage}[c]{0.32\linewidth}
        \begin{minipage}[c]{1\linewidth}
            \begin{tikzpicture}
                \node (x) at (0.0\linewidth,0.0\linewidth) {\includegraphics[width=0.6\linewidth]{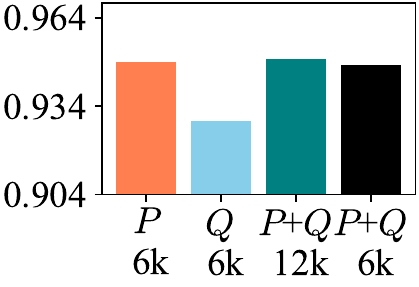}};
                \node[above = -0.03\linewidth of x, xshift=0.2\linewidth] {\footnotesize Contrasts};
                \node (a) [right = -0.04\linewidth of x, yshift=0.07\linewidth]  {\includegraphics[width=0.26\linewidth, ]{figs/imgs/CORPD_FBK-Skyra.jpg}};
                \node[below = -0.03\linewidth of a, align=center, font=\scriptsize] {\textbf{$P$: PD}};
                \node[rotate=90] [left = 0.01\linewidth of x, xshift=0.22\linewidth] {\scriptsize SSIM on $P$};
                
                \node (y) [below = -0.05\linewidth of x]  {\includegraphics[width=0.6\linewidth]{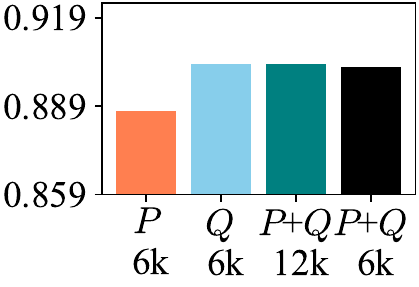}};
                \node (b) [right = -0.04\linewidth of y, yshift=0.07\linewidth]  {\includegraphics[width=0.26\linewidth, ]{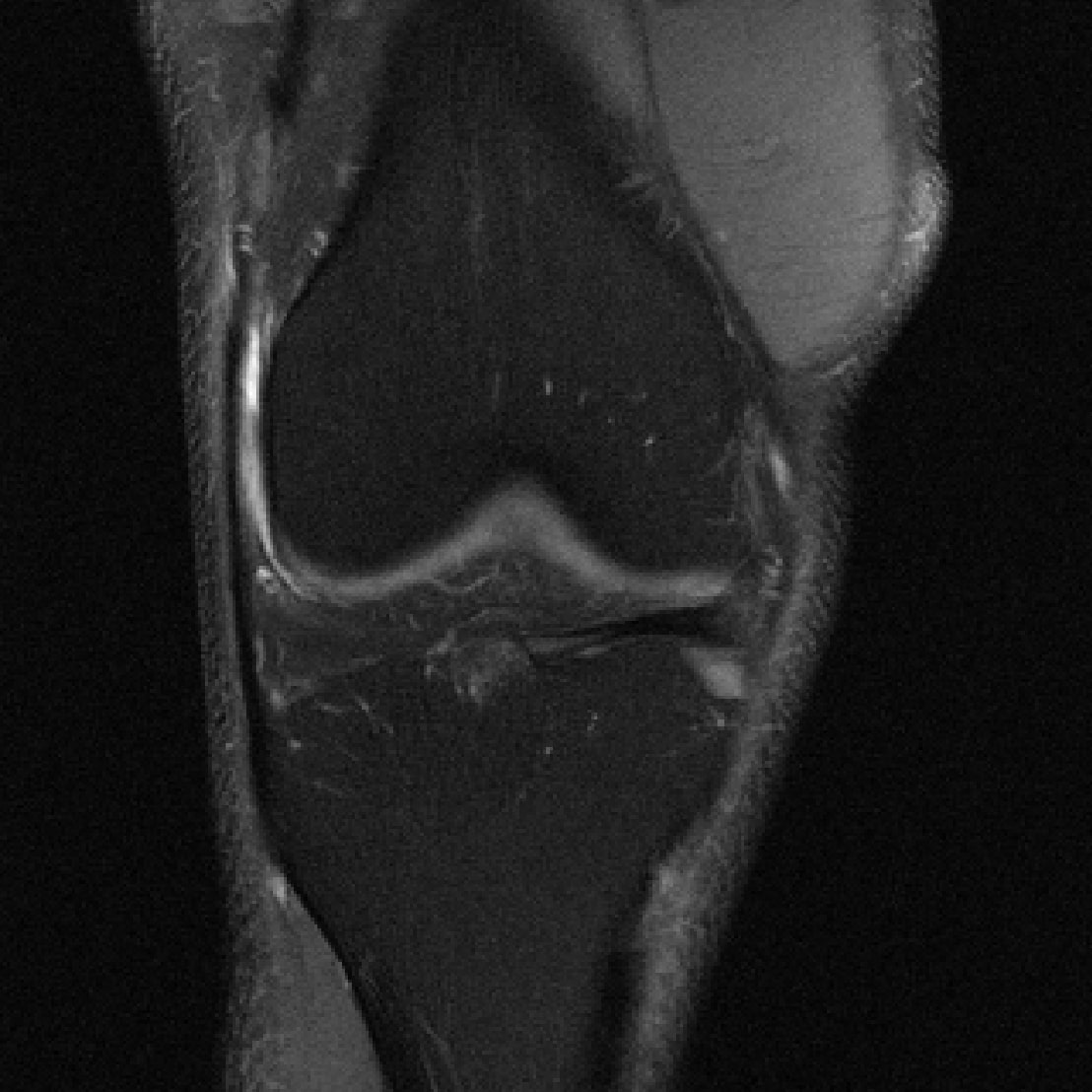}};
                \node[below = -0.03\linewidth of b, align=center, font=\scriptsize] {\textbf{$Q$: PDFS}};
                \node[rotate=90] [left = 0.01\linewidth of y, xshift=0.22\linewidth] {\scriptsize SSIM on $Q$};
            \end{tikzpicture}
        \end{minipage}
    \end{minipage}\hfill
     \begin{minipage}[c]{0.32\linewidth}
        \begin{minipage}[c]{1\linewidth}
            \begin{tikzpicture}
                \node (x) at (0.0\linewidth,0.0\linewidth) {\includegraphics[width=0.6\linewidth]{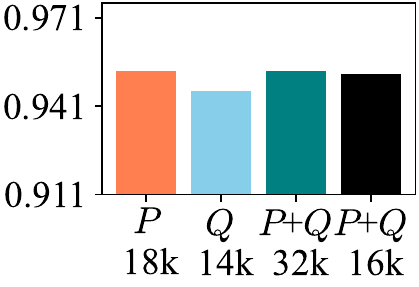}};
                \node[above = -0.03\linewidth of x, xshift=0.2\linewidth] {\footnotesize Magnetic field};
                \node (a) [right = -0.04\linewidth of x, yshift=0.07\linewidth]  {\includegraphics[width=0.26\linewidth, ]{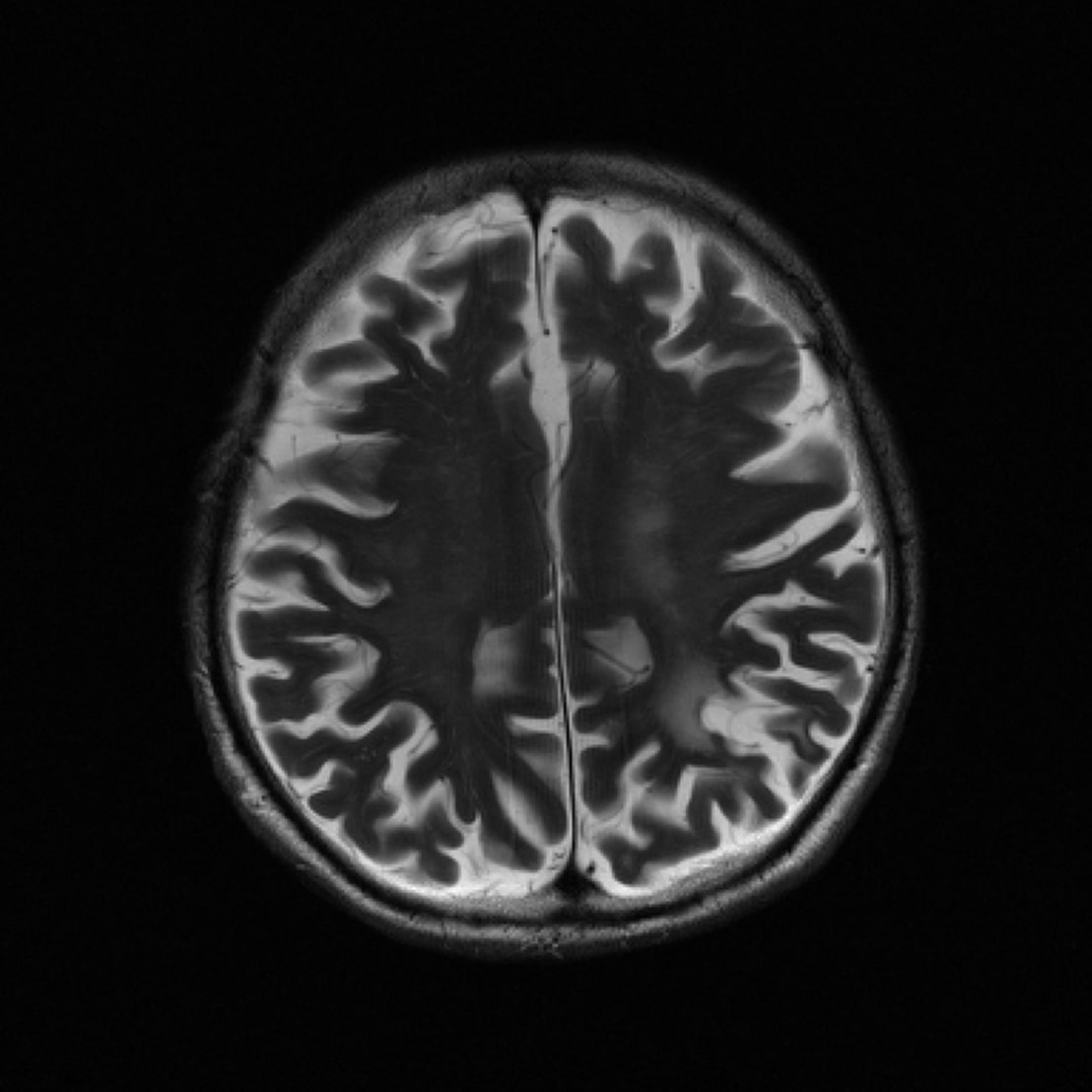}};
                \node[below = -0.03\linewidth of a, align=center, font=\scriptsize] {\textbf{$P$: 3.0T}};
                \node[rotate=90] [left = 0.01\linewidth of x, xshift=0.22\linewidth] {\scriptsize SSIM on $P$};
                
                \node (y) [below = -0.05\linewidth of x]  {\includegraphics[width=0.6\linewidth]{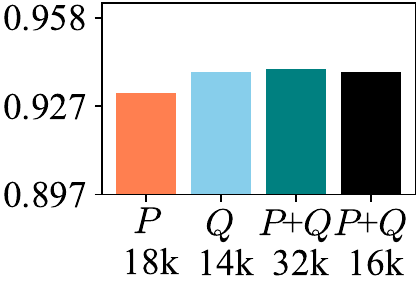}};
                \node (b) [right = -0.04\linewidth of y, yshift=0.07\linewidth]  {\includegraphics[width=0.26\linewidth, ]{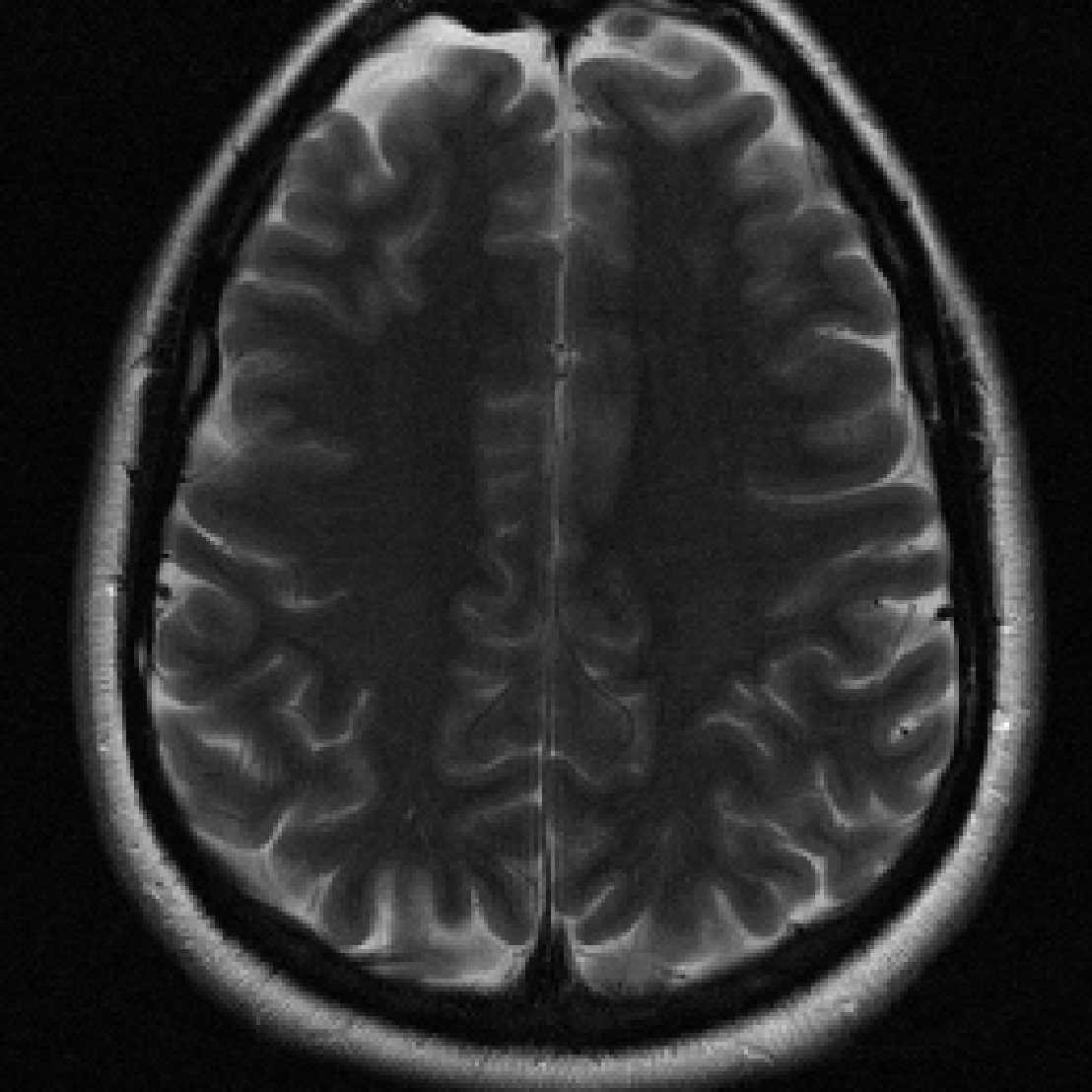}};
                \node[below = -0.03\linewidth of b, align=center, font=\scriptsize] {\textbf{$Q$: 1.5T}};
                \node[rotate=90] [left = 0.01\linewidth of y, xshift=0.22\linewidth] {\scriptsize SSIM on $Q$};
            \end{tikzpicture}
        \end{minipage}
    \end{minipage}\hfill
    \caption{
     The {\color{orange} \textbf{orange}} and {\color{skyblue} \textbf{blue}} bars are the U-net models trained exclusively on data from {\color{orange} $P$} ({\color{orange} $\setD_P$}) and {\color{skyblue} $Q$} ({\color{skyblue} $\setD_Q$}), respectively, and the {\color{teal} \textbf{teal}} bars are the models trained on both sets {\color{teal}$\setD_P \cup \setD_Q$}. 
    As a reference point, the {\color{black} \textbf{black}} bars are the performance of models trained on random samples of $\setD_P \cup \setD_Q$ of \textbf{half the size}. Below each bar is the total number of training images. We are in the high-data regime where increasing the dataset further gives only minor improvements. 
    For all distributions, the joint model trained on $P$ and $Q$ performs as well on $P$ and $Q$ as the models trained individually for each of those distributions. 
    }
    \label{fig:sep_unet}
    \end{minipage}\\
    \begin{minipage}[c]{1\textwidth}
        \begin{minipage}[c]{0.32\textwidth}
            \begin{minipage}[c]{1\textwidth}
                \begin{tikzpicture}
                    \node (x) at (0.0\textwidth,0.0\textwidth) {\includegraphics[width=0.6\textwidth]{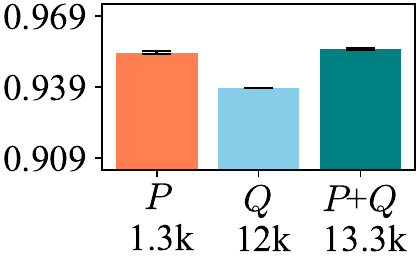}};
                    \node[above = -0.03\textwidth of x, xshift=0.2\textwidth] {\footnotesize Anatomy};
                    \node (a) [right = -0.04\textwidth of x, yshift=0.07\textwidth] {\includegraphics[width=0.22\textwidth, ]{figs/imgs/AXFLAIR-Skyra.jpg}};
                    \node[below = -0.03\textwidth of a, align=center, font=\scriptsize] {\textbf{$P$: Brain}};
                    \node[rotate=90] [left = 0.01\textwidth of x, xshift=0.22\textwidth] {\scriptsize SSIM on $P$};

                    \node (y) [below = -0.05\textwidth of x]  {\includegraphics[width=0.6\textwidth]{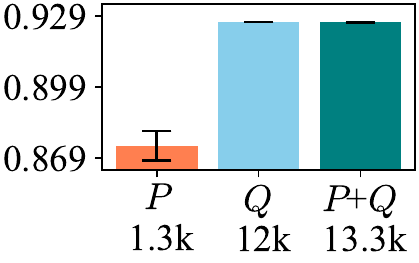}};
                    \node (b) [right = -0.04\textwidth of y, yshift=0.07\textwidth]  {\includegraphics[width=0.22\textwidth, ]{figs/imgs/CORPD_FBK-Skyra.jpg}};
                    \node[below = -0.03\textwidth of b, align=center, font=\scriptsize] {\textbf{$Q$: Knee}};
                    \node[rotate=90] [left = 0.01\textwidth of y, xshift=0.22\textwidth] {\scriptsize SSIM on $Q$};
                \end{tikzpicture}
            \end{minipage}
        \end{minipage}\hfill
        \begin{minipage}[c]{0.32\textwidth}
            \begin{minipage}[c]{1\textwidth}
                \begin{tikzpicture}
                    \node (x) at (0.0\textwidth,0.0\textwidth) {\includegraphics[width=0.6\textwidth]{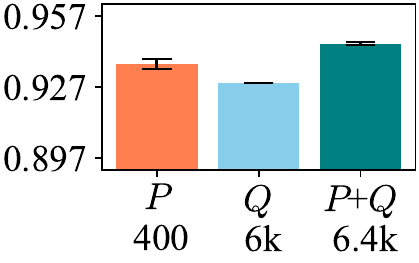}};
                    \node[above = -0.03\textwidth of x, xshift=0.2\textwidth] {\footnotesize Contrasts};
                    \node (a) [right = -0.04\textwidth of x, yshift=0.07\textwidth]  {\includegraphics[width=0.22\textwidth, ]{figs/imgs/CORPD_FBK-Skyra.jpg}};
                    \node[below = -0.03\textwidth of a, align=center, font=\scriptsize] {\textbf{$P$: PD}};
                    \node[rotate=90] [left = 0.01\textwidth of x, xshift=0.22\textwidth] {\scriptsize SSIM on $P$};
                    
                    \node (y) [below = -0.05\textwidth of x]  {\includegraphics[width=0.6\textwidth]{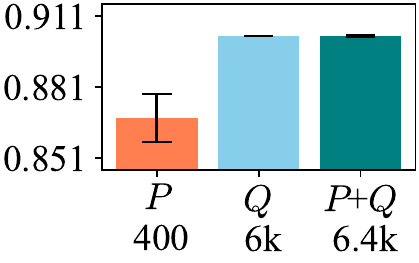}};
                    \node (b) [right = -0.04\textwidth of y, yshift=0.07\textwidth]  {\includegraphics[width=0.22\textwidth, ]{figs/imgs/CORPDFS_FBK-Skyra.jpg}};
                    \node[below = -0.03\textwidth of b, align=center, font=\scriptsize] {\textbf{$Q$: PDFS}};
                    \node[rotate=90] [left = 0.01\textwidth of y, xshift=0.22\textwidth] {\scriptsize SSIM on $Q$};
                \end{tikzpicture}
            \end{minipage}
        \end{minipage}\hfill
         \begin{minipage}[c]{0.32\textwidth}
            \begin{tikzpicture}
                \node (x) at (0.0\textwidth,0.0\textwidth) {\includegraphics[width=0.6\textwidth]{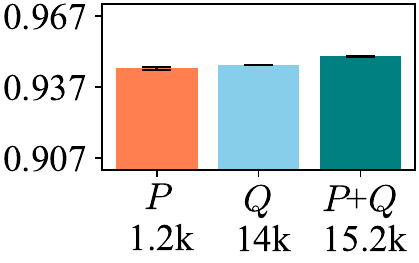}};
                \node[above = -0.03\textwidth of x, xshift=0.2\textwidth] {\footnotesize Magnetic field};
                \node (a) [right = -0.04\textwidth of x, yshift=0.07\textwidth]  {\includegraphics[width=0.22\textwidth, ]{figs/imgs/AXT2-Prisma_fit.jpg}};
                \node[below = -0.03\textwidth of a, align=center, font=\scriptsize] {\textbf{$P$: 3.0T}};
                \node[rotate=90] [left = 0.01\textwidth of x, xshift=0.22\textwidth] {\scriptsize SSIM on $P$};
                
                \node (y) [below = -0.05\textwidth of x]  {\includegraphics[width=0.6\textwidth]{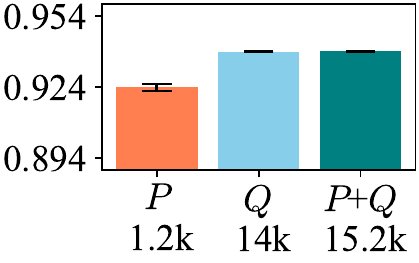}};
                \node (b) [right = -0.04\textwidth of y, yshift=0.07\textwidth]  {\includegraphics[width=0.22\textwidth, ]{figs/imgs/AXT2-Aera.jpg}};
                \node[below = -0.03\textwidth of b, align=center, font=\scriptsize] {\textbf{$Q$: 1.5T}};
                \node[rotate=90] [left = 0.01\textwidth of y, xshift=0.22\textwidth] {\scriptsize SSIM on $Q$};
            \end{tikzpicture}
        \end{minipage}
    \caption{Training a single model (here U-net) on a slightly skewed dataset does not harm performance on the individual data distributions. The number below each bar is the number of training examples used. We report the mean $\pm$ two standard deviations from five runs, each with a different random seed for sampling training data from $P$ and model initialization. 
    We note for training sets exceeding 3k images there is next to no variation (see Figure~\ref{fig:sep_unet_ci} in the Appendix); therefore, we only have error bars for this experiment which includes training runs on small datasets. 
    }
    \label{fig:sep_skewed}
    \end{minipage}
\end{figure}

\section{Training a Single Model or Separate Models on Different Distributions} \label{sec:separate}
We start with studying whether training a model on data from a diverse set of distributions compromises the performance on the individual distributions. In its simplest instance, the question is whether a model for image reconstruction trained on data from both distributions $P$ and $Q$ performs as well on distributions $P$ and $Q$ as a model trained on $P$ and applied on $P$ and a model trained on $Q$ and applied on $Q$. 

In general, this depends on the distributions $P$ and $Q$, and on the estimator. For example, consider a simple toy denoising problem, where the data from distribution $P$ is generated as $\vy = \vx + \ve$, with $\vx$ is drawn i.i.d. from the unit sphere of a subspace, and $\ve$ is drawn from a zero-mean Gaussian with co-variance matrix $\sigma_P\mI$. Data for distribution $Q$ is generated equally, but the noise is drawn from a zero-mean distribution with different noise variance, i.e., $\ve \sim \mc N(0,\sigma_Q^2 \mI)$ with $\sigma_P^2 \neq \sigma_Q^2$. 
Then the optimal linear estimator learned from data drawn from both distribution $P$ and $Q$ is sub-optimal for both distributions $P$ and $Q$. However, there exists a non-linear estimator that is as good as the optimal linear estimator on distribution $P$ and distribution $Q$. 

In addition, conventional approaches to MRI such as $\ell_1$-regularized least-squares need to be tuned individually on different distributions to achieve best performance, as discussed in Appendix~\ref{app:hyperpl1}. 

Thus it is unclear whether it is preferable to train a neural network for MRI on diverse data from many distributions or to train several networks and use them for each individual distribution. For example, is it better to train a network specific for knees and another one for brains or to train a single network on knees and brains together? 
Here, we find that training a network on several distributions simultaneously does not compromise performance on the individual distribution relative to training one model for each distribution.  

\vsparagraph{Experiments for training a joint or separate models.}
We consider two distributions $P$ and $Q$, and train U-nets~\citep{ronnebergerUNetConvolutionalNetworks2015a}, ViTs~\citep{dosovitskiyImageWorth16x162021b} and end-to-end VarNets~\citep{sriramEndtoEndVariationalNetworks2020b} on data $\setD_P$ from distributions $P$ and on data $D_Q$ from distribution $Q$ separately. We also train the same models on data from $P$ and $Q$, i.e., $\setD_P \cup \setD_Q$. We then evaluate on separate test sets from distribution $P$ and $Q$.  
We consider the end-to-end VarNet because it is a state-of-the-art model for accelerated MRI, and consider the U-net and ViT as popular baseline models. This diverse selection of architectures (unrolled, convolutional, transformer) aims to demonstrate that our qualitative results are independent of the specific architectural choice.
We consider the following choices for $\setD_P$ and $\setD_Q$, which are subsets of the fastMRI dataset specified in Figure~\ref{fig:sources}:
\begin{itemize}
    \item \textbf{Anatomies.} $P$ are knees scans collected with 6 different combinations of image contrasts and scanners and $Q$ are the brain scans collected with 10 different combinations of image contrasts and scanners. 
    \item \textbf{Contrasts.} We select $P$ as PD-weighted knee images from 3 different scanners and $Q$ are PDFS-weighted knee images from the same 3 scanners. 
    \item \textbf{Magnetic field.} Here, we pick $P$ to contain all 3.0T scanners and $Q$ to contain all 1.5T scanners regardless of anatomy or image contrast.
\end{itemize}

Figure~\ref{fig:sep_unet} shows for U-net that the models trained on both $P$ and $Q$ achieve essentially the same performance on both $P$ and $Q$ as the individual models. 
The model trained on both $P+Q$ uses more examples than the model trained on $P$ and $Q$ individually. To rule out the possibility that the joint model is only as good as the individual models since it is trained on more examples, we also trained a model on $P+Q$ with half the number of examples (obtained by randomly subsampling). Again, the model performs essentially equally well as the other models. We refer to Appendix~\ref{app:dataprep_fm} and~\ref{app:models_training} for details regarding the setup.

Results for VarNet and ViT are qualitatively the same as the results in Figure~\ref{fig:sep_unet} for U-net (see Appendix~\ref{app:single_vs_separate}), and indicate that our findings are architecture-independent.

Thus, separating datasets into data from individual distributions and training individual models does not yield benefits, unlike for $\ell_1$-regularized least squares 
or the toy-subspace example. 

\vsparagraph{Experiments for training a joint or separate models on skewed data.}
Next, we consider skewed data, i.e., the training set $\setD_P$ is by a factor of about $10$ smaller than the training set $\setD_Q$. The choices for distributions $P$ and $Q$ are as in the previous experiment.
Figure~\ref{fig:sep_skewed} shows that even for data skewed by a factor 10, the performance on distributions $P$ and $Q$ of models (U-net) trained on both distributions is comparable to the models trained on the individual distributions. 

\begin{figure}[t]
    \centering
    \begin{minipage}[c]{0.01\textwidth}
        \begin{tikzpicture}
            \node[rotate=90] at (-0.2,0) {\scriptsize SSIM on $Q$};
        \end{tikzpicture}
    \end{minipage}
    \begin{minipage}[c]{0.32\textwidth}
        \begin{tikzpicture}
            \node (x) at (0.0\textwidth,0.0\textwidth) {\includegraphics[width=0.63\textwidth]{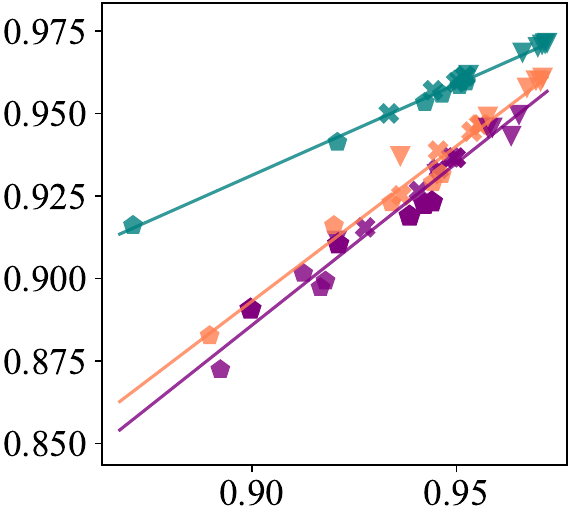}};
            \node[above = -0.04\textwidth of x, xshift=0.05\textwidth] {\scriptsize Anatomy shift};
            \node[below = -0.04\textwidth of x, xshift=0.05\textwidth] {\scriptsize SSIM on $P_\text{best}$};
            \node (a) [right = 0.04\textwidth of x, yshift=0.18\textwidth]  {\includegraphics[width=0.2\textwidth, ]{figs/imgs/CORPD_FBK-Skyra.jpg}};   
            \node (b) [right = 0.01\textwidth of x, yshift=0.15\textwidth]  {\includegraphics[width=0.2\textwidth, ]{figs/imgs/CORPDFS_FBK-Skyra.jpg}};
            \node[above = 0.0\textwidth of b, align=center, font=\scriptsize] {$P$: Knee};
            \node (c) [right = 0.01\textwidth of x, yshift=-0.15\textwidth] {\includegraphics[width=0.2\textwidth, ]{figs/imgs/AXT2-Aera.jpg}};
            \node[above = -0.04\textwidth of c] {\scriptsize $Q$: Brain};
        \end{tikzpicture}
    \end{minipage}
    \begin{minipage}[c]{0.32\textwidth}
        \begin{tikzpicture}
            \node (x) at (0.0\textwidth,0.0\textwidth) {\includegraphics[width=0.6\textwidth]{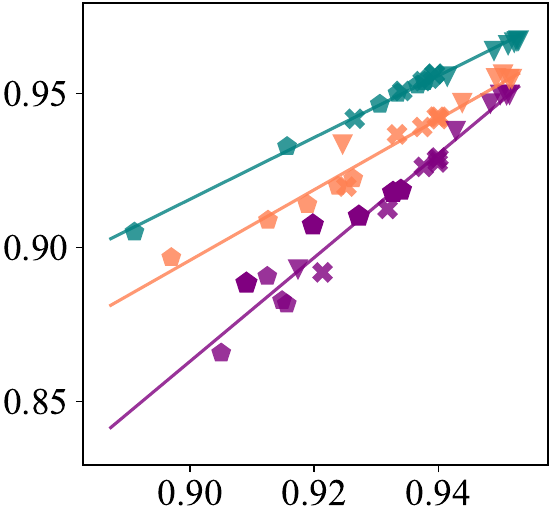}};
            \node[above = -0.04\textwidth of x, xshift=0.05\textwidth] {\scriptsize Contrast shift};
            \node[below = -0.04\textwidth of x, xshift=0.05\textwidth] {\scriptsize SSIM on $P_\text{best}$};
            \node (a) [right = 0.04\textwidth of x, yshift=0.18\textwidth]  {\includegraphics[width=0.2\textwidth, ]{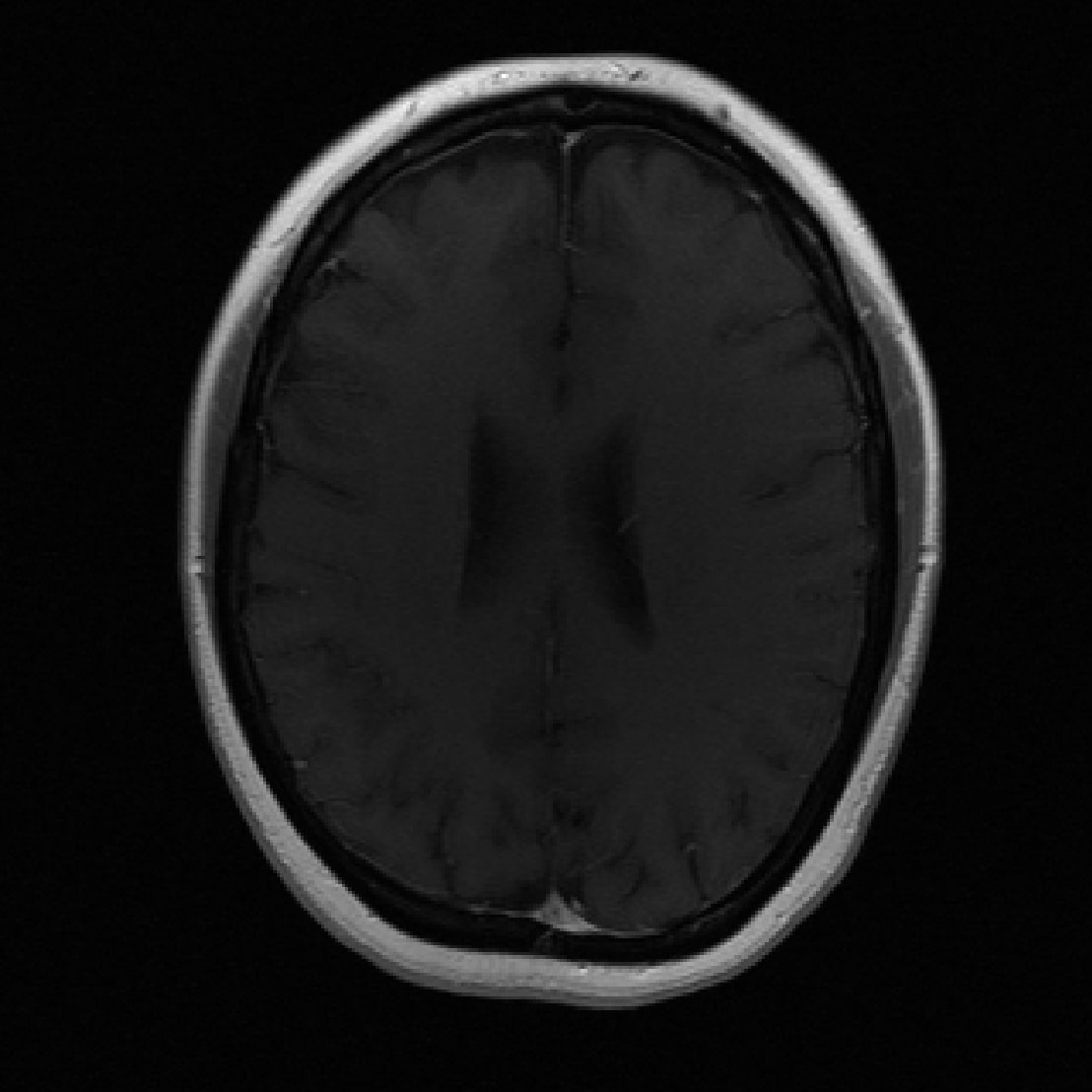}};   
            \node (b) [right = 0.01\textwidth of x, yshift=0.15\textwidth]  {\includegraphics[width=0.2\textwidth, ]{figs/imgs/AXFLAIR-Skyra.jpg}};
            \node[above = 0.0\textwidth of b, align=center, font=\scriptsize] {$P$: non-T2};
            \node (c) [right = 0.01\textwidth of x, yshift=-0.15\textwidth] {\includegraphics[width=0.2\textwidth, ]{figs/imgs/AXT2-Aera.jpg}};
            \node[above = -0.04\textwidth of c] {\scriptsize $Q$: T2};
        \end{tikzpicture}
    \end{minipage}
    \begin{minipage}[c]{0.32\textwidth}
        \begin{tikzpicture}
            \node (x) at (0.0\textwidth,0.0\textwidth) {\includegraphics[width=0.6\textwidth]{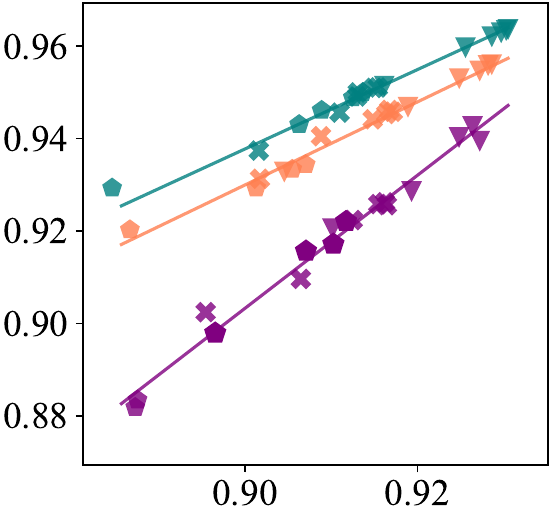}};
            \node[above = -0.04\textwidth of x, xshift=0.05\textwidth] {\scriptsize Magnetic field shift};
            \node[below = -0.04\textwidth of x, xshift=0.05\textwidth] {\scriptsize SSIM on $P_\text{best}$};
            \node (a) [right = 0.04\textwidth of x, yshift=0.18\textwidth]  {\includegraphics[width=0.2\textwidth, ]{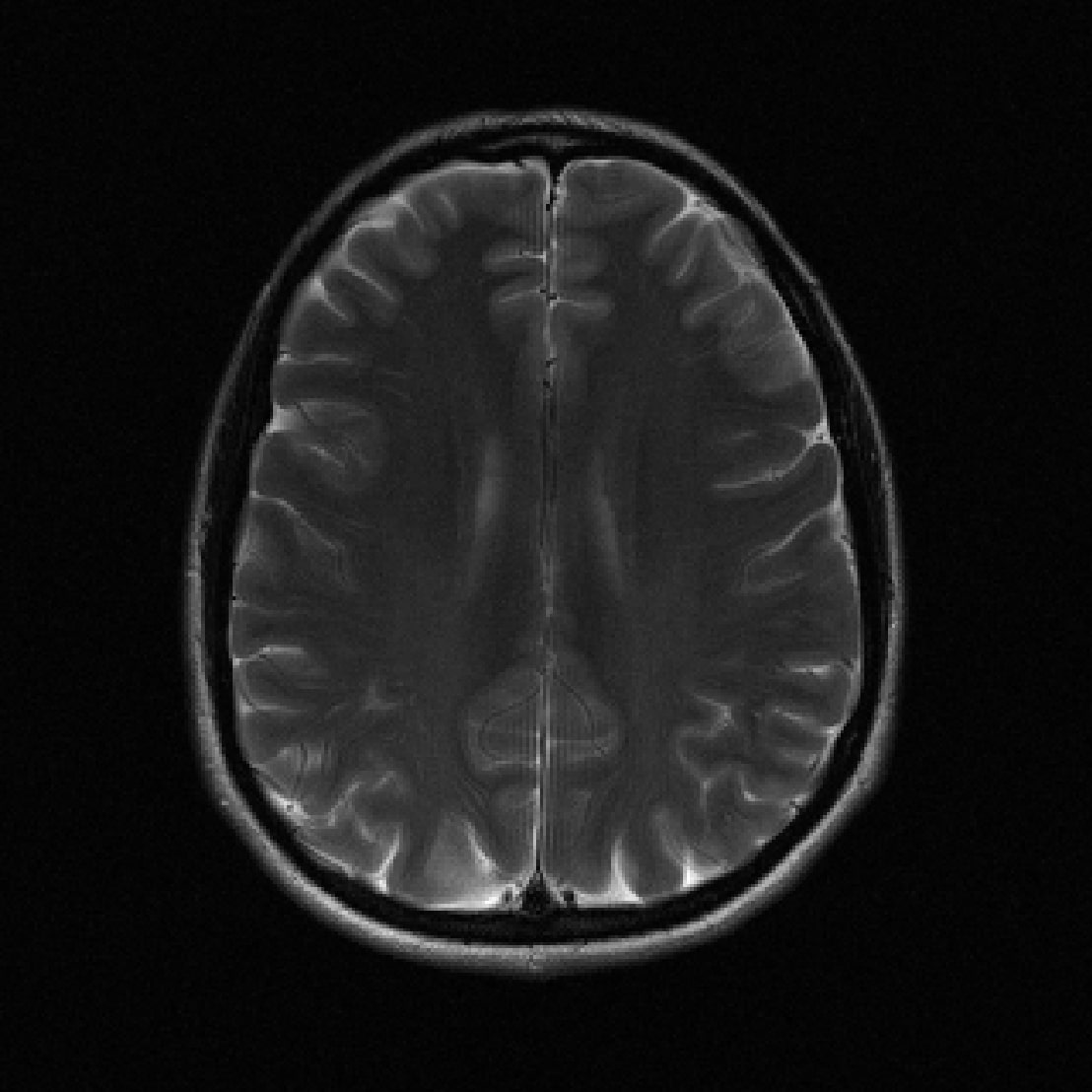}};   
            \node (b) [right = 0.01\textwidth of x, yshift=0.15\textwidth]  {\includegraphics[width=0.2\textwidth, ]{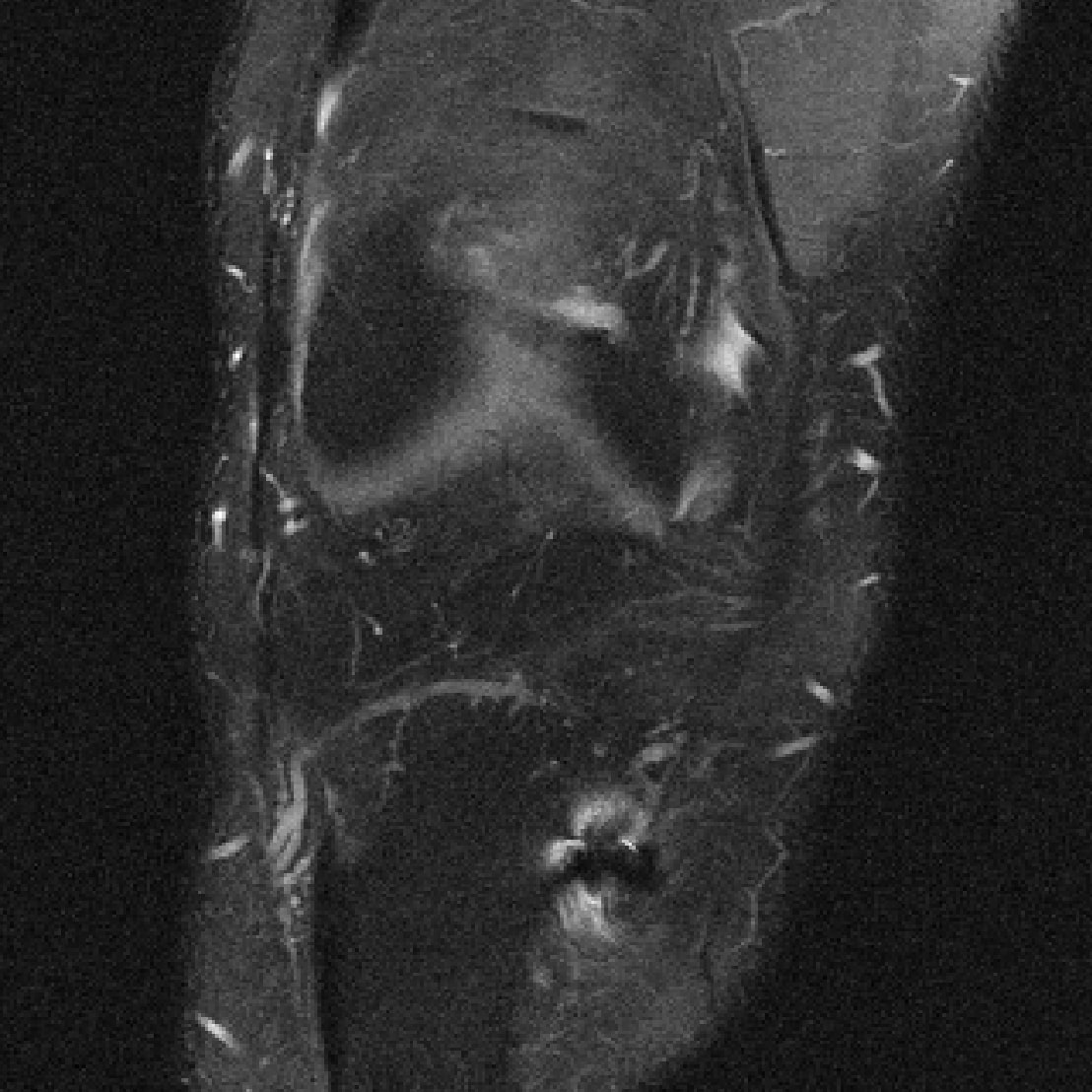}};
            \node[above = 0.0\textwidth of b, align=right, font=\scriptsize] {$P$: 1.5T};
            \node (c) [right = 0.01\textwidth of x, yshift=-0.15\textwidth] {\includegraphics[width=0.2\textwidth, ]{figs/imgs/CORPDFS_FBK-Skyra.jpg}}; 
            \node[above = -0.04\textwidth of c] {\scriptsize $Q$: 3T};
        \end{tikzpicture}
    \end{minipage}\\
    \begin{minipage}[c]{0.95\textwidth}
        \hspace{0.03\textwidth}\frame{\includegraphics[width=0.9\textwidth]{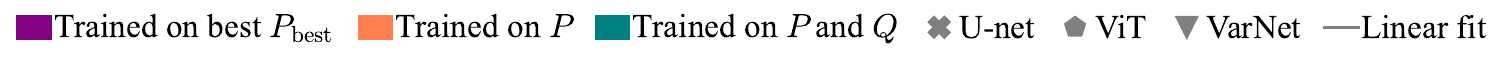}}
    \end{minipage}
    \caption{For a distribution-shift from distributions $P = \{P_1, \ldots, P_m\}$ to distribution $Q$, we compare robustness of models trained on $P$ ({\color{orange}\textbf{orange}}) to baselines trained on the best single distribution $P_\text{best}$ ({\color{violet}\textbf{violet}}). As additional reference, we also report models trained on both $P$ and $Q$ to imitate ideally robust models ({\color{teal}\textbf{teal}}). 
    For the three distribution-shifts shown, training on the more diverse dataset $P$ is beneficial compared to training on $P_\text{best}$ alone.
    }
    \label{fig:robust_diverse_non-diverse}
\end{figure}
\section{Data Diversity Enhances Robustness Towards Distribution-Shifts} \label{sec:robustness}
We now study how training on diverse data affects the out-of-distribution performance of a model. 
\citet{nguyenQualityNotQuantity2022a} note that for image recognition two outcomes can be expected from training a model jointly on two distributions $P_1$ and $P_2$, and evaluating on another distribution $Q$. Assuming that training a model on $P_1$ gives better performance on $Q$ than training a model on $P_2$, the model trained on $P_1$ and $P_2$ could perform (i) at least as well as a model trained on $P_1$ or (ii) performs worse than the model trained on $P_1$ but better than a model trained on $P_2$. 
Here, we find that for accelerated MRI, training a model on diverse data improves a model's out-of-distribution performance, i.e., case (i).

\paragraph{Measuring robustness.}
Our goal is to measure the expected robustness gain by training models on diverse data, and we would like this measure to be independent of the model itself. Hence, we measure robustness with the notion of `effective robustness' by~\citet{taoriMeasuringRobustnessNatural2020a}. 
We evaluate models on a standard in-distribution test set (i.e., data from the same source that generated the training data) and on an out-of-distribution test set. 
We then plot the out-of-distribution performance for different models as a function of the in-distribution performance, see Figure~\ref{fig:robust_diverse_non-diverse}. 

It can be seen that the in- and out-of-distribution performance of models trained on data from one distribution,
(e.g., in-distribution data \textcolor{violet}{\textbf{violet}}) is well described by a linear fit. 
If a robustness intervention only moves a model along the line, it doesn’t increase out-of-distribution performance beyond what’s expected for a given in-distribution performance and therefore does not yield effective robustness. Thus, a dataset yields more effective robustness if models trained on it lie above the \textcolor{violet}{\textbf{violet}} line, since such models have higher out-of-distribution performance than what's expected for a fixed in-distribution performance.

\vsparagraph{Experiment.}
We are given data from two distributions $P$ and $Q$, where distribution $P$ can be split up into $m$ sub-distributions $P_1,\ldots,P_m$. We consider the following choices for the two distributions, all based on the knee and brain fastMRI datasets illustrated in Figure~\ref{fig:sources}:
\begin{itemize}
    \item {\bf Anatomy shift:} $P_1,\ldots,P_6$ is knee data collected with all 6 different combinations of image contrasts and scanners, and $Q$ are the different brain datasets collected with 8 different combinations of image contrasts (FLAIR, T1, T1POST, T2) and scanners (Skyra, Prisma, Aera, Biograph mMR).
    \item {\bf Contrast shift:} $P_1,\ldots,P_5$ are all FLAIR, T1POST, or T1 brain images and $Q$ are T2 brain data. 
    \item {\bf Magnetic field shift:} $P_1,\ldots,P_7$ are brain and knee data collected with 1.5T scanners (Aera, Avanto) and $Q$ are brain and knee data collected with 3T scanners (Skyra, Prisma, Biograph mMR).
\end{itemize}

For each of the distributions $P_1,\ldots,P_m$ we construct a training set with 2048 images and a test set with 128 images. 

We then train U-nets on each of the distributions $P_1, \ldots, P_m$ separately and select from these distributions the distribution $P_\text{best}$ that maximizes the performance of the U-net on a test set from the distribution $Q$.

Now, we train a variety of different model architectures including the U-net, end-to-end VarNet~\citep{sriramEndtoEndVariationalNetworks2020b}, and vision transformer (ViT) for image reconstruction~\citep{linVisionTransformersEnable2022} on data from the distribution $P_\text{best}$, data from the distribution $P$ (which contains $P_\text{best}$), and data from the distribution $P$ and $Q$. 
We also sample different models by early stopping and by decreasing the training set size by four. 
We plot the performance of the models evaluated on the distribution $Q$ as a function of their performance evaluated on the distribution $P_\text{best}$. The configurations of our models are the same as in Section~\ref{sec:separate}.

\begin{figure}[t!]
    \centering
    \begin{minipage}{0.75\linewidth}
    \centering
        \begin{tikzpicture}
            \node (x) at (0,0) {\includegraphics[width=0.24\linewidth]{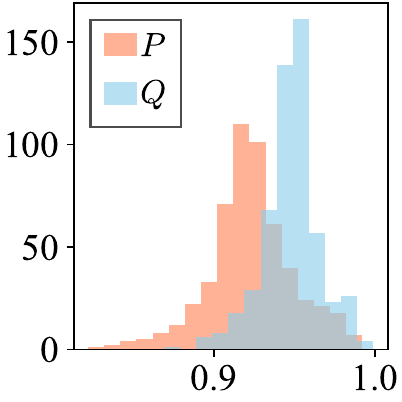}};
            \node[below = -0.01\linewidth of x] {\small Similarity to test set of $Q$};
        \node[left = 0.007\linewidth of x, rotate=90, font=\small, xshift=0.06\linewidth] {Count};
        \node (y) [right = 0.05\linewidth of x] {\includegraphics[width=0.25\linewidth]{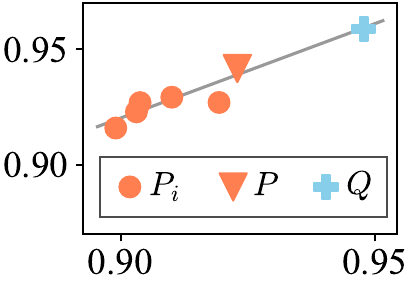}};
        \node[below = -0.01\linewidth of y, font=\small, align=center] {Similarity to test set of $Q$ \\ (Average over test set)};
        \node[left = 0.007\linewidth of y, rotate=90, font=\small, xshift=0.07\linewidth] {SSIM on Q};
        \node (a) [right = 0.04\linewidth of y, yshift=0.06\linewidth]  {\includegraphics[width=0.1\linewidth, ]{figs/imgs/AXT1POST-Aera.jpg}};   
        \node (b) [right = 0.03\linewidth of y, yshift=0.05\linewidth]  {\includegraphics[width=0.1\linewidth, ]{figs/imgs/AXFLAIR-Skyra.jpg}};
        \node[above = 0.0\linewidth of b, align=center, font=\small] {$P$: non-T2};
        \node (c) [right = 0.03\linewidth of y, yshift=-0.08\linewidth] {\includegraphics[width=0.1\linewidth, ]{figs/imgs/AXT2-Aera.jpg}};
        \node[above = -0.015\linewidth of c] {\small $Q$: T2};
        \end{tikzpicture}
    \end{minipage}
    \caption{We compute the cosine similarity in CLIP feature-space between a test sample and the nearest neighbor from the training set and relate this similarity to model performance. \textbf{Left:} Histograms of nearest neighbor similarity for each test sample of $Q$ to the training set from distributions $P$ or $Q$. \textbf{Right:} Strong correlation between nearest neighbor similarity and performance. Compared to datasets from distributions $P_i$, a more diverse dataset from distribution $P=\{P_1, \ldots, P_m\}$ increases both similarity to the out-of-distribution test set and model (U-net) performance.}
    \label{fig:clip_similarity}
\end{figure}

Figure~\ref{fig:robust_diverse_non-diverse} shows that the models {\color{orange}\textbf{trained on $P$}} are outperformed by models {\color{teal}\textbf{trained on $P$ and $Q$}} when evaluated on $Q$, as expected, since a model {\color{teal}\textbf{trained on $P$ and $Q$}} is an ideal robust baseline (as it contains data from $Q$). 
The difference of the {\color{teal}\textbf{trained on $P$ and $Q$}}-line and the {\color{violet}\textbf{trained on $P_\text{best}$}}-line is a measure of the severity of the distribution-shift, as it indicates the loss in performance when a model trained on $P_\text{best}$ is evaluated on $Q$. Comparing the difference between the line for the models {\color{orange}\textbf{trained on $P$}} and the line for models {\color{violet}\textbf{trained on $P_\text{best}$}} shows that effective robustness is improved by training on a diverse dataset, even when compared to distribution $P_\text{best}$ which is the most beneficial distribution for performance on $Q$. Moreover, we find similar results for distribution-shifts in the acceleration factor and in the number of receiver coils as detailed in Appendix~\ref{app:fwd_map_shift}.

\vsparagraph{Robustness improvements are related to increasing similarity of train and test distributions.}
A plausible explanation for the observed performance gains through training on a more diverse dataset is that a diverse dataset increases the similarity between samples in the training and target test set, and thus the target test set is less out-of-distribution. To test this hypothesis, we use CLIP-similarity, similar to how~\citet{mayilvahananDoesCLIPGeneralization2024} used CLIP-similarity to study CLIP's robustness in image classification. 

Specifically, for each test sample we find the nearest neighbor to a training sample in terms of the cosine similarity of CLIP features, and then take the mean of the histogram of similarities as the similarity of the test set to a training set. The left plot in Figure~\ref{fig:clip_similarity}, illustrates this measure for a contrast shift. 
The histograms for $P$ and $Q$ show the distribution of the similarity-scores between the test set and a training set. As expected, the training set from the same distribution ($Q$) shows higher similarity than a training set from a different distribution ($P$). The right plot extends this analysis to sub-distributions $P_i$ within $P$, showing that a more diverse training set $P$ is more similar to the test set, and most importantly this metric correlates well with reconstruction performance. More details are in Appendix~\ref{app:clip_distance}.

\begin{figure}[t]
    \centering
    \begin{minipage}{0.55\textwidth}
        \begin{tikzpicture}
            \node (x) at (0.0\textwidth,0.0\textwidth) {\includegraphics[width=0.33\textwidth]{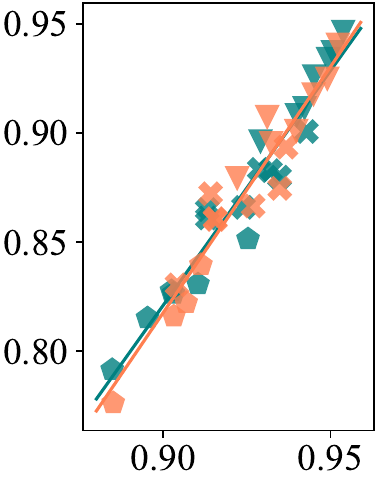}};
            \node[below = -0.02\textwidth of x, xshift=0.05\textwidth] {\scriptsize SSIM on $P$};
            \node[above = -0.02\textwidth of x, xshift=0.0\textwidth] {\scriptsize ${\color{red}\bm{A}} \leq 1\%$ of image size };
            \node[left = 0.01\textwidth of x, rotate=90, xshift=0.15\textwidth] {\scriptsize SSIM on $Q$};
            \node[right = -0.008\textwidth of x] (y) {\includegraphics[width=0.33\textwidth]{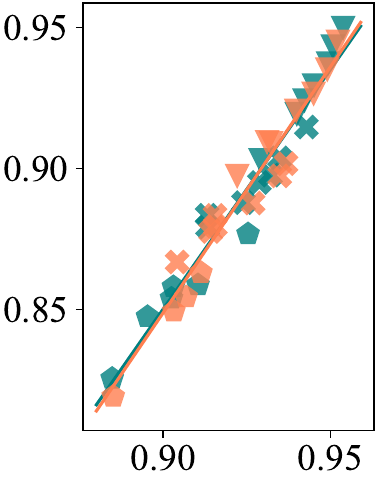}};
            \node[below = -0.02\textwidth of y, xshift=0.05\textwidth] {\scriptsize SSIM on $P$};
            \node[above = -0.02\textwidth of y, xshift=0.0\textwidth] {\scriptsize ${\color{red}\bm{A}} > 1\%$ of image size};
            \node[below = 0.035\textwidth of x, xshift=0.3\textwidth] (l) {\frame{\includegraphics[width=1\textwidth]{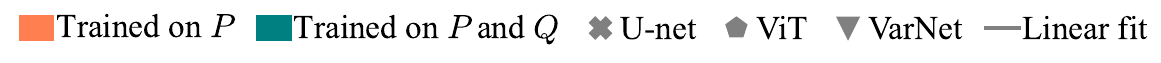}}};
            \node (a) [right = 0.01\textwidth of y, yshift=0.1\textwidth]  {\includegraphics[width=0.18\textwidth]{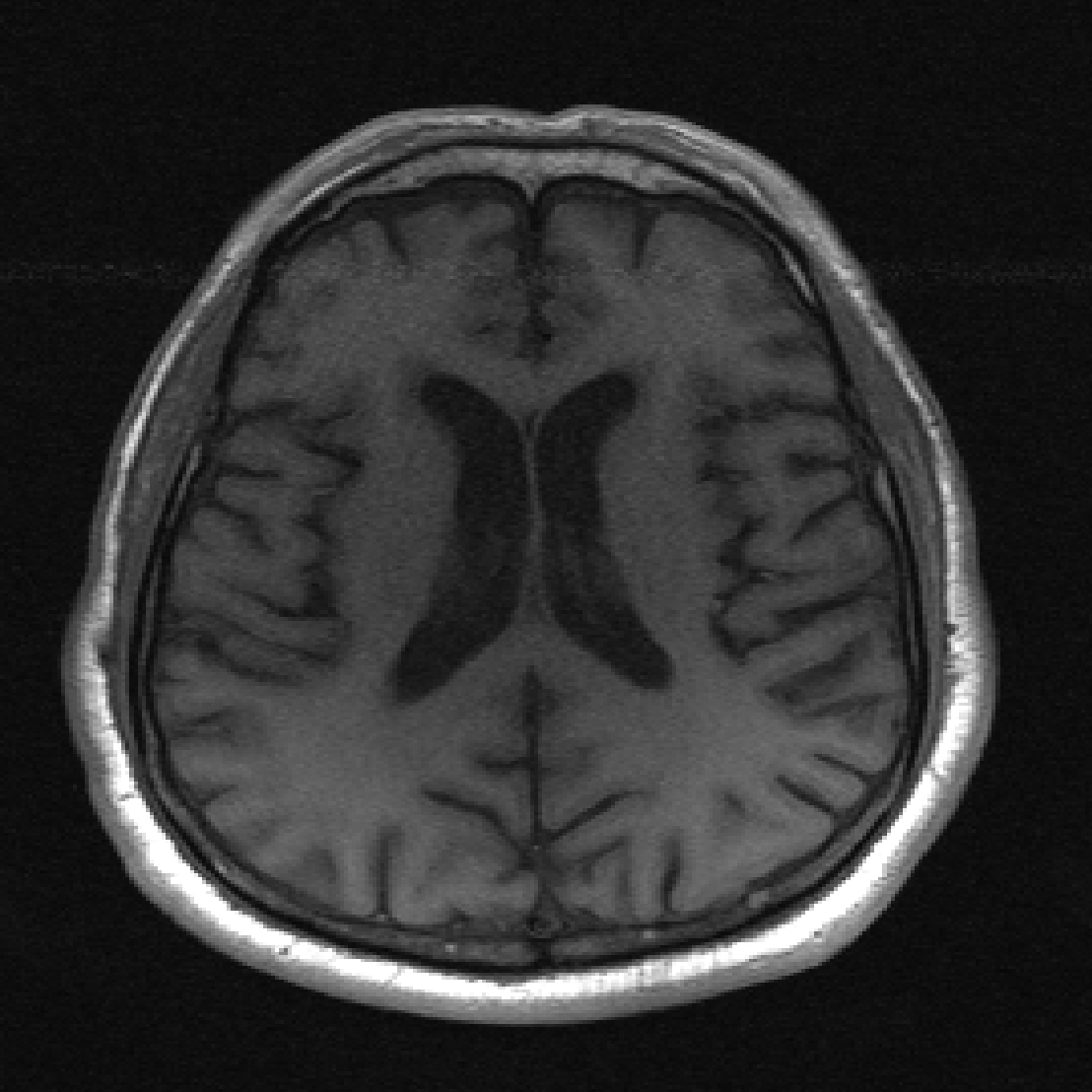}};   
            \node[above = -0.01\textwidth of a, align=center, font=\small, xshift=0.013\textwidth] {$P$: w/o Pathology};
            \node (c) [right = 0.01\textwidth of y, yshift=-0.16\textwidth] {\includegraphics[width=0.18\textwidth]{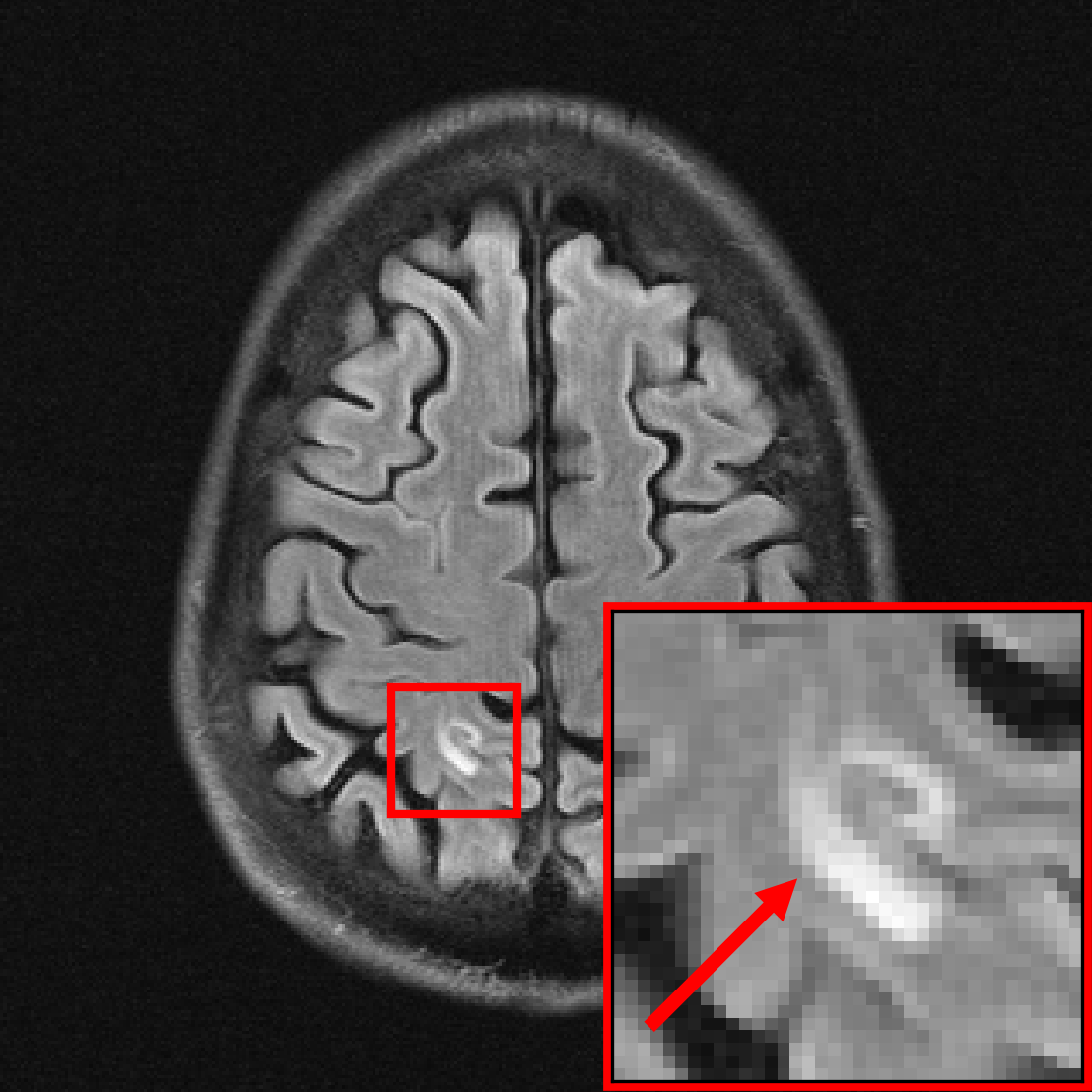}};
            \node[above = -0.01\textwidth of c] {\small $Q$: Pathology};
            \node[below = -0.145\textwidth of c, xshift=0.074\textwidth] {\color{red}\footnotesize $\bm{A}$};
        \end{tikzpicture}
    \captionof{figure}{
    Models trained on images without pathologies can reconstruct pathologies as well as models trained on images with pathologies. SSIM is calculated for the pathology region ({\color{red}$\bm{A}$}) for small (\textbf{left}) and large (\textbf{right}) pathologies.
    }
    \label{fig:pathology}
    \end{minipage}\hfill
    \begin{minipage}{0.4\textwidth}
        \begin{tikzpicture}
            \node (x) at (0.0\textwidth,0.0\textwidth) {\includegraphics[width=0.29\textwidth]{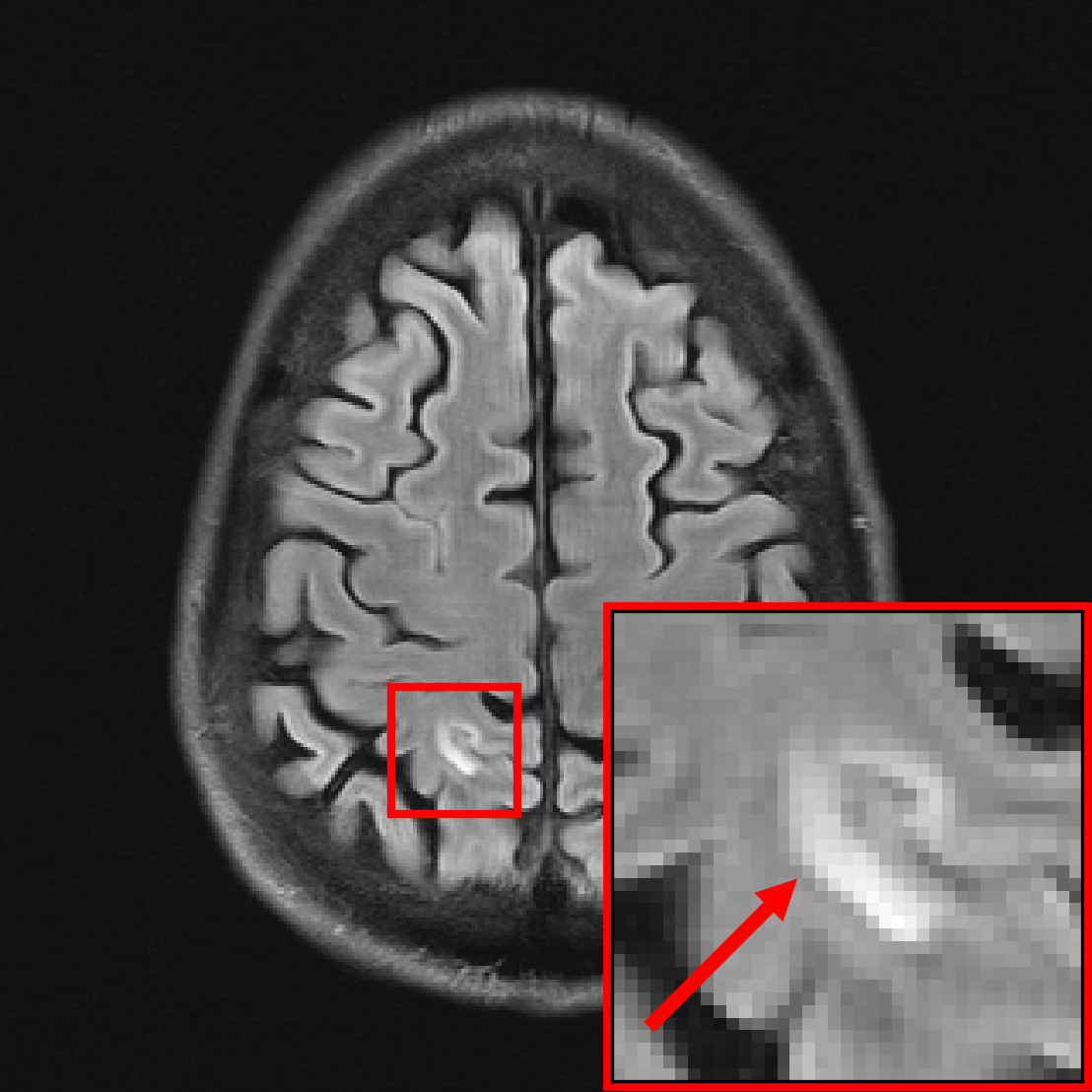}};
            \node[below = 0\textwidth of x] {\includegraphics[width=0.29\textwidth]{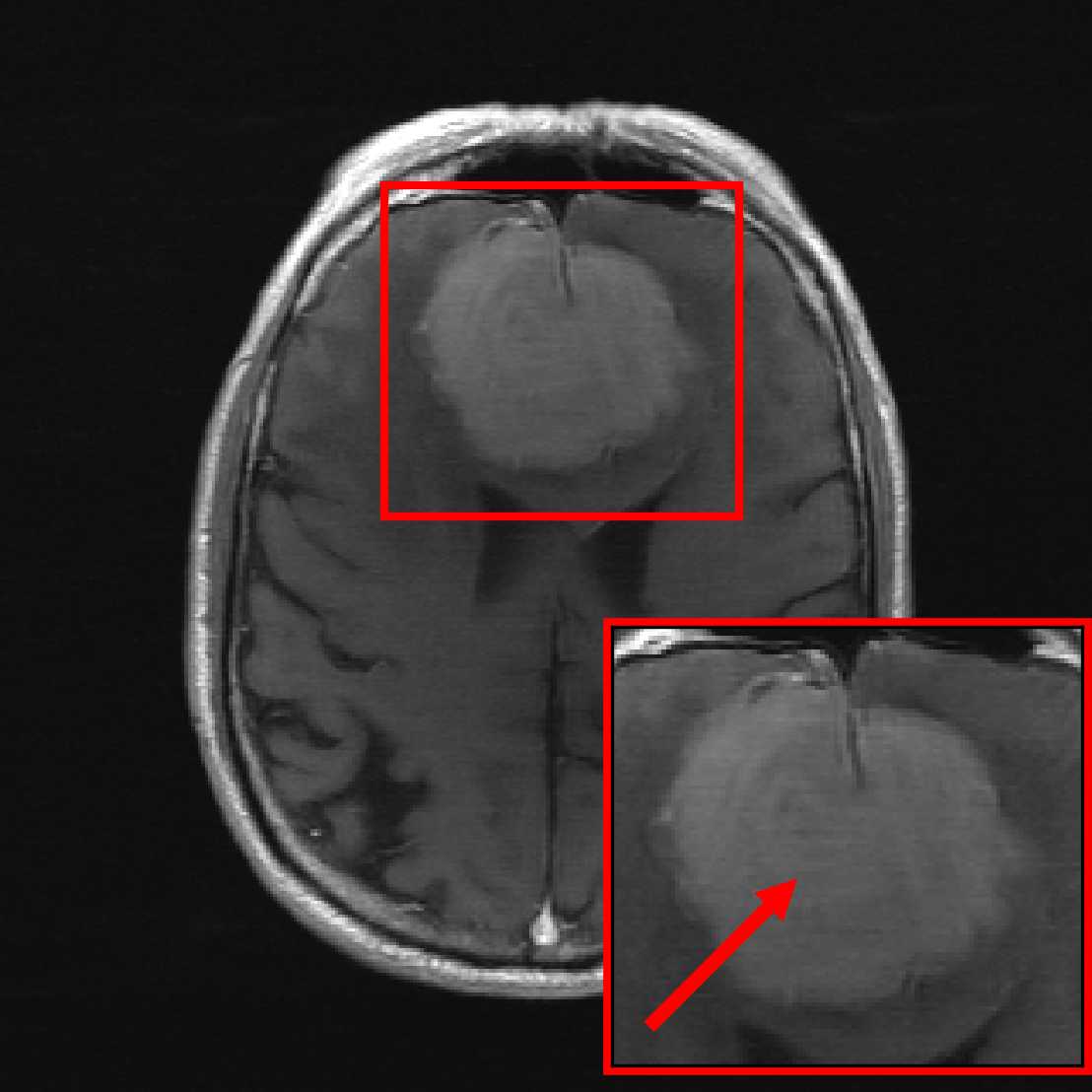}};
            \node[above = 0\textwidth of x] {\small Trained on $P$};
            \node (a) [right = 0.\textwidth of x] {\includegraphics[width=0.29\textwidth]{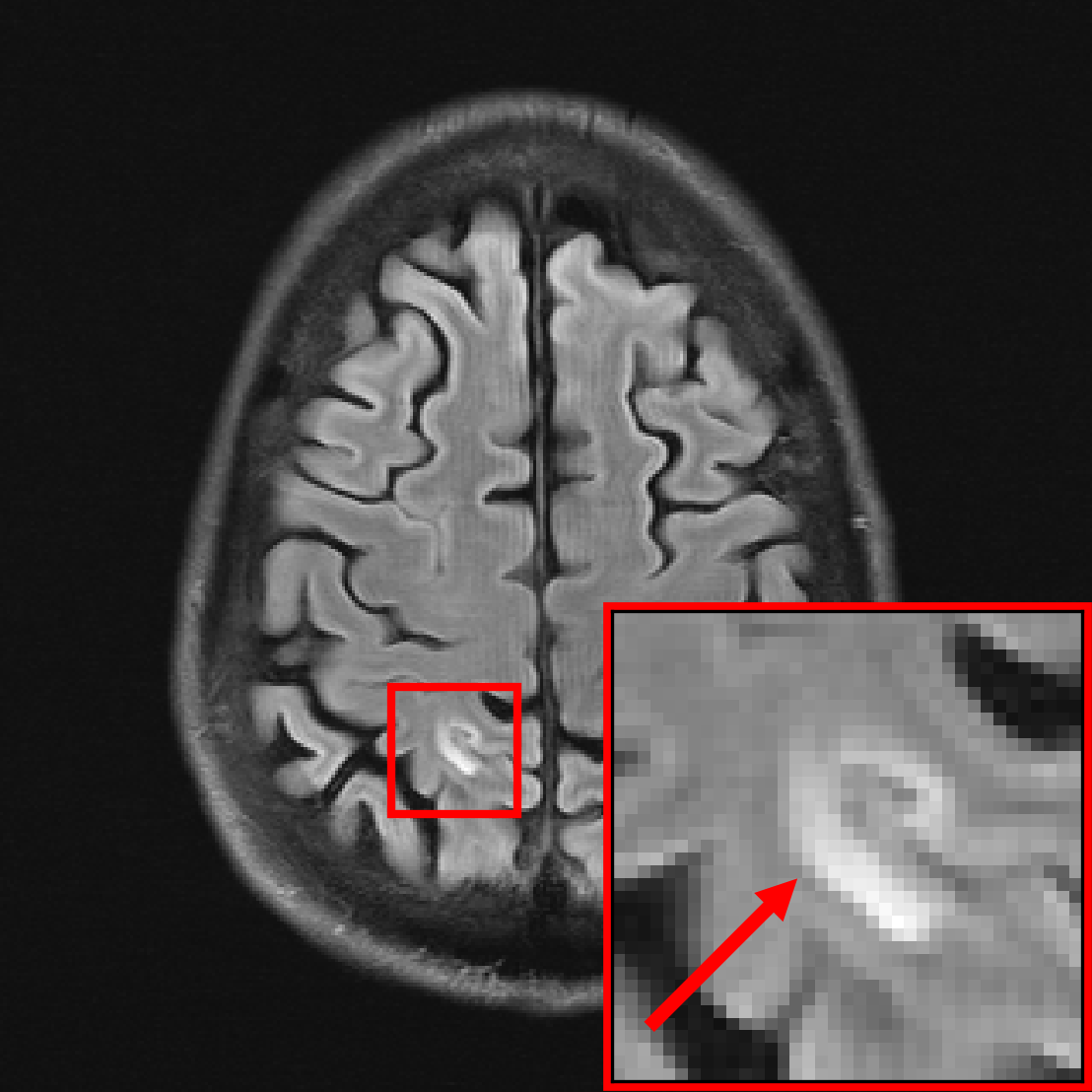}};
            \node[below = 0\textwidth of a] {\includegraphics[width=0.29\textwidth]{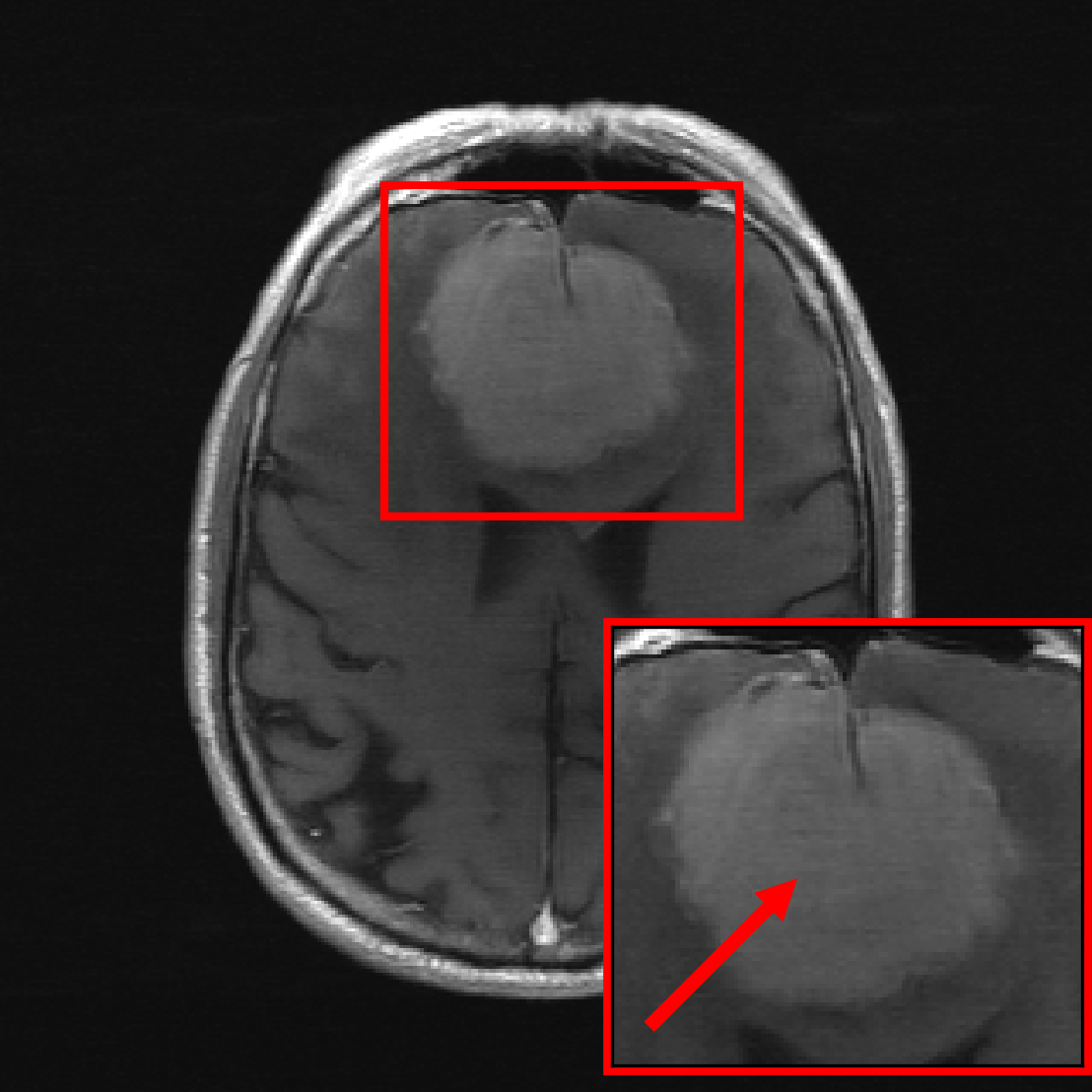}};
            \node[above = 0.0\textwidth of a, yshift=-0.01\textwidth] {\small on $P+Q$};
            \node (b) [right = 0\textwidth of a] {\includegraphics[width=0.29\textwidth]{figs/pathology/lacunar_infarct_ref.png}};
            \node[below = 0\textwidth of b] {\includegraphics[width=0.29\textwidth]{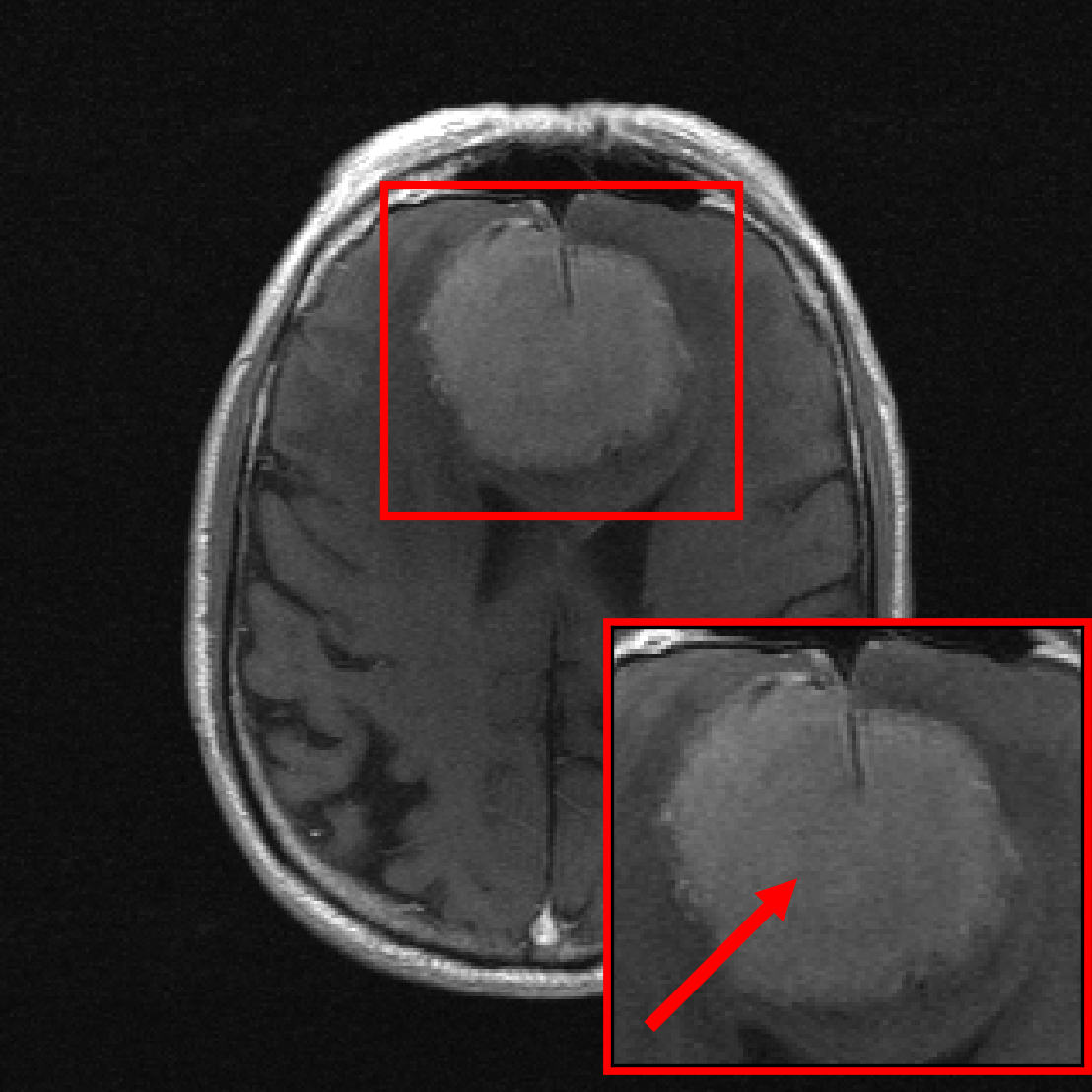}};
            \node[above = 0.0\textwidth of b] {\small Ground truth};
        \end{tikzpicture}
    \captionof{figure}{Reconstructions by a VarNet trained on images without pathologies ($P$) and with pathologies ($P+Q$) of an image containing a small-sized 
    pathology.}
    \label{fig:pathology_reconstructions}
    \end{minipage}
\end{figure}

\section{Reconstruction of Pathologies Using Data From Healthy Subjects} \label{sec:pathology}
In this section, we investigate the distribution-shift from healthy to non-healthy subjects by measuring how well models reconstruct images containing a pathology if no pathologies are contained in the training set. 
We find that models trained on fastMRI data without pathologies reconstruct fastMRI data with pathologies as accurately as the same models trained on fastMRI data with pathologies. 

\vsparagraph{Experiment.} We rely on the fastMRI+ annotations~\citep{zhaoFastMRIClinicalPathology2022} to partition the fastMRI brain dataset into sets of images with and without pathologies. The annotations cover various pathologies in the fastMRI dataset.
We extract a set of volumes without pathologies by selecting all scans with the fastMRI+ label ``Normal for age", and select images with pathologies by taking all images with slice-level annotations of a pathology. The training set contains 4.5k images without pathologies ($P$) and 2.5k images with pathologies ($Q$). We train U-nets, ViTs, and VarNets on $P$ and on $P+Q$, and sample different models by varying the training set size by factors of 2, 4 and 8, and by early stopping. While the training set from distribution $P$ does not contain images with pathologies, $P$ is a diverse distribution containing data from different scanners and image contrasts. 

Figure~\ref{fig:pathology} shows the models' performance on $Q$ relative to their performance on $P$. Reconstructions are evaluated only on the region containing the pathology, where we distinguish between small pathologies ($\leq 1\%$ of the total image size) and large pathologies ($>1\%$ of the total image size) to see potential dependencies on the size of the pathology.

We see that the models {\color{orange}\textbf{trained on $P$}} show essentially the same performance (SSIM) as models {\color{teal}\textbf{trained on $P+Q$}} regardless of pathology size. The results indicate that models trained on images without pathologies can reconstruct pathologies as accurately as models trained on images with pathologies.
This is  further illustrated in Figure~\ref{fig:pathology_reconstructions}, where we show reconstructions given by the VarNet of images with a pathology: The model recovers the pathology well even though no pathologies are in the training set. We provide additional results, reconstruction examples, and discussion in Appendix~\ref{app:pathology_eval}, including a more nuanced evaluation of the SSIM values for VarNet (Figure~\ref{fig:pathology_scatter}).

\begin{figure}[t]
    \centering
    \begin{minipage}[c]{0.01\textwidth}
        \begin{tikzpicture}
            \node (a) [rotate=90, font=\scriptsize] at (0.\textwidth,0) {SSIM on $Q$};
            \node[below = 5\textwidth of a,  font=\scriptsize] {\phantom{S}};
        \end{tikzpicture}
    \end{minipage}
    \begin{minipage}[c]{0.32\textwidth}
        \begin{tikzpicture}
            \node (x) at (0.0\textwidth,0.0\textwidth) {\includegraphics[width=0.6\textwidth]{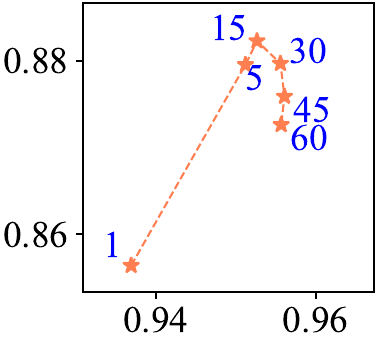}};
            \node[below = -0.04\textwidth of x, xshift=0.05\textwidth] {\scriptsize SSIM on $P$};
            \node (a) [right = 0.04\textwidth of x, yshift=0.18\textwidth]  {\includegraphics[width=0.2\textwidth, ]{figs/imgs/AXT2-Prisma_fit.jpg}};   
            \node (b) [right = 0.01\textwidth of x, yshift=0.15\textwidth]  {\includegraphics[width=0.2\textwidth, ]{figs/imgs/AXT2-Aera.jpg}};
            \node[above = 0.0\textwidth of b, align=center, font=\scriptsize] {$P$: fm brain, T2 \phantom{1}};
            \node (c) [right = 0.01\textwidth of x, yshift=-0.15\textwidth] {\includegraphics[width=0.2\textwidth, ]{figs/imgs/CORPD_FBK-Skyra.jpg}};
            \node[above = -0.04\textwidth of c] {\scriptsize $Q$: fm knee};
        \end{tikzpicture}
    \end{minipage}
    \begin{minipage}[c]{0.32\textwidth}
        \begin{tikzpicture}
            \node (x) at (0.0\textwidth,0.0\textwidth) {\includegraphics[width=0.6\textwidth]{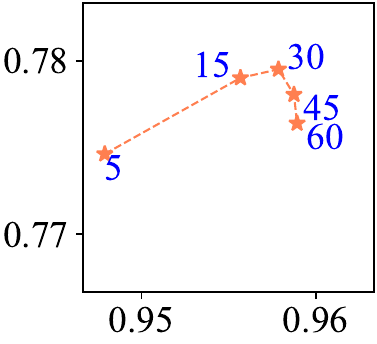}};
            \node[below = -0.04\textwidth of x, xshift=0.05\textwidth] {\scriptsize SSIM on $P$};
            \node (a) [right = 0.04\textwidth of x, yshift=0.18\textwidth]  {\includegraphics[width=0.2\textwidth, ]{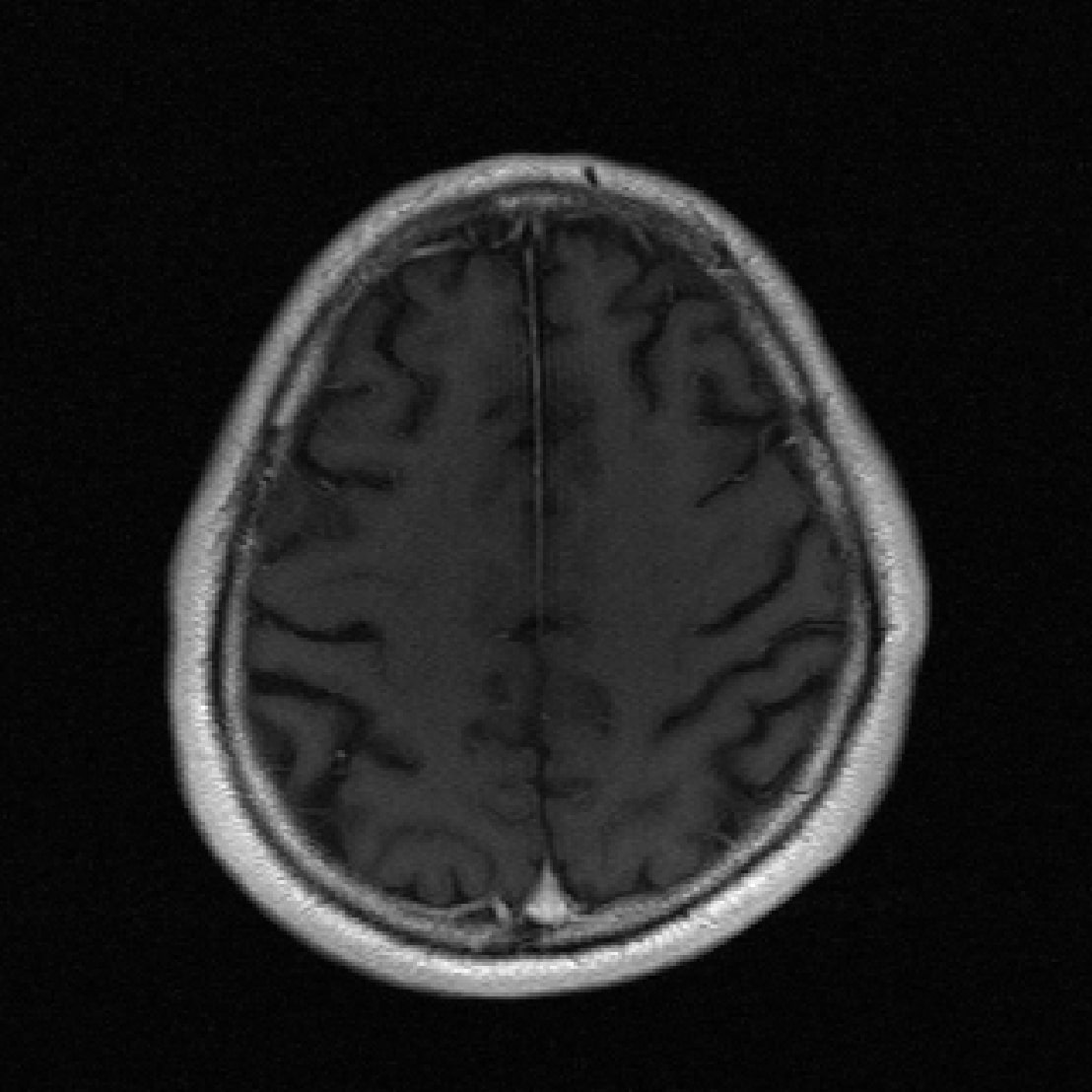}};   
            \node (b) [right = 0.01\textwidth of x, yshift=0.15\textwidth]  {\includegraphics[width=0.2\textwidth, ]{figs/imgs/AXFLAIR-Skyra.jpg}};
            \node[above = 0.0\textwidth of b, align=center, font=\scriptsize] {$P$: fm brain};
            \node (c) [right = 0.01\textwidth of x, yshift=-0.15\textwidth] {\includegraphics[width=0.2\textwidth]{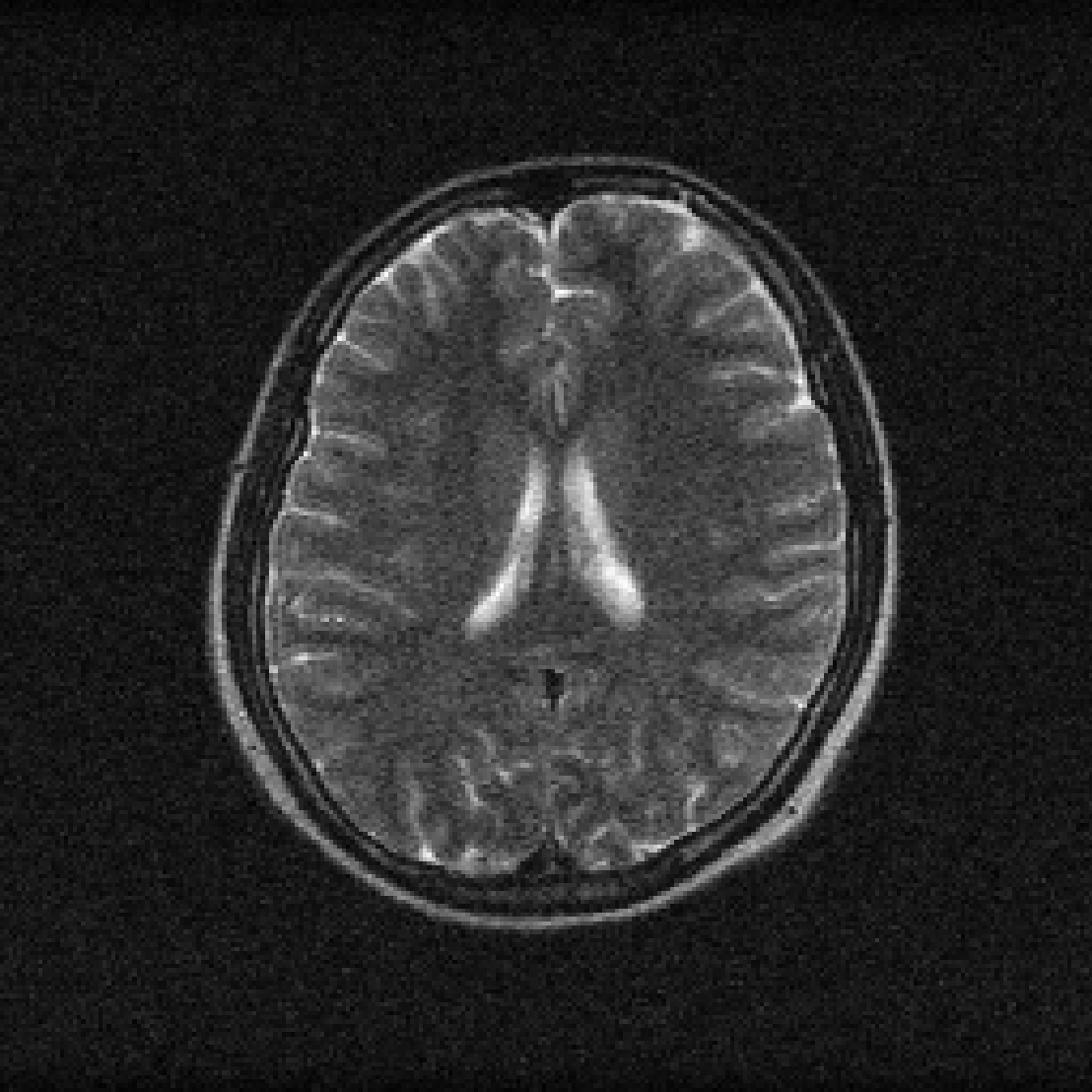}};
            \node[above = -0.04\textwidth of c] {\scriptsize $Q$: M4Raw};
        \end{tikzpicture}
    \end{minipage}
    \begin{minipage}[c]{0.32\textwidth}
        \begin{tikzpicture}
            \node (x) at (0.0\textwidth,0.0\textwidth) {\includegraphics[width=0.6\textwidth]{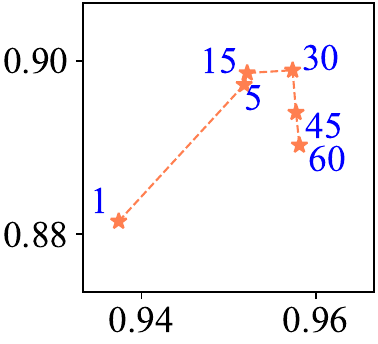}};
            \node[below = -0.04\textwidth of x, xshift=0.05\textwidth] {\scriptsize SSIM on $P$};
            \node (a) [right = 0.04\textwidth of x, yshift=0.18\textwidth]  {\includegraphics[width=0.2\textwidth, ]{figs/imgs/CORPD_FBK-Skyra.jpg}};   
            \node (b) [right = 0.01\textwidth of x, yshift=0.15\textwidth]  {\includegraphics[width=0.2\textwidth, ]{figs/imgs/AXT2-Prisma_fit.jpg}};
            \node[above = 0.0\textwidth of b, align=right, font=\scriptsize] {$P$: fm brain+knee};
            \node (c) [right = 0.01\textwidth of x, yshift=-0.15\textwidth] {\includegraphics[width=0.2\textwidth]{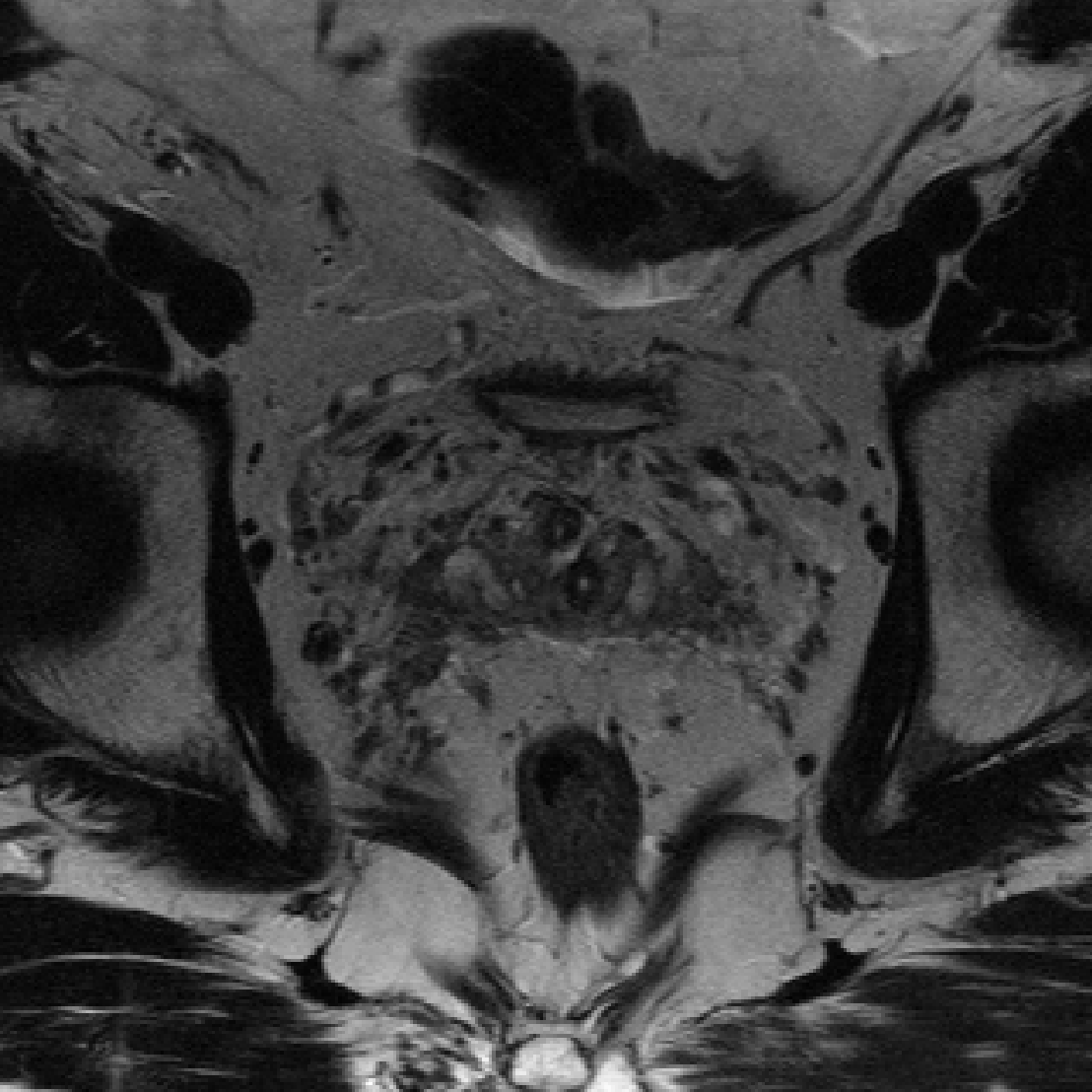}}; 
            \node[above = -0.04\textwidth of c] {\scriptsize $Q$: fm prostate};
        \end{tikzpicture}
    \end{minipage}
    \caption{For a distribution-shift from distribution $P$ to distribution $Q$, the in- and out-of-distribution performance is plotted as a function of training epochs (1 to 60). At the beginning of training, out-of-distribution performance increases together with in-distribution performance. Then, out-of-distribution performance starts to drop while in-distribution performance continues to marginally increase.}
    \label{fig:dist_overfit}
\end{figure}
\section{Distributional Overfitting}\label{sec:dist_overfit}
We observed that when training for long, while in-distribution performance continues to improve slightly, out-of-distribution performance can deteriorate.  We refer to this as distributional overfitting. Unlike conventional overfitting, where a model's in-distribution performance declines after prolonged training, distributional overfitting involves a decline in out-of-distribution performance while in-distribution performance continues to improve (slightly). A similar observation has been made in the context of fine-tuning CLIP models~\citep{wortsmanRobustFinetuningZeroshot2022}.

Figure~\ref{fig:dist_overfit} illustrates distributional overfitting on two distribution-shifts. Each plot depicts the in and out-of-distribution ($P$ and $Q$) performance of an U-net as a function of trained epochs. For example, in the left plot $P$ is fastMRI T2-weighted brain data and $Q$ is fastMRI knee data. We observe as training progresses, initially, the model's in-distribution and out-of-distribution performance both improve. However, after epoch 15, out-of-distribution performance deteriorates, despite marginal improvements in in-distribution performance. 
In Appendix~\ref{app:dist_overfit}, we show that distributional overfitting also occurs for VarNet and ViT, and when using other optimizers.

This finding indicates that early stopping before conventional overfitting sets in, can help to improve model robustness with minimal impact on in-distribution performance.

\begin{figure}[t]
    \centering
    \begin{minipage}{0.95\textwidth}
    \centering
    \begin{tikzpicture}
        \node (a) at (0.0\textwidth,0.0\textwidth){\includegraphics[width=0.25\textwidth]{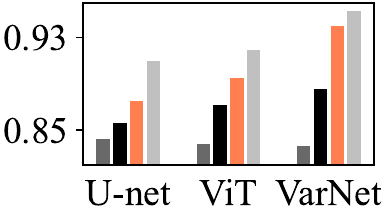}};
        \node[above = -0.01\textwidth of a, xshift=0.02\textwidth, font=\small] {$\mathcal{D}_Q$: CC-359, sagittal};
        \node[left = 0.01\textwidth of a, rotate=90, font=\small, xshift=0.045\textwidth] {SSIM};
        \node[right = -0.008\textwidth of a] (b) {\includegraphics[width=0.25\textwidth]{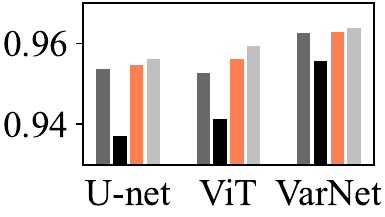}};
        \node[above = -0.01\textwidth of b, xshift=0.02\textwidth, font=\small] {$\mathcal{D}_Q$: NYU dataset};
        \node[right = 0.07\textwidth of b] (c) {\includegraphics[width=0.25\textwidth]{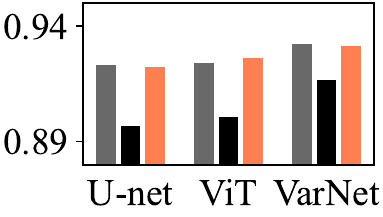}};
        \node[above = -0.01\textwidth of c, xshift=0.02\textwidth, font=\small] {fastMRI knee};
        \node[left = 0.01\textwidth of c, rotate=90, font=\small, xshift=0.045\textwidth] {SSIM};
        \node[below = 0.03\textwidth of a] (d)
        {\includegraphics[width=0.25\textwidth]{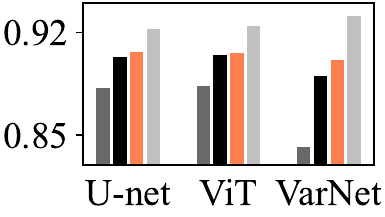}};
        \node[left = 0.01\textwidth of d, rotate=90, font=\small, xshift=0.045\textwidth] {SSIM};
        \node[above = -0.01\textwidth of d, xshift=0.02\textwidth, font=\small]  {$\mathcal{D}_Q$: M4Raw GRE};
        \node[right = -0.008\textwidth of d] (e) {\includegraphics[width=0.25\textwidth]{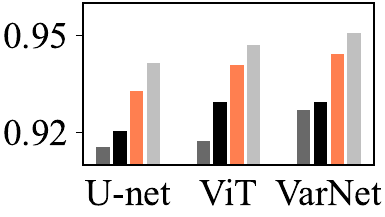}};
        \node[above = -0.01\textwidth of e, xshift=0.02\textwidth, font=\small] {$\mathcal{D}_Q$: Stanford 2D};
        \node[right = 0.07\textwidth of e] (f) {\includegraphics[width=0.25\textwidth]{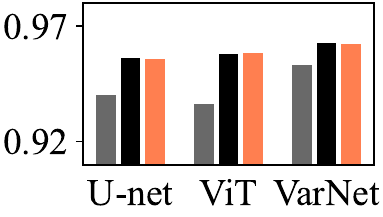}};
        \node[above = -0.01\textwidth of f, xshift=0.02\textwidth, font=\small] {fastMRI brain};
        \node[left = 0.01\textwidth of f, rotate=90, font=\small, xshift=0.045\textwidth] {SSIM};
        \node[below = 0.01\textwidth of d, xshift=0.28\textwidth] (l)
        {\frame{\includegraphics[width=0.95\textwidth]{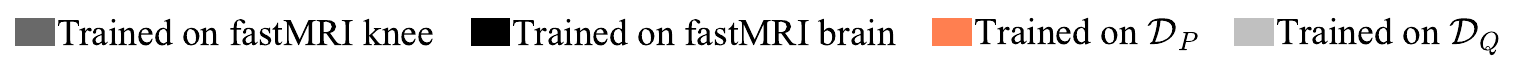}}};
    \end{tikzpicture}
    \end{minipage}
    \caption{
    Training on a diverse collection of datasets improves robustness under distribution-shifts. A model trained on the diverse set of datasets $\mathcal{D}_P$ can significantly outperform models trained on fastMRI data when evaluated on out-of-distribution data $\mathcal{D}_Q$, while maintaining the same performance on fastMRI data.}
    \label{fig:diverse_datasets}
\end{figure}
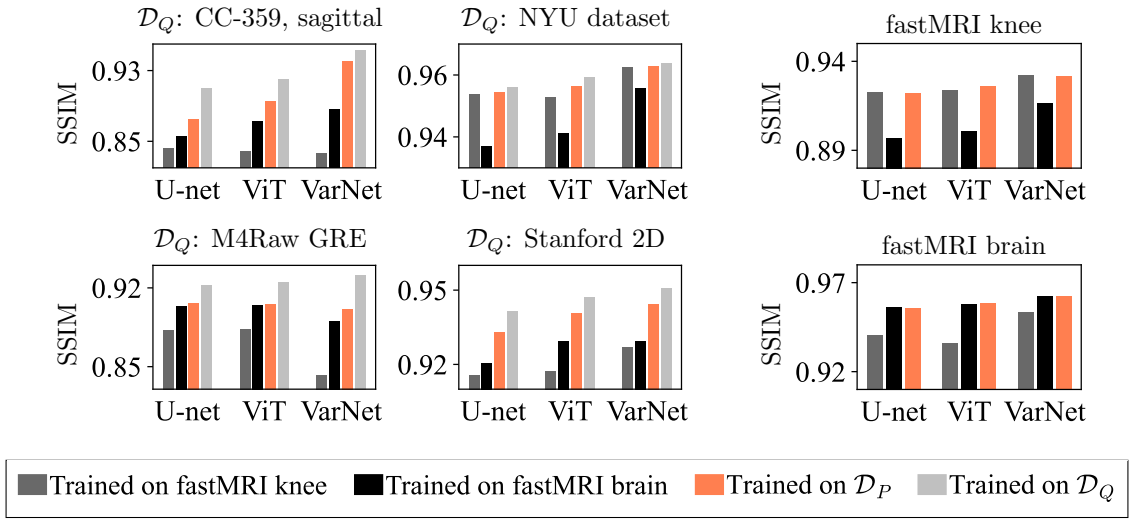

\section{Robust Models for Accelerated MRI}\label{sec:robust_models}
The results from the previous sections based on the fastMRI dataset suggest that training a single model on a diverse 
dataset consisting of several data distributions is beneficial to out-of-distribution performance without sacrificing in-distribution performance on individual distributions. 

We now move beyond the fastMRI dataset and demonstrate that this finding continues to hold on a large collection of datasets, enabling significant out-of-distribution performance improvements. We train a single large model for 4-fold accelerated 2D MRI on a \textbf{diverse collection of 13 datasets} including the fastMRI brain and knee datasets, and evaluate on 4 out-of-distribution datasets (the descriptions of the sets are in Table~\ref{tbl:datasets}). 
The resulting model, when compared to models trained only on the fastMRI dataset, shows significant robustness improvements while maintaining its performance on the fastMRI dataset.  

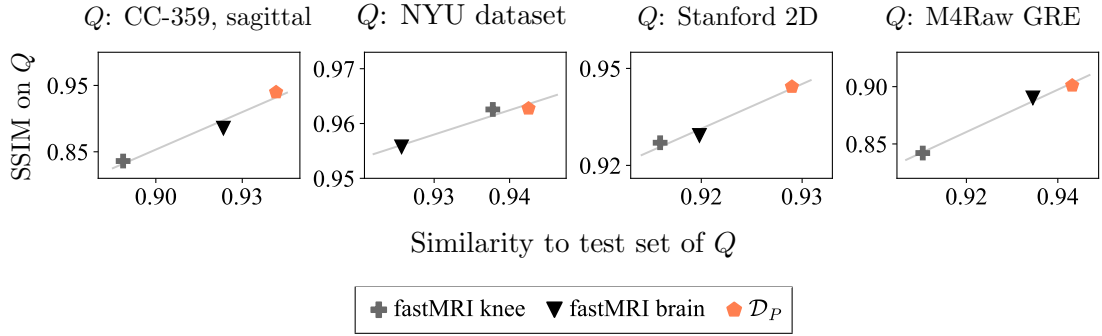
\begin{figure}[t]
    \centering
    \begin{minipage}{0.9\textwidth}
    \begin{tikzpicture}
    \node (11) at (0.0\linewidth,0.0\linewidth) {\includegraphics[width=0.23\linewidth]{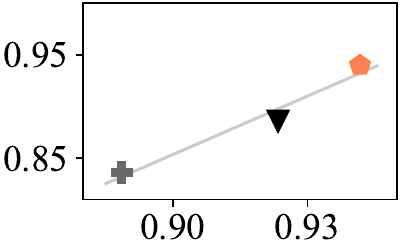}};
    \node (12) [right = -0.01\linewidth of 11] {\includegraphics[width=0.23\linewidth]{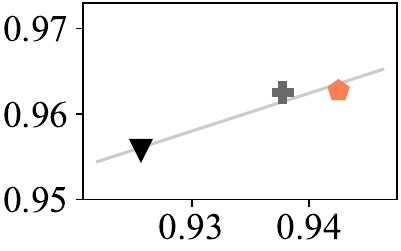}};
    \node (21) [right = -0.01\linewidth of 12] {\includegraphics[width=0.23\linewidth]{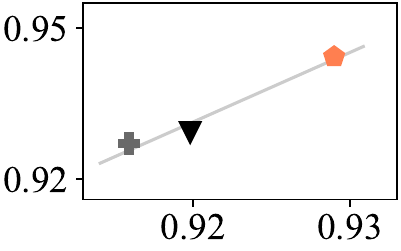}};
    \node (22) [right = -0.01\linewidth of 21] {\includegraphics[width=0.23\linewidth]{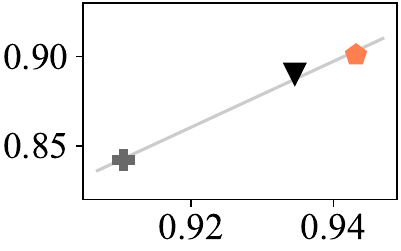}};
    \node (23) [below = 0.05\linewidth of 12, xshift=0.12\linewidth] {\frame{\includegraphics[width=0.39\linewidth]{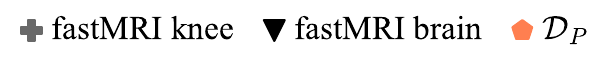}}};
    \node[above = 0.0\linewidth of 11, xshift=0.022\linewidth, font=\small] {$Q$: CC-359, sagittal};
    \node[above = 0.0\linewidth of 12, xshift=0.02\linewidth] {$Q$: NYU dataset};
    \node[above = 0.0\linewidth of 21, xshift=0.02\linewidth, font=\small] {$Q$: Stanford 2D};
    \node[above = 0.0\linewidth of 22, xshift=0.01\linewidth, font=\small] {$Q$: M4Raw GRE};
    \node[rotate=90] [left = 0.01\linewidth of 11, xshift=0.08\linewidth, font=\small] {SSIM on $Q$};
    \node[below = 0.0\linewidth of 12, xshift=0.12\linewidth] {Similarity to test set of $Q$};
    \end{tikzpicture}
    \end{minipage}
    \caption{Nearest-neighbor CLIP similarity between a training set and out-of-distribution ($Q$) test set correlate well with the performance of VarNet.}
    \label{fig:similarity_vs_performance}
\end{figure}
\vsparagraph{Experiment.} 
We train an U-net, ViT, and an end-to-end VarNet on the collection of the first 13 datasets listed in Table~\ref{tbl:datasets}. We denote this collection of datasets by $\mathcal{D}_P$. For the fastMRI knee and brain datasets, we exclude the fastMRI knee validation set and the fastMRI brain test set from the training set, as we use those for testing. The total number of training slices after the data preparation is 413k. 
For each model family, we also train a model on fastMRI knee, and one on fastMRI brain as baselines.
To mitigate the risk of distributional overfitting, we early stop training when the improvement on the fastMRI knee dataset becomes marginal. Further experimental details are in Appendix~\ref{app:dataprep_non-fm} and~\ref{app:robust_models_training}.

The fastMRI knee validation and fastMRI brain test set are used to measure in-distribution performance. 
We measure out-of-distribution performance on CC-359 sagittal view~\citep{souzaOpenMultivendorMultifieldstrength2018}, Stanford 2D~\cite{chengStanford2DFSE2018}, M4Raw GRE~\citep{lyuM4RawMulticontrastMultirepetition2023a}, and  NYU data~\citep{hammernikLearningVariationalNetwork2018a}.
These datasets constitute a distribution-shift relative to the training data with respect to vendors, anatomic views, anatomies, time-frame of data collection, anatomical views, MRI sequences, contrasts and combinations thereof and therefore enable a broad robustness evaluation. As a further reference point we also train models on the out-of-distribution datasets to quantify the robustness gap.

Figure~\ref{fig:diverse_datasets} shows for all architectures considered, the model trained on the collection of datasets $\mathcal{D}_P$ significantly outperforms the models trained on fastMRI data when evaluated on out-of-distribution data, without compromising performance on fastMRI data. For example, on the CC-359 sagittal view dataset, the VarNet trained on $\mathcal{D}_P$ almost closes the distribution-shift performance gap (i.e., the gap to the model trained on the out-of-distribution data). We refer to Figure~\ref{fig:P_recon} and Figure~\ref{fig:reconstruction_examples} (in the Appendix) for reconstruction examples.

\vsparagraph{Robustness improvements are related to increasing similarity of train and test distributions.} In Figure~\ref{fig:similarity_vs_performance}, we compute the CLIP similarity between the training sets and the out-of-distribution test sets and observe a strong correlation between similarity and performance. This supports the idea that diverse datasets enhance similarity to out-of-distribution data, leading to improved performance. 

We further validate our findings in Appendix~\ref{app:additional_metrics}, where we evaluate model performance using the deep feature metrics LPIPS~\citep{zhangUnreasonableEffectivenessDeep2018} and DISTS~\citep{dingImageQualityAssessment2022}, because these metrics were found to align well with radiologist evaluations~\citep{adamsonUsingDeepFeature2023, kastryulinImageQualityAssessment2023}. The qualitative results are consistent with the SSIM results in this section: As shown in Table~\ref{tbl:deep_metrics}, the models trained on the large collection of datasets yield better scores on the out-of-distribution datasets than the fastMRI baselines while achieving similar scores on the fastMRI datasets. In addition, we provide an error analysis by quantifying the prominence of artifacts in the reconstructions (Table~\ref{tbl:error_metrics}) and find that models trained on the diverse dataset $\mathcal{D}_P$ produce less pronounced artifacts compared to the fastMRI baseline.

The results in this section reinforce our earlier findings, affirming that large and diverse MRI training sets can significantly enhance robustness without compromising in-distribution performance.

\section{Conclusion and Limitations}
While our research shows that diverse training sets significantly enhances out-of-distribution robustness for deep learning models for accelerated 2D MRI, training a model on a diverse dataset often doesn't close the distribution-shift performance gap, i.e., the gap between the model and the same idealized model trained on the out-of-distribution data (see Figure~\ref{fig:robust_diverse_non-diverse} and~\ref{fig:diverse_datasets}). 
Nevertheless, as datasets grow in size and diversity, training networks on larger and even more diverse data might progressively narrow the distribution-shift performance gap (as suggested in Figure~\ref{fig:clip_similarity} and~\ref{fig:similarity_vs_performance}). However, in practice it might be difficult or expensive to collect diverse and large datasets. 

Besides demonstrating the effect of diverse training data, our work shows that care must be taken when training models for long as this can yield to a less robust model due to distributional overfitting. This finding also emphasizes the importance of evaluating on out-of-distribution data.

\section*{Acknowledgements}
We thank Stefan Ruschke for helpful discussions. 

The authors acknowledge the financial support by the Federal Ministry of Education and Research of Germany in the programme of “Souverän. Digital. Vernetzt.“. Joint project 6G-life, project identification number: 16KISK002, as well as from Deutsche Forschungsgemeinschaft (DFG, German Research Foundation) - 456465471, 464123524.

\section*{Impact Statement}
The findings of this work highlight the importance of dataset design and in particular data diversity for deep learning based accelerated MRI. As this work encourages the utilization of large and diverse datasets for training robust deep learning models, ethical considerations around data privacy, consent, and security become crucial. While this work relies on public datasets intended for research purposes, practitioners should prioritize the ethical handling of patient data and ensure compliance with data protection regulations. Our code is available at \url{https://github.com/MLI-lab/mri_data_diversity}.

\printbibliography

\appendix

\section{$\ell_1$-Regularized Least-Squares Requires Different Hyperparameters on Different Distributions}
\label{app:hyperpl1}
The traditional approach for accelerated MRI is $\ell_1$-regularized least-squares \citep{lustigSparseMRIApplication2007a}.
While $\ell_1$-regularized least-squares is not considered data-driven, the regularization hyperparameter it typically chosen in a data-driven manner. 
For different distributions like different anatomies or contrasts, the regularization parameter takes on different values and thus the method needs to be tuned separately for different distributions. 
This can be seen for example from Table 4 of~\citet{zbontarFastMRIOpenDataset2019a}. 

To demonstrate this, we performed wavelet-based $\ell_1$-regularized least-squares on the single-coil knee version of the fastMRI dataset~\citep{zbontarFastMRIOpenDataset2019a} using 100 images from distribution $P$: PD Knee Skyra, 3.0T and 100 from distribution $Q$: PDFS Knee Aera, 1.5T. Using a regularization weight $\lambda = 0.01$ on distribution $P$ gives a SSIM of $0.792$, while $\lambda=0.001$ yields subpar SSIM of $0.788$. Contrary, on distribution $Q$, $\lambda=0.01$ only yields $0.602$, while $\lambda=0.001$ yields SSIM $0.609$. Thus, using the same model (i.e., the same regularization parameter for both distributions) is suboptimal for $\ell_1$-regularized. least squares. We used BART \url{https://mrirecon.github.io/bart/} for running $\ell_1$-regularized least squares.


\begin{figure}[t!]
    \centering
    \begin{minipage}[c]{1\linewidth}
        \par\noindent\rule{\linewidth}{0.4pt}
        \footnotesize \textbf{VarNet}
    \end{minipage}\\
    \begin{minipage}[c]{0.32\textwidth}
        \begin{minipage}[c]{1\textwidth}
            \begin{tikzpicture}
                \node (x) at (0.0\textwidth,0.0\textwidth) {\includegraphics[width=0.6\textwidth]{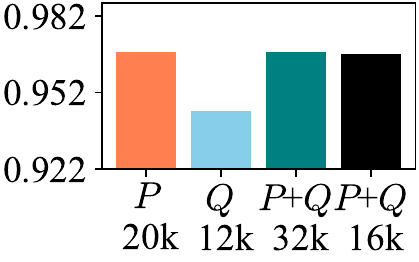}};
                \node[above = -0.03\textwidth of x, xshift=0.2\textwidth] {\footnotesize Anatomy};
                \node (a) [right = -0.04\textwidth of x, yshift=0.07\textwidth] {\includegraphics[width=0.22\textwidth, ]{figs/imgs/AXFLAIR-Skyra.jpg}};
                \node[below = -0.03\textwidth of a, align=center, font=\scriptsize] {\textbf{$P$: Brain}};
                \node[rotate=90] [left = 0.01\textwidth of x, xshift=0.22\textwidth] {\scriptsize SSIM on $P$};
                
                \node (y) [below = -0.05\textwidth of x]  {\includegraphics[width=0.6\textwidth]{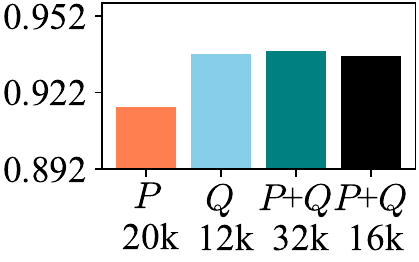}};
                \node (b) [right = -0.04\textwidth of y, yshift=0.07\textwidth]  {\includegraphics[width=0.22\textwidth, ]{figs/imgs/CORPD_FBK-Skyra.jpg}};
                \node[below = -0.03\textwidth of b, align=center, font=\scriptsize] {\textbf{$Q$: Knee}};
                \node[rotate=90] [left = 0.01\textwidth of y, xshift=0.22\textwidth] {\scriptsize SSIM on $Q$};
            \end{tikzpicture}
        \end{minipage}
    \end{minipage}\hfill
    \begin{minipage}[c]{0.32\textwidth}
        \begin{minipage}[c]{1\textwidth}
            \begin{tikzpicture}
                \node (x) at (0.0\textwidth,0.0\textwidth) {\includegraphics[width=0.6\textwidth]{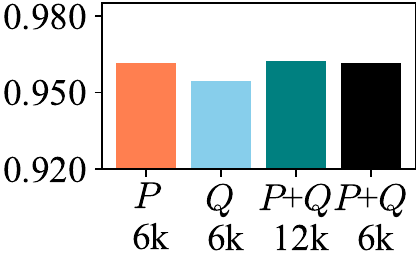}};
                \node[above = -0.03\textwidth of x, xshift=0.2\textwidth] {\footnotesize Contrasts};
                \node (a) [right = -0.04\textwidth of x, yshift=0.07\textwidth]  {\includegraphics[width=0.22\textwidth, ]{figs/imgs/CORPD_FBK-Skyra.jpg}};
                \node[below = -0.03\textwidth of a, align=center, font=\scriptsize] {\textbf{$P$: PD}};
                \node[rotate=90] [left = 0.01\textwidth of x, xshift=0.22\textwidth] {\scriptsize SSIM on $P$};
                
                \node (y) [below = -0.05\textwidth of x]  {\includegraphics[width=0.6\textwidth]{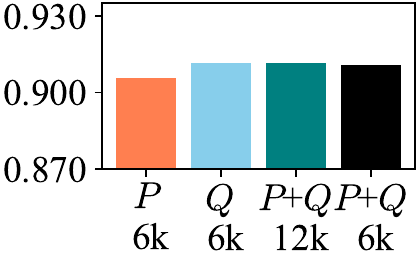}};
                \node (b) [right = -0.04\textwidth of y, yshift=0.07\textwidth]  {\includegraphics[width=0.22\textwidth, ]{figs/imgs/CORPDFS_FBK-Skyra.jpg}};
                \node[below = -0.03\textwidth of b, align=center, font=\scriptsize] {\textbf{$Q$: PDFS}};
                \node[rotate=90] [left = 0.01\textwidth of y, xshift=0.22\textwidth] {\scriptsize SSIM on $Q$};
            \end{tikzpicture}
        \end{minipage}
    \end{minipage}\hfill
     \begin{minipage}[c]{0.32\textwidth}
        \begin{minipage}[c]{1\textwidth}
            \begin{tikzpicture}
                \node (x) at (0.0\textwidth,0.0\textwidth) {\includegraphics[width=0.6\textwidth]{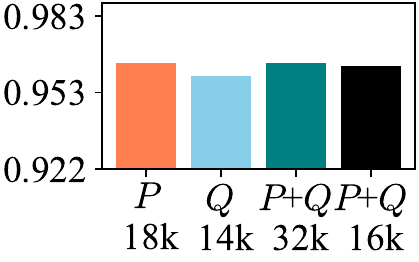}};
                \node[above = -0.03\textwidth of x, xshift=0.2\textwidth] {\footnotesize Magnetic field};
                \node (a) [right = -0.04\textwidth of x, yshift=0.07\textwidth]  {\includegraphics[width=0.22\textwidth, ]{figs/imgs/AXT2-Prisma_fit.jpg}};
                \node[below = -0.03\textwidth of a, align=center, font=\scriptsize] {\textbf{$P$: 3.0T}};
                \node[rotate=90] [left = 0.01\textwidth of x, xshift=0.22\textwidth] {\scriptsize SSIM on $P$};
                
                \node (y) [below = -0.05\textwidth of x]  {\includegraphics[width=0.6\textwidth]{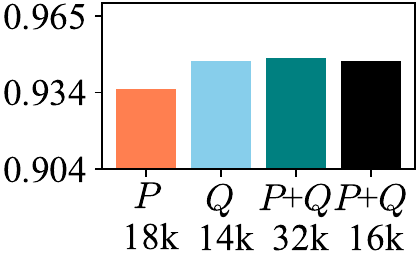}};
                \node (b) [right = -0.04\textwidth of y, yshift=0.07\textwidth]  {\includegraphics[width=0.22\textwidth, ]{figs/imgs/AXT2-Aera.jpg}};
                \node[below = -0.03\textwidth of b, align=center, font=\scriptsize] {\textbf{$Q$: 1.5T}};
                \node[rotate=90] [left = 0.01\textwidth of y, xshift=0.22\textwidth] {\scriptsize SSIM on $Q$};
            \end{tikzpicture}
        \end{minipage}
    \end{minipage}\hfill\\
    \begin{minipage}[c]{1\linewidth}
        \par\noindent\rule{\linewidth}{0.4pt}
        \footnotesize \textbf{ViT}
    \end{minipage}\\
    \begin{minipage}[c]{0.32\linewidth}
        \begin{minipage}[c]{1\linewidth}
            \begin{tikzpicture}
                \node (x) at (0.0\linewidth,0.0\linewidth) {\includegraphics[width=0.6\linewidth]{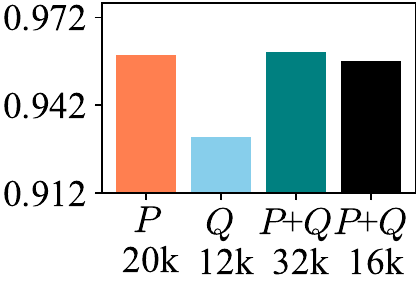}};
                \node[above = -0.03\linewidth of x, xshift=0.2\linewidth] {\footnotesize Anatomy};
                \node (a) [right = -0.04\linewidth of x, yshift=0.07\linewidth] {\includegraphics[width=0.26\linewidth, ]{figs/imgs/AXFLAIR-Skyra.jpg}};
                \node[below = -0.03\linewidth of a, align=center, font=\scriptsize] {\textbf{$P$: Brain}};
                \node[rotate=90] [left = 0.01\linewidth of x, xshift=0.22\linewidth] {\scriptsize SSIM on $P$};
                
                \node (y) [below = -0.05\linewidth of x]  {\includegraphics[width=0.6\linewidth]{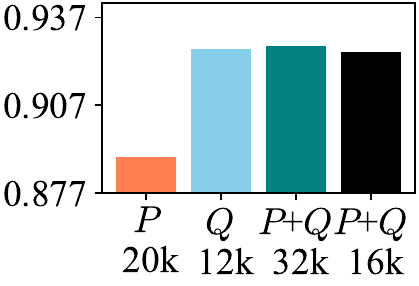}};
                \node (b) [right = -0.04\linewidth of y, yshift=0.07\linewidth]  {\includegraphics[width=0.26\linewidth, ]{figs/imgs/CORPD_FBK-Skyra.jpg}};
                \node[below = -0.03\linewidth of b, align=center, font=\scriptsize] {\textbf{$Q$: Knee}};
                \node[rotate=90] [left = 0.01\linewidth of y, xshift=0.22\linewidth] {\scriptsize SSIM on $Q$};
            \end{tikzpicture}
        \end{minipage}
    \end{minipage}\hfill
    \begin{minipage}[c]{0.32\linewidth}
        \begin{minipage}[c]{1\linewidth}
            \begin{tikzpicture}
                \node (x) at (0.0\linewidth,0.0\linewidth) {\includegraphics[width=0.6\linewidth]{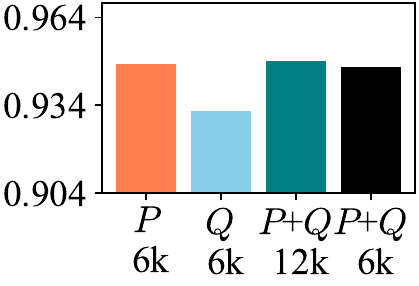}};
                \node[above = -0.03\linewidth of x, xshift=0.2\linewidth] {\footnotesize Contrasts};
                \node (a) [right = -0.04\linewidth of x, yshift=0.07\linewidth]  {\includegraphics[width=0.26\linewidth, ]{figs/imgs/CORPD_FBK-Skyra.jpg}};
                \node[below = -0.03\linewidth of a, align=center, font=\scriptsize] {\textbf{$P$: PD}};
                \node[rotate=90] [left = 0.01\linewidth of x, xshift=0.22\linewidth] {\scriptsize SSIM on $P$};
                
                \node (y) [below = -0.05\linewidth of x]  {\includegraphics[width=0.6\linewidth]{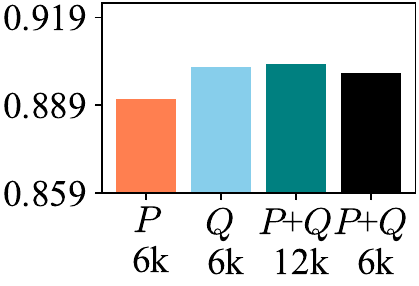}};
                \node (b) [right = -0.04\linewidth of y, yshift=0.07\linewidth]  {\includegraphics[width=0.26\linewidth, ]{figs/imgs/CORPDFS_FBK-Skyra.jpg}};
                \node[below = -0.03\linewidth of b, align=center, font=\scriptsize] {\textbf{$Q$: PDFS}};
                \node[rotate=90] [left = 0.01\linewidth of y, xshift=0.22\linewidth] {\scriptsize SSIM on $Q$};
            \end{tikzpicture}
        \end{minipage}
    \end{minipage}\hfill
     \begin{minipage}[c]{0.32\linewidth}
        \begin{minipage}[c]{1\linewidth}
            \begin{tikzpicture}
                \node (x) at (0.0\linewidth,0.0\linewidth) {\includegraphics[width=0.6\linewidth]{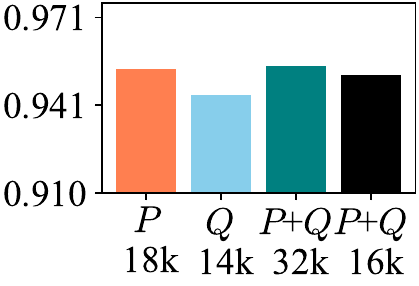}};
                \node[above = -0.03\linewidth of x, xshift=0.2\linewidth] {\footnotesize Magnetic field};
                \node (a) [right = -0.04\linewidth of x, yshift=0.07\linewidth]  {\includegraphics[width=0.26\linewidth, ]{figs/imgs/AXT2-Prisma_fit.jpg}};
                \node[below = -0.03\linewidth of a, align=center, font=\scriptsize] {\textbf{$P$: 3.0T}};
                \node[rotate=90] [left = 0.01\linewidth of x, xshift=0.22\linewidth] {\scriptsize SSIM on $P$};
                
                \node (y) [below = -0.05\linewidth of x]  {\includegraphics[width=0.6\linewidth]{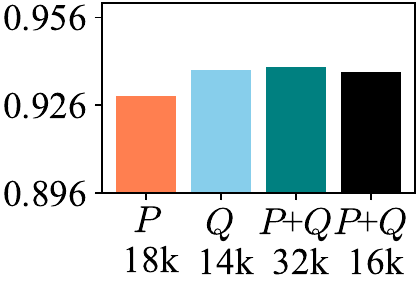}};
                \node (b) [right = -0.04\linewidth of y, yshift=0.07\linewidth]  {\includegraphics[width=0.26\linewidth, ]{figs/imgs/AXT2-Aera.jpg}};
                \node[below = -0.03\linewidth of b, align=center, font=\scriptsize] {\textbf{$Q$: 1.5T}};
                \node[rotate=90] [left = 0.01\linewidth of y, xshift=0.22\linewidth] {\scriptsize SSIM on $Q$};
            \end{tikzpicture}
        \end{minipage}
    \end{minipage}\hfill\\
    \caption{ 
     The {\color{orange} \textbf{orange}} and {\color{skyblue} \textbf{blue}} bars are the VarNet (\textbf{Top}) and ViT (\textbf{Bottom}) trained exclusively on data from {\color{orange} $P$} ({\color{orange} $\setD_P$}) and {\color{skyblue} $Q$} ({\color{skyblue} $\setD_Q$}), respectively, and the {\color{teal} \textbf{teal}} bars are the models trained on both sets {\color{teal}$\setD_P \cup \setD_Q$}. 
    As a reference point, the {\color{black} \textbf{black}} bars are the performance of models trained on random samples of $\setD_P \cup \setD_Q$ of \textbf{half the size}. The number below each bar is the total number of training images. It can be seen that we are in the high-data regime where increasing the dataset further gives minor improvements. 
    For all distributions, the joint model trained on $P$ and $Q$ performs as well on $P$ and $Q$ as the models trained individually for each of those distributions.
    }
    \label{fig:sep_varnet_vit}
\end{figure}
\begin{figure}[t!]
\begin{minipage}[c]{1\linewidth}
    \begin{minipage}[c]{1\linewidth}
        \par\noindent\rule{\linewidth}{0.4pt}
        \footnotesize \textbf{Small dataset}
    \end{minipage}\\
    \begin{minipage}[c]{0.32\linewidth}
        \begin{minipage}[c]{1\linewidth}
            \begin{tikzpicture}
                \node (x) at (0.0\linewidth,0.0\linewidth) {\includegraphics[width=0.6\linewidth]{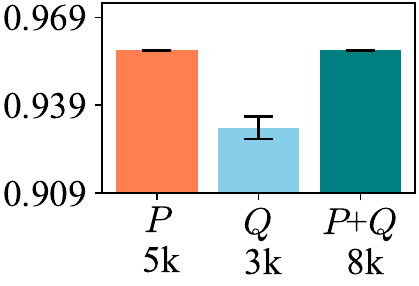}};
                \node[above = -0.03\linewidth of x, xshift=0.2\linewidth] {\footnotesize Anatomy};
                \node (a) [right = -0.04\linewidth of x, yshift=0.07\linewidth] {\includegraphics[width=0.26\linewidth, ]{figs/imgs/AXFLAIR-Skyra.jpg}};
                \node[below = -0.03\linewidth of a, align=center, font=\scriptsize] {\textbf{$P$: Brain}};
                \node[rotate=90] [left = 0.01\linewidth of x, xshift=0.22\linewidth] {\scriptsize SSIM on $P$};
                
                \node (y) [below = -0.05\linewidth of x]  {\includegraphics[width=0.6\linewidth]{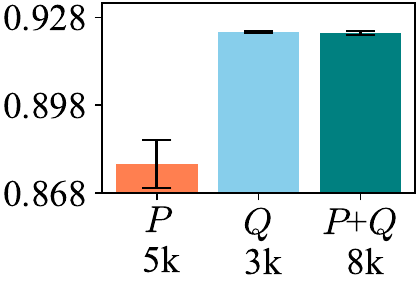}};
                \node (b) [right = -0.04\linewidth of y, yshift=0.07\linewidth]  {\includegraphics[width=0.26\linewidth, ]{figs/imgs/CORPD_FBK-Skyra.jpg}};
                \node[below = -0.03\linewidth of b, align=center, font=\scriptsize] {\textbf{$Q$: Knee}};
                \node[rotate=90] [left = 0.01\linewidth of y, xshift=0.22\linewidth] {\scriptsize SSIM on $Q$};
            \end{tikzpicture}
        \end{minipage}
    \end{minipage}
    \hfill
    \begin{minipage}[c]{0.32\linewidth}
        \begin{minipage}[c]{1\linewidth}
            \begin{tikzpicture}
                \node (x) at (0.0\linewidth,0.0\linewidth) {\includegraphics[width=0.6\linewidth]{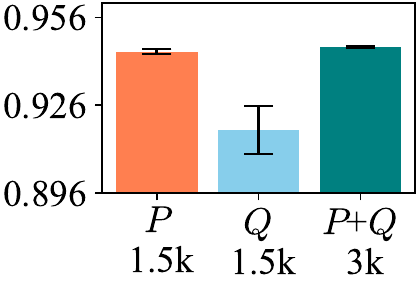}};
                \node[above = -0.03\linewidth of x, xshift=0.2\linewidth] {\footnotesize Contrasts};
                \node (a) [right = -0.04\linewidth of x, yshift=0.07\linewidth]  {\includegraphics[width=0.26\linewidth, ]{figs/imgs/CORPD_FBK-Skyra.jpg}};
                \node[below = -0.03\linewidth of a, align=center, font=\scriptsize] {\textbf{$P$: PD}};
                \node[rotate=90] [left = 0.01\linewidth of x, xshift=0.22\linewidth] {\scriptsize SSIM on $P$};
                
                \node (y) [below = -0.05\linewidth of x]  {\includegraphics[width=0.6\linewidth]{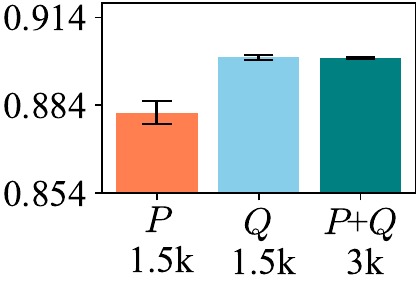}};
                \node (b) [right = -0.04\linewidth of y, yshift=0.07\linewidth]  {\includegraphics[width=0.26\linewidth, ]{figs/imgs/CORPDFS_FBK-Skyra.jpg}};
                \node[below = -0.03\linewidth of b, align=center, font=\scriptsize] {\textbf{$Q$: PDFS}};
                \node[rotate=90] [left = 0.01\linewidth of y, xshift=0.22\linewidth] {\scriptsize SSIM on $Q$};
            \end{tikzpicture}
        \end{minipage}
    \end{minipage}\hfill
     \begin{minipage}[c]{0.32\linewidth}
        \begin{minipage}[c]{1\linewidth}
            \begin{tikzpicture}
                \node (x) at (0.0\linewidth,0.0\linewidth) {\includegraphics[width=0.6\linewidth]{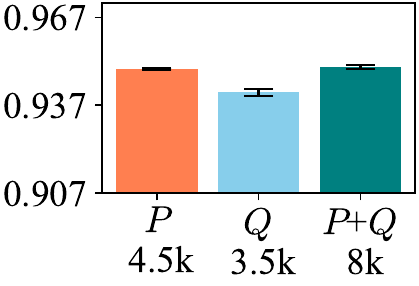}};
                \node[above = -0.03\linewidth of x, xshift=0.2\linewidth] {\footnotesize Magnetic field};
                \node (a) [right = -0.04\linewidth of x, yshift=0.07\linewidth]  {\includegraphics[width=0.26\linewidth, ]{figs/imgs/AXT2-Prisma_fit.jpg}};
                \node[below = -0.03\linewidth of a, align=center, font=\scriptsize] {\textbf{$P$: 3.0T}};
                \node[rotate=90] [left = 0.01\linewidth of x, xshift=0.22\linewidth] {\scriptsize SSIM on $P$};
                
                \node (y) [below = -0.05\linewidth of x]  {\includegraphics[width=0.6\linewidth]{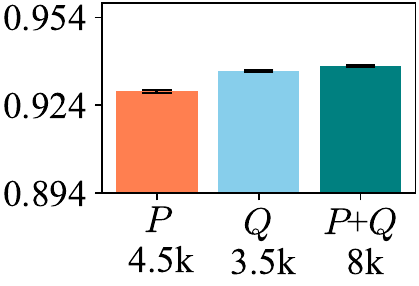}};
                \node (b) [right = -0.04\linewidth of y, yshift=0.07\linewidth]  {\includegraphics[width=0.26\linewidth, ]{figs/imgs/AXT2-Aera.jpg}};
                \node[below = -0.03\linewidth of b, align=center, font=\scriptsize] {\textbf{$Q$: 1.5T}};
                \node[rotate=90] [left = 0.01\linewidth of y, xshift=0.22\linewidth] {\scriptsize SSIM on $Q$};
            \end{tikzpicture}
        \end{minipage}
    \end{minipage}\hfill
    \begin{minipage}[c]{1\linewidth}
        \par\noindent\rule{\linewidth}{0.4pt}
        \footnotesize \textbf{Small model}
    \end{minipage}\\
    \begin{minipage}[c]{0.32\linewidth}
        \begin{minipage}[c]{1\linewidth}
            \begin{tikzpicture}
                \node (x) at (0.0\linewidth,0.0\linewidth) {\includegraphics[width=0.6\linewidth]{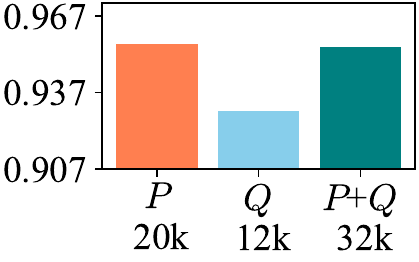}};
                \node[above = -0.03\linewidth of x, xshift=0.2\linewidth] {\footnotesize Anatomy};
                \node (a) [right = -0.04\linewidth of x, yshift=0.07\linewidth] {\includegraphics[width=0.26\linewidth, ]{figs/imgs/AXFLAIR-Skyra.jpg}};
                \node[below = -0.03\linewidth of a, align=center, font=\scriptsize] {\textbf{$P$: Brain}};
                \node[rotate=90] [left = 0.01\linewidth of x, xshift=0.22\linewidth] {\scriptsize SSIM on $P$};
                
                \node (y) [below = -0.05\linewidth of x]  {\includegraphics[width=0.6\linewidth]{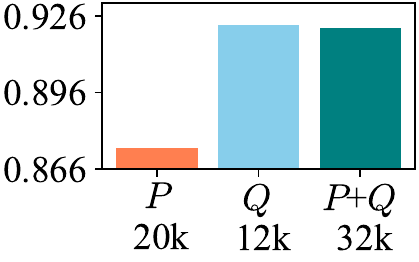}};
                \node (b) [right = -0.04\linewidth of y, yshift=0.07\linewidth]  {\includegraphics[width=0.26\linewidth, ]{figs/imgs/CORPD_FBK-Skyra.jpg}};
                \node[below = -0.03\linewidth of b, align=center, font=\scriptsize] {\textbf{$Q$: Knee}};
                \node[rotate=90] [left = 0.01\linewidth of y, xshift=0.22\linewidth] {\scriptsize SSIM on $Q$};
            \end{tikzpicture}
        \end{minipage}
    \end{minipage}\hfill
    \begin{minipage}[c]{0.32\linewidth}
        \begin{minipage}[c]{1\linewidth}
            \begin{tikzpicture}
                \node (x) at (0.0\linewidth,0.0\linewidth) {\includegraphics[width=0.6\linewidth]{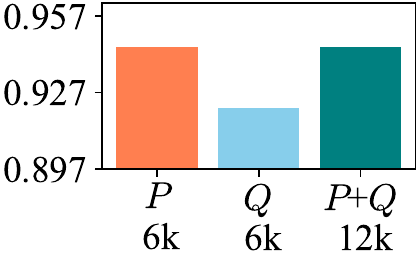}};
                \node[above = -0.03\linewidth of x, xshift=0.2\linewidth] {\footnotesize Contrasts};
                \node (a) [right = -0.04\linewidth of x, yshift=0.07\linewidth]  {\includegraphics[width=0.26\linewidth, ]{figs/imgs/CORPD_FBK-Skyra.jpg}};
                \node[below = -0.03\linewidth of a, align=center, font=\scriptsize] {\textbf{$P$: PD}};
                \node[rotate=90] [left = 0.01\linewidth of x, xshift=0.22\linewidth] {\scriptsize SSIM on $P$};
                
                \node (y) [below = -0.05\linewidth of x]  {\includegraphics[width=0.6\linewidth]{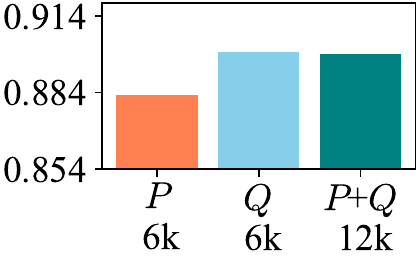}};
                \node (b) [right = -0.04\linewidth of y, yshift=0.07\linewidth]  {\includegraphics[width=0.26\linewidth, ]{figs/imgs/CORPDFS_FBK-Skyra.jpg}};
                \node[below = -0.03\linewidth of b, align=center, font=\scriptsize] {\textbf{$Q$: PDFS}};
                \node[rotate=90] [left = 0.01\linewidth of y, xshift=0.22\linewidth] {\scriptsize SSIM on $Q$};
            \end{tikzpicture}
        \end{minipage}
    \end{minipage}\hfill
     \begin{minipage}[c]{0.32\linewidth}
        \begin{minipage}[c]{1\linewidth}
            \begin{tikzpicture}
                \node (x) at (0.0\linewidth,0.0\linewidth) {\includegraphics[width=0.6\linewidth]{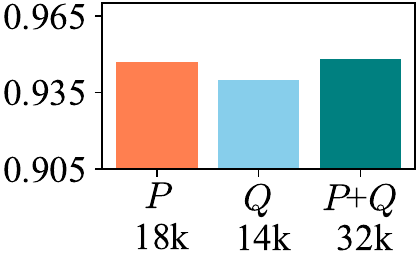}};
                \node[above = -0.03\linewidth of x, xshift=0.2\linewidth] {\footnotesize Magnetic field};
                \node (a) [right = -0.04\linewidth of x, yshift=0.07\linewidth]  {\includegraphics[width=0.26\linewidth, ]{figs/imgs/AXT2-Prisma_fit.jpg}};
                \node[below = -0.03\linewidth of a, align=center, font=\scriptsize] {\textbf{$P$: 3.0T}};
                \node[rotate=90] [left = 0.01\linewidth of x, xshift=0.22\linewidth] {\scriptsize SSIM on $P$};
                
                \node (y) [below = -0.05\linewidth of x]  {\includegraphics[width=0.6\linewidth]{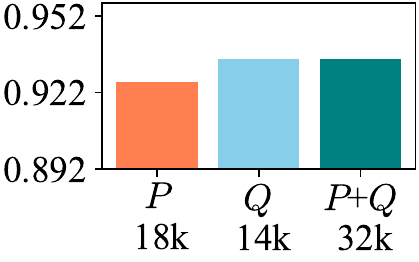}};
                \node (b) [right = -0.04\linewidth of y, yshift=0.07\linewidth]  {\includegraphics[width=0.26\linewidth, ]{figs/imgs/AXT2-Aera.jpg}};
                \node[below = -0.03\linewidth of b, align=center, font=\scriptsize] {\textbf{$Q$: 1.5T}};
                \node[rotate=90] [left = 0.01\linewidth of y, xshift=0.22\linewidth] {\scriptsize SSIM on $Q$};
            \end{tikzpicture}
        \end{minipage}
    \end{minipage}\hfill
    \caption{
     Also for smaller sized datasets (\textbf{Top}) or small models (U-net with 757k parameters, \textbf{Bottom}), a single model performs at least as good as separate models. We report the mean $\pm$ two standard deviations from five runs with the U-net, each with a different random seed for sampling training data and model initialization. Note that when training on datasets with more than 3k images, there is next to no variation.
    }
    \label{fig:sep_unet_ci}
\end{minipage}
\end{figure}
\section{Experimental Details and Additional Results for Section~\ref{sec:separate}}
\subsection{Data Preparation}\label{app:dataprep_fm}

For each of the distributions in Figure~\ref{fig:sources}, we randomly sample volumes from the fastMRI training set for training so that the total number of slices is around 2048, and we  randomly sample from the validation set for testing so that the number of test slices is around 128. Training sets of combination of distributions are then constructed by aggregating the training data from the individual distributions. For example, if we consider distribution $P$ to contain all six knee distributions from Figure~\ref{fig:sources}, then the corresponding training set has $6 \cdot 2048$ training images. Likewise, if $Q$ is for example all T2-weighted brain images the corresponding training set has $5 \cdot 2048$ training images.

\subsection{Models, Training, and Evaluation}\label{app:models_training}
Our configuration of the end-to-end VarNet~\citep{sriramEndtoEndVariationalNetworks2020b} contains 8 cascades, each containing an U-net with 4 pooling layers and 12 channels in the first pooling layer. The sensitivity-map U-net of the VarNet has 4 pooling layers and 9 channels in the first pooling layer. The code for the model is taken from \href{https://github.com/facebookresearch/fastMRI}{fastMRI's GitHub repository}.

The U-nets used in the experiments have 4 pooling layers and 32 channels in the first pooling layer. The implementation of the model is taken from the \href{https://github.com/facebookresearch/fastMRI}{fastMRI GitHub repository}. Our configuration of the vision transformer~\cite{dosovitskiyImageWorth16x162021b} for image reconstruction is the ViT-S configuration from~\citet{linVisionTransformersEnable2022}, and the code is taken from the \href{https://github.com/MLI-lab/transformers_for_imaging}{paper's GitHub repository}. 
As input data for the U-net and ViT, we first fill missing k-space values with zeros, then apply 2D-IFFT, followed by a root-sum-of-squares (RSS) reconstructions to combine all the coil images into one single image, and lastly normalize it to zero-mean and unit-variance. The mean and variance are added and multiplied back to the model output, respectively. This is a standard prepossessing step, see for example \href{https://github.com/facebookresearch/fastMRI}{fastMRI's GitHub repository}. The models are trained end-to-end with the objective to maximize SSIM between output and ground-truth.

For any model and any choice of distributions $P$ or $Q$, the models are trained to maximize SSIM between model output and RSS target for a total of 60 epochs and we use the Adam optimizer with $\beta_1= 0.9$ and $\beta_2=0.999$. The mini-batch size is set to 1. We use linear learning-rate warm-up until a learning-rate of 1e-3 is reached and linearly decay the learning rate to 4e-5. The warm up period amounts to 1\% of the total number of gradient steps. Gradients are clipped at a global $\ell_2$-norm of 1. During training, we randomly sample a different undersampling mask for each mini-batch independently. During evaluation, for each volume we generated an undersampling mask randomly, and this mask is then used for all slices within the volume.

The maximal learning-rate for each model is tuned based on a grid search on the values \{1.3e-3, 1e-3, 7e-4, 4e-4\} and training on a random subset (2k slices) of the fastMRI dataset. We found negligible differences between learning rates \{1.3e-3, 1e-3, 7e-4\} and therefore keep the learning rate to 1e-3 for simplicity. We also performed the same grid search on fastMRI subsets for PD-weighted knee and PDFS knee scans and made the same observations.

\subsection{Results for VarNet and ViT} \label{app:single_vs_separate}

In the main body we presented results for the U-net, here we present results for the VarNet and ViT. 
In Figure~\ref{fig:sep_varnet_vit}, we see that, like for the U-net discussed in the main body, for the VarNet and ViT training on two distributions gives the same performance as separate models trained on the individual distributions. Moreover, Figure~\ref{fig:sep_unet_ci} shows the same experiment for the U-net on smaller datasets and smaller models, where the same observation can be made.

\section{Distribution-Shifts Induced by Changes of the Forward Model} \label{app:fwd_map_shift}
In the main body, we considered distribution-shifts primarily related to the images, such as different contrast and  anatomies. In this section, we consider two distribution-shifts that are induced by changes of the forward model, i.e., the relation of measurements and object to be imaged. We consider shifts in the acceleration factor and a distribution-shift related to the number of coils. 

\begin{table}[t!]
    \centering
    \small
    \captionof{table}{
    A model trained on data with various acceleration factors (2, 4, 8, 16-fold) performs comparable as models trained individually for each acceleration factor. 
    Additionally, the model trained on various acceleration factors can enhance performance for unseen acceleration factors (3-fold). Models were trained and evaluated on the fastMRI PDFS knee subset.
    }
    \label{tbl:comb_accel}
    \begin{adjustbox}{max width=\linewidth}
    \begin{tabular}{l r r r r r}
    \toprule
    \diagbox[width=0.12\linewidth]{Train}{Test} & 2-fold & 4-fold & 8-fold & 16-fold & 3-fold \\
    \midrule
    2-fold & 0.945 & --- & --- & --- & 0.906 \\
    4-fold & --- & 0.903 & --- & --- & 0.899 \\
    8-fold & --- & --- & 0.867 & --- & 0.834 \\
    16-fold & --- & --- & --- & 0.828 & 0.758 \\
    All of above & 0.944 & 0.902 & 0.866 & 0.829 & 0.912 \\
    \midrule
    3-fold & --- & --- & --- & --- & 0.921 \\
    \bottomrule
    \end{tabular}
    \end{adjustbox}
\end{table}
\begin{figure}[t!]
    \centering
    \begin{minipage}[c]{0.01\linewidth}
        \begin{tikzpicture}
            \node[rotate=90] at (-0.2,0) {\scriptsize SSIM on $Q$};
        \end{tikzpicture}
    \end{minipage}
    \begin{minipage}[c]{0.36\linewidth}
        \begin{tikzpicture}
            \node (x) at (0.0\linewidth,0.0\linewidth) 
             {\includegraphics[width=0.6\linewidth]{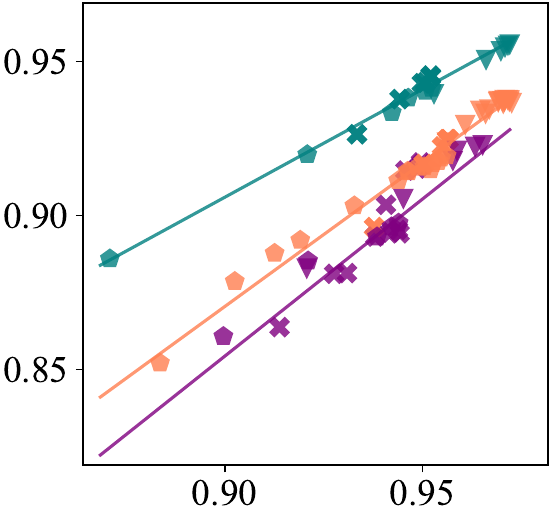}};
            \node[above = -0.04\linewidth of x, xshift=0.05\linewidth] {\scriptsize Coil and anatomy shift};
            \node[below = -0.04\linewidth of x, xshift=0.05\linewidth] {\scriptsize SSIM on $P_\text{best}$};
            \node (a) [right = 0.04\linewidth of x, yshift=0.18\linewidth]  {\includegraphics[width=0.2\linewidth, ]{figs/imgs/CORPD_FBK-Skyra.jpg}};   
            \node (b) [right = 0.02\linewidth of x, yshift=0.15\linewidth]  {\includegraphics[width=0.2\linewidth, ]{figs/imgs/CORPDFS_FBK-Skyra.jpg}};
            \node[above = 0.0\linewidth of b, align=center, font=\scriptsize] {$P$: 15 coils, \\Knee};
            \node (c) [right = 0.01\linewidth of x, yshift=-0.2\linewidth] {\includegraphics[width=0.2\linewidth, ]{figs/imgs/AXT2-Avanto.jpg}};
            \node[above = -0.04\linewidth of c, align=center, font=\scriptsize] {$Q$: 4 coils, \\Brain};
        \end{tikzpicture}
    \end{minipage}\hspace{0.01\linewidth}
    \begin{minipage}[c]{0.2\linewidth}
        \frame{\includegraphics[width=0.9\linewidth]{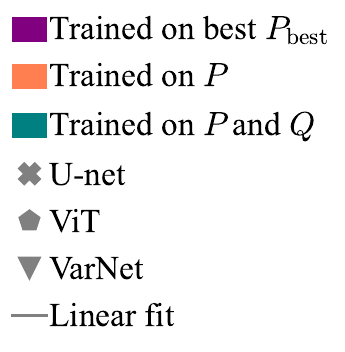}}
    \end{minipage}
    \caption{Training on a diverse dataset ($P$) increases effective robustness under distribution-shifts related to the number of coils.}
    \label{fig:coil_shift}
\end{figure}
\subsection{Results for Training With Multiple Acceleration Factors} \label{app:single_vs_separate_accel}
In the main body, we presented reults for a single accelerated factor of 4. 

We now train a (U-net) model simultaneously on data with 2-fold, 4-fold, 8-fold, and 16-fold acceleration factors, and analyze how the performance compares to U-net models trained for each acceleration separately.  
We train only on the fastMRI PDFS knee scans (see Figure~\ref{fig:sources}). We report the performance in SSIM for each model and for each acceleration factor in Table~\ref{tbl:comb_accel}. We also evaluate each model on 3-fold acceleration to see how combining different accelerations affects robustness towards a distribution-shifts related to the acceleration factor.

In Table~\ref{tbl:comb_accel}, we see that the model trained on all four acceleration factors simultaneously (5th row) yields similar performance to the models trained individually on each acceleration factor (the differences in performance are within 0.001 SSIM, which is negligible). 

For the out-of-distribution setup (i.e., evaluation on 3-fold acceleration), we observe that the model trained on all accelerations performs by 0.006 SSIM better relative to the best separately trained model (2-fold acceleration). 

Taken together, these two observations suggest that training on combinations of different acceleration factors can increase the effective robustness of a model towards distribution-shifts related to changes to the acceleration factor slightly.
\subsection{Distribution-Shifts Related to Number of Receiver Coils}\label{app:coil_shift}

We now consider a shift in the number of receiver coils. For distribution $P$, we select all knees scans collected with the 6 different combinations of image contrasts and scanners (see Figure~\ref{fig:sources}). All knee scans are collected with 15 coils. For distribution $Q$, we select the brain scans from the scanner Avanto since measurements from this scanner are collected with 4 coils.

For this distribution-shift we noticed that models, in particular VarNet which estimates sensitivity maps, struggles to accurately predict the mean value of the images resulting in a noticeable drops in SSIM. However, this degradation was hardly noticeable when looking at the reconstructions. Given that radiologists routinely adjust the brightness and contrast of MRI images during inspection through a process known as windowing~\citep{ishidaDigitalImageProcessing1984}, we normalize the model output and target to have the same mean and variance during evaluation. The results are depicted in Figure~\ref{fig:coil_shift}, where we see that training on a diverse set $P$ increases effective robustness on this distribution-shift.

\section{Measuring Similarity Between Two Datasets With CLIP Similarity} \label{app:clip_distance}

A possible explanation for performance improvements on out-of-distribution data is that a more diverse training set is more likely to contain data that is similar to out-of-distribution data and thus diverse data is `less' out-of-distribution. Training on data that is `less' out of distribution is in turn expected to increase performance. 

There are several measures for the the similarity of two image dataset. We adapt the method from \citet{mayilvahananDoesCLIPGeneralization2024}.
\citet{mayilvahananDoesCLIPGeneralization2024} demonstrate that similarity between training set and test set plays a crucial role in explaining robustness of CLIP~\cite{radfordLearningTransferableVisual2021a} models for image recognition. The method utilizes a pre-trained CLIP model to compute features of images in CLIP's image embedding space. Then given a training set and test set, this measure computes the nearest neighbor in the training set for each sample in the test set based on the cosine similarity of the CLIP image features resulting in a distribution of similarity scores---one for each sample in the test set. We then compute the mean of the histograms of those nearest neighbors. 

For this study, we rely on the pre-trained CLIP models by \citet{chertiReproducibleScalingLaws2023} and use the ViT-B/16 \citep{dosovitskiyImageWorth16x162021b} pre-trained on DataComp-1B \citep{gadreDataCompSearchNext2023a}, a large and diverse high-quality dataset that includes medical images. From the root-sum-of-square ground truth images, we first randomly extract smaller image patches of size 80$\times$80 and discard those that mainly contain background and noise. CLIP features are then computed for the remaining image patches.

\begin{figure}[t!]
\centering
    \begin{minipage}{1\linewidth}
    \centering
        \includegraphics[width=0.32\linewidth]{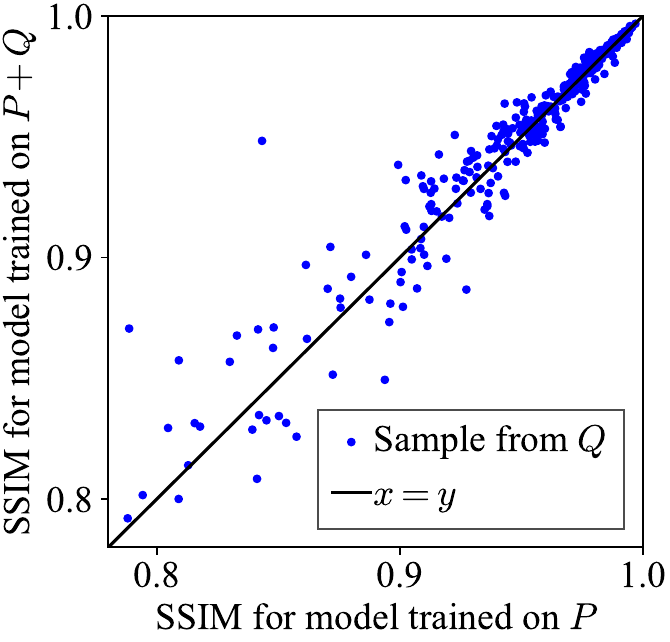}
    \caption{Reconstruction performance for each small pathology reconstructed with a VarNet trained only on data without pathologies (i.e., $P$) relative to the performance of a VarNet trained on data with and without pathologies ($P$ and $Q$). SSIM is measured only within the region containing the pathology.
    The majority of pathologies are reconstructed similarly well by both models, however in the regime where SSIM is low some images are reconstructed better by the $P$-model and others are reconstructed better by the $P+Q$-model.} 
    \label{fig:pathology_scatter}
    \end{minipage}\\
    \begin{minipage}{1\linewidth}
    \centering
        \begin{tikzpicture}
            \node (x) at (0.0\linewidth,0.0\linewidth) {\includegraphics[width=0.35\linewidth]{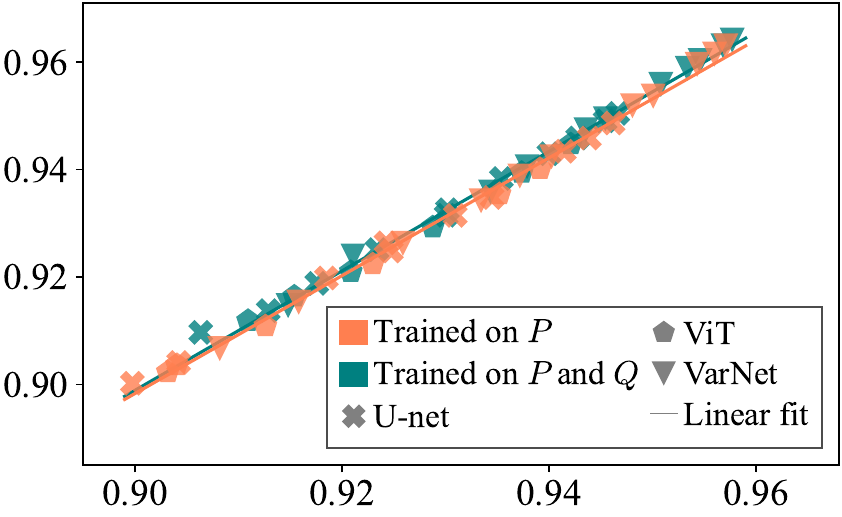}};
            \node[below = -0.01\linewidth of x, xshift=0.025\linewidth] {\small SSIM on $P$};
            \node[left = 0.005\linewidth of x, rotate=90, xshift=0.075\linewidth] {\small SSIM on $Q$};
            \node (a) [right = 0.0\linewidth of x, yshift=0.05\linewidth]  {\includegraphics[width=0.09\linewidth]{figs/pathology/normal_for_age.png}};   
            \node[above = -0.005\linewidth of a, align=center, font=\small] {$P$: w/o Pathology};
            \node (c) [right = 0.0\linewidth of x, yshift=-0.08\linewidth] {\includegraphics[width=0.09\linewidth]{figs/pathology/lacunar_infarct_ref.png}};
            \node[above = -0.005\linewidth of c] {\small $Q$: Pathology};
        \end{tikzpicture}
    \caption{Models trained only on images without pathologies and models trained on images with pathologies have similar global SSIM. Different models are sampled by varying the training set size by factors of 2, 4 and 8, and by early stopping.}
    \label{fig:pathology_global}
    \end{minipage}
\end{figure}
\begin{figure}[t!]
    \newcommand{\w}{0.33\linewidth}
    \centering
    \begin{minipage}{0.7\linewidth}
        \begin{tikzpicture}
            \node (x0) at (0.0\linewidth,0.0\linewidth) {\includegraphics[width=\w]{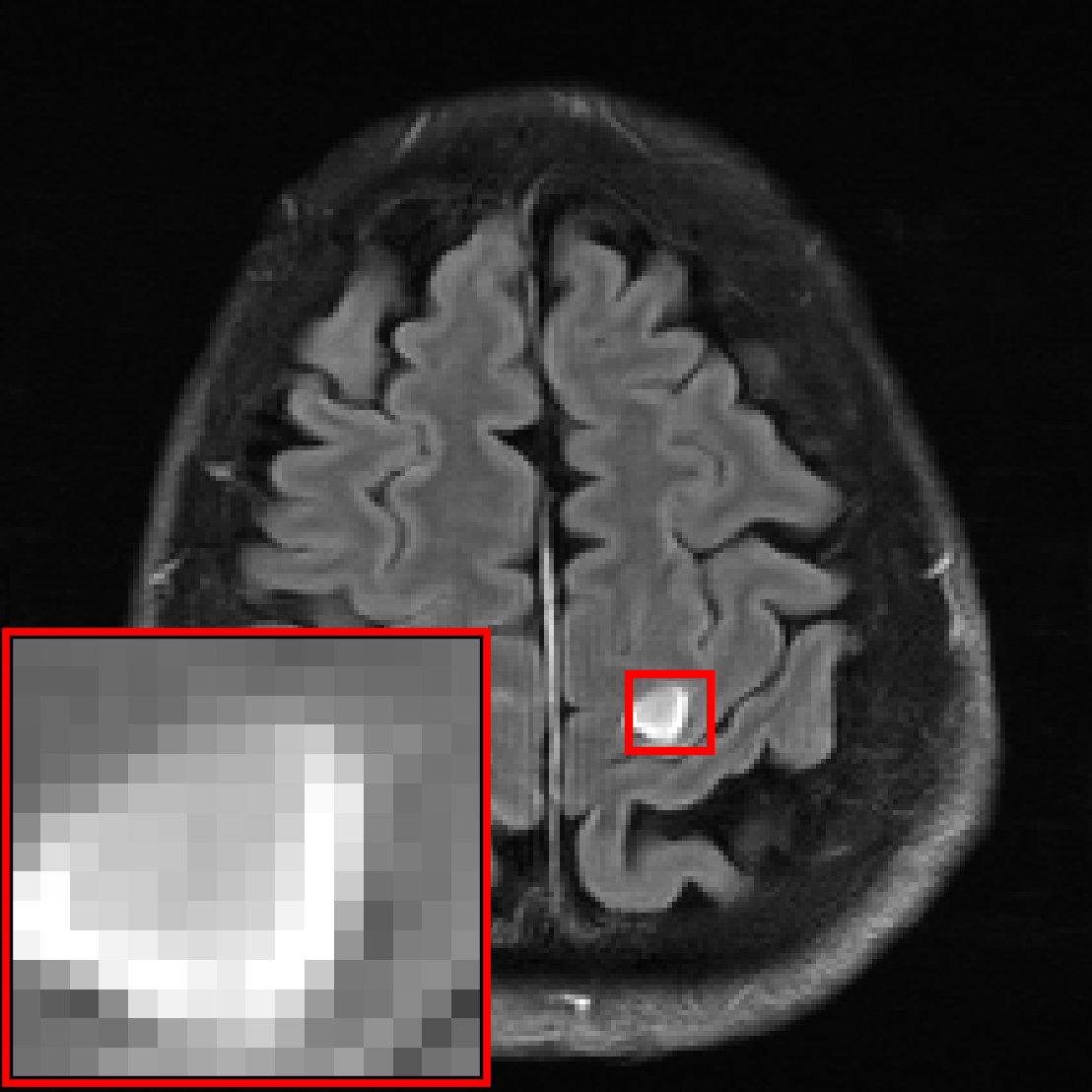}};
            \node (x1)[below = 0\linewidth of x0] {\includegraphics[width=\w]{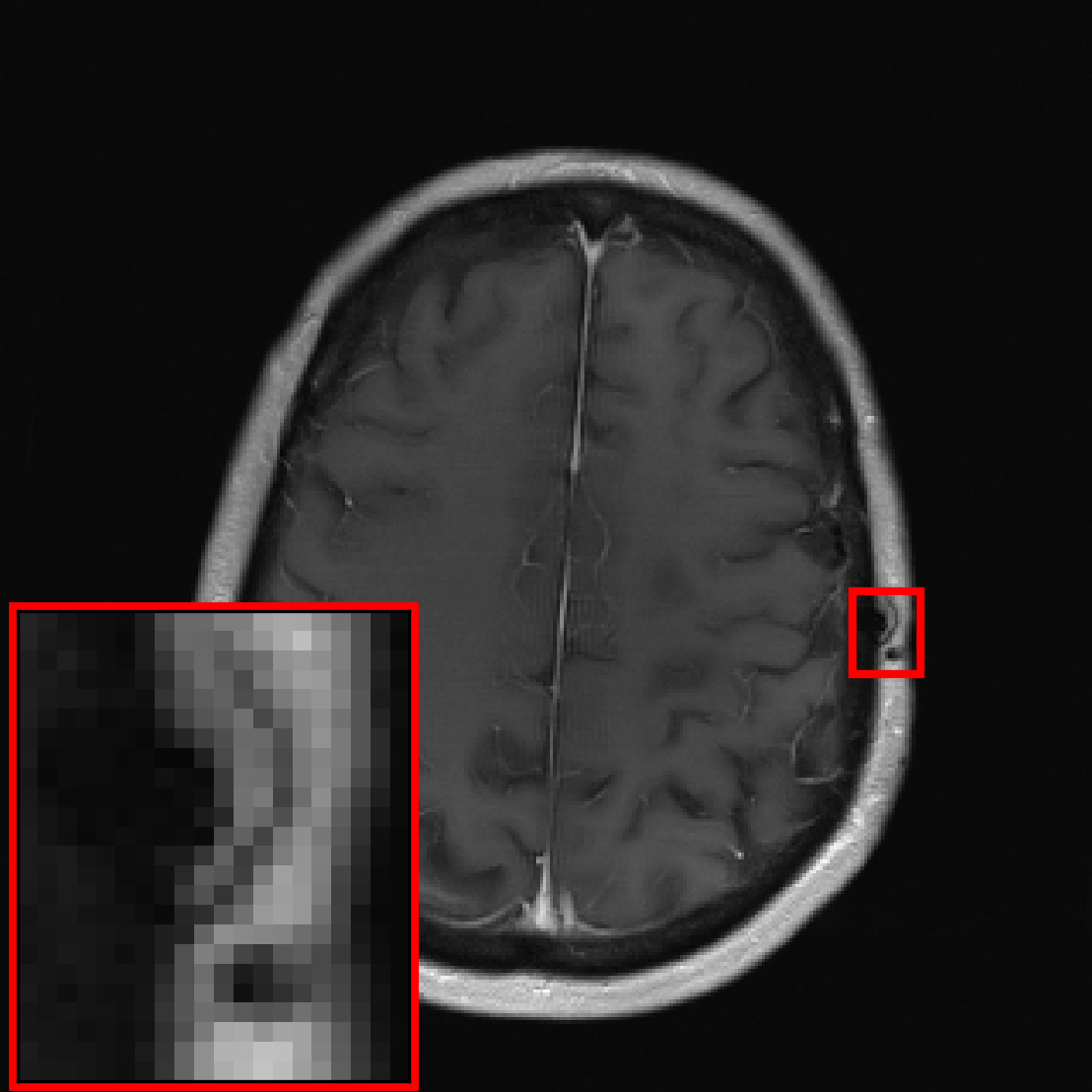}};
            \node (x2)[below = 0\linewidth of x1] {\includegraphics[width=\w]{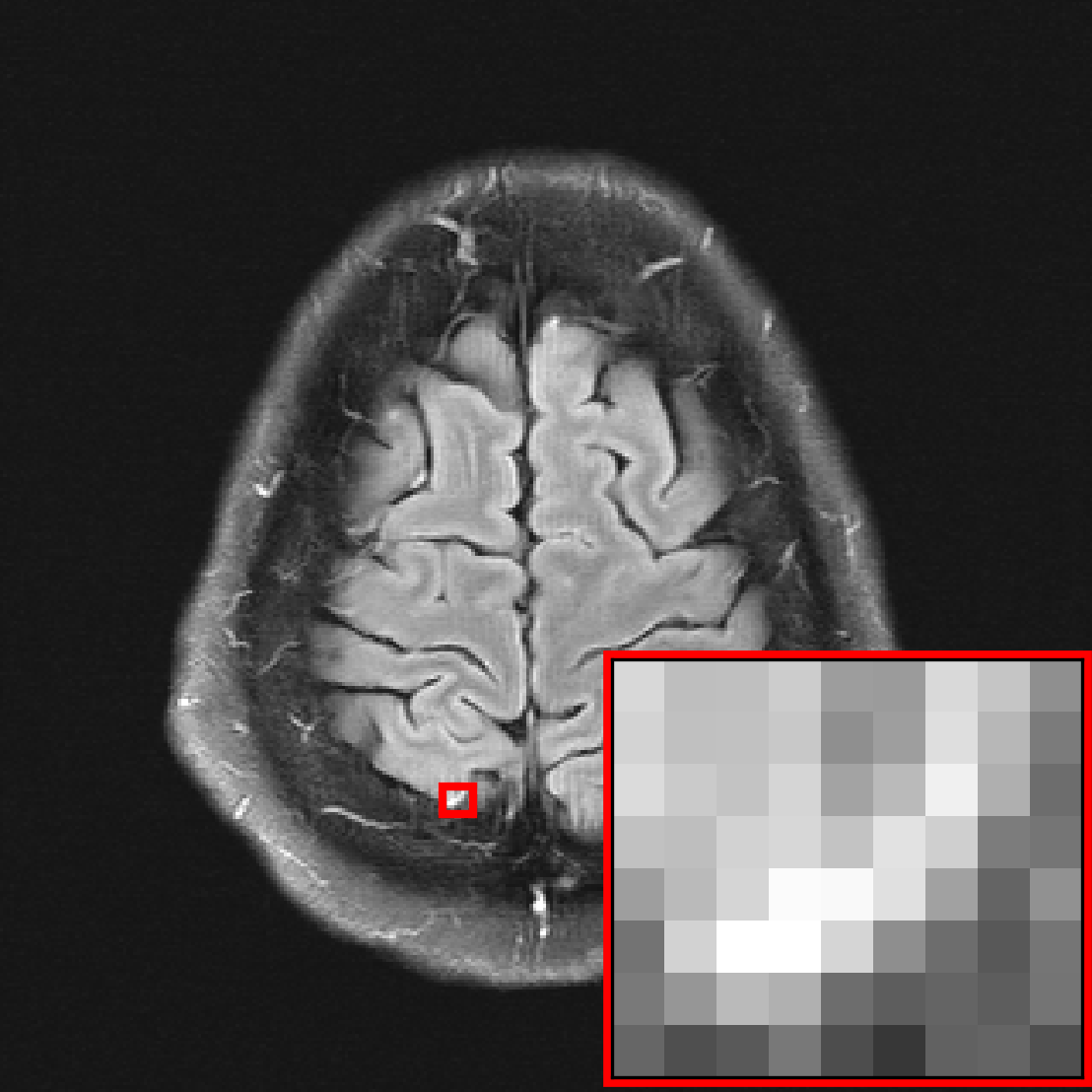}};
            
            \node[above = 0\linewidth of x0] {Trained on $P$};
            
            \node (y0) [right = 0.\linewidth of x0]
            {\includegraphics[width=\w]{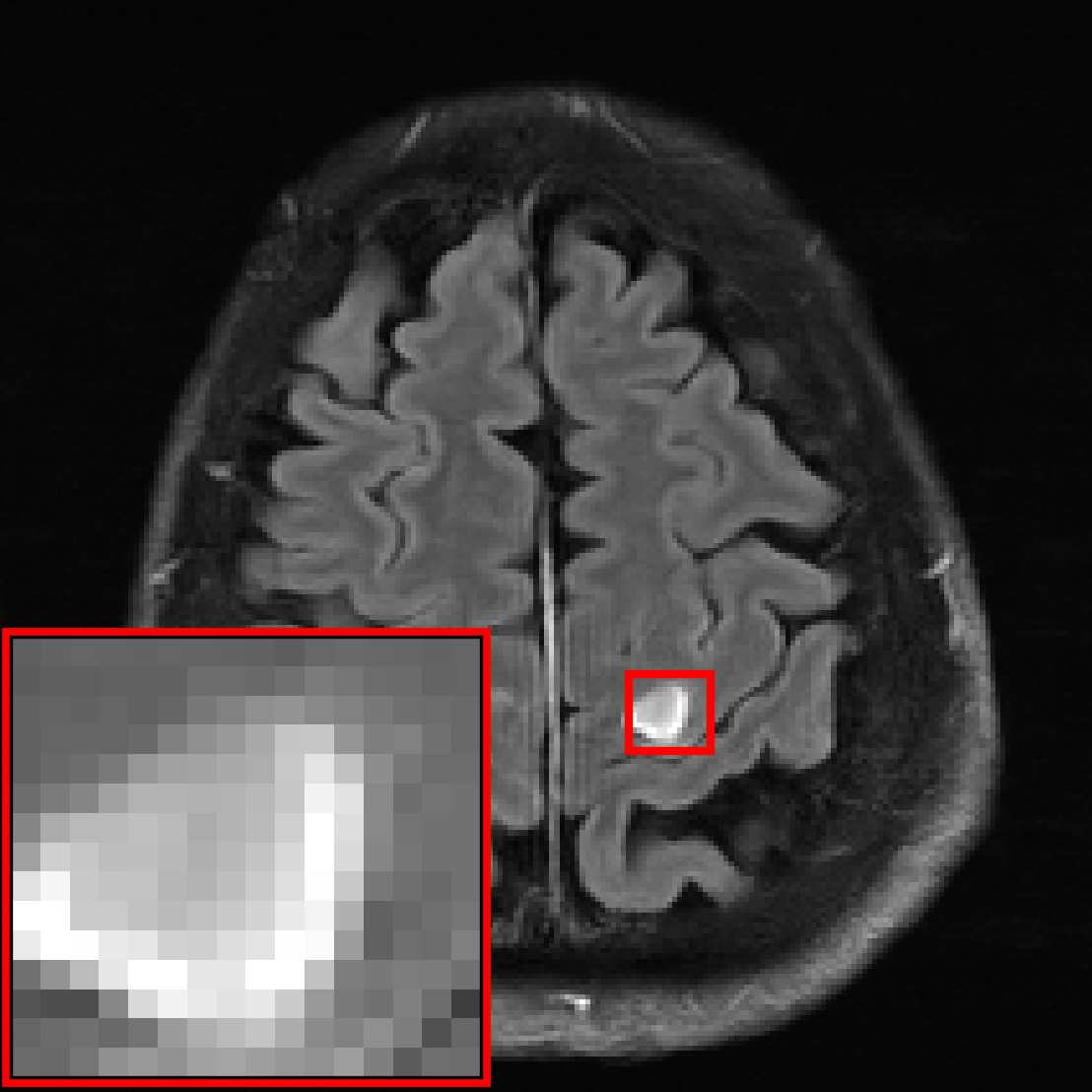}};
            \node (y1)[below = 0\linewidth of y0] {\includegraphics[width=\w]{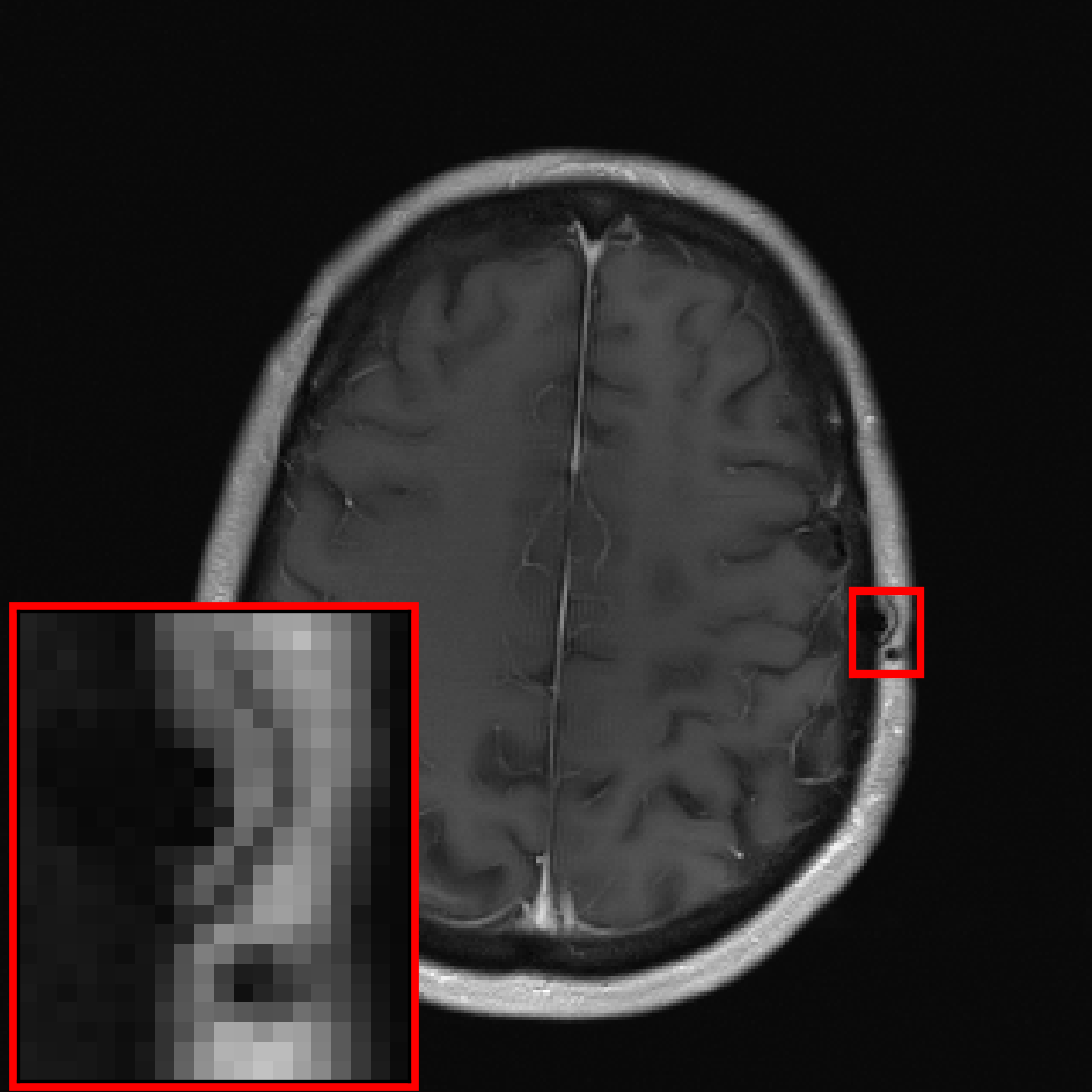}};
            \node (y2)[below = 0\linewidth of y1] {\includegraphics[width=\w]{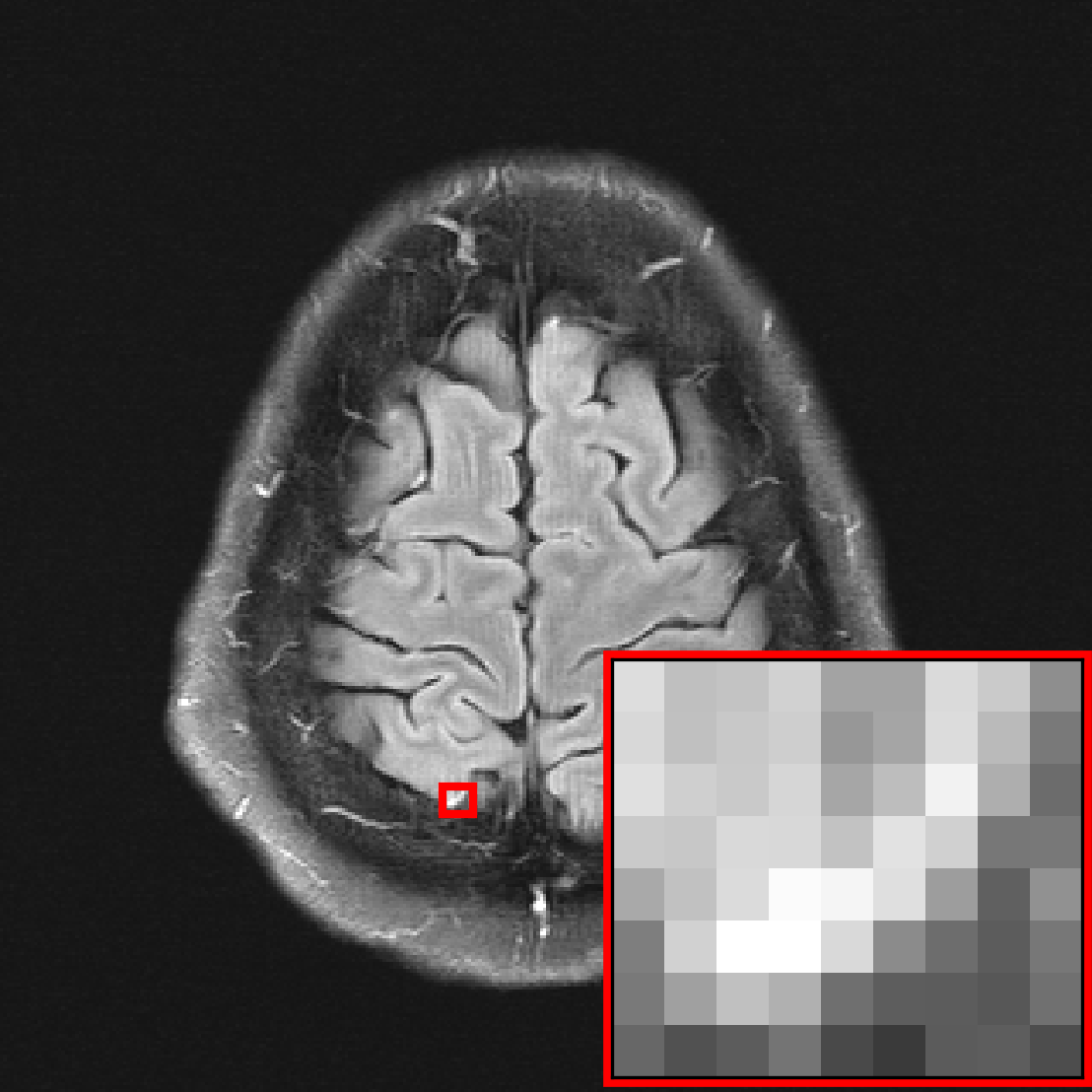}};
            
            \node[above = 0.0\linewidth of y0, yshift=-0.01\linewidth] {Trained on $P+Q$};
            
            \node (z0) [right = 0.\linewidth of y0] {\includegraphics[width=\w]{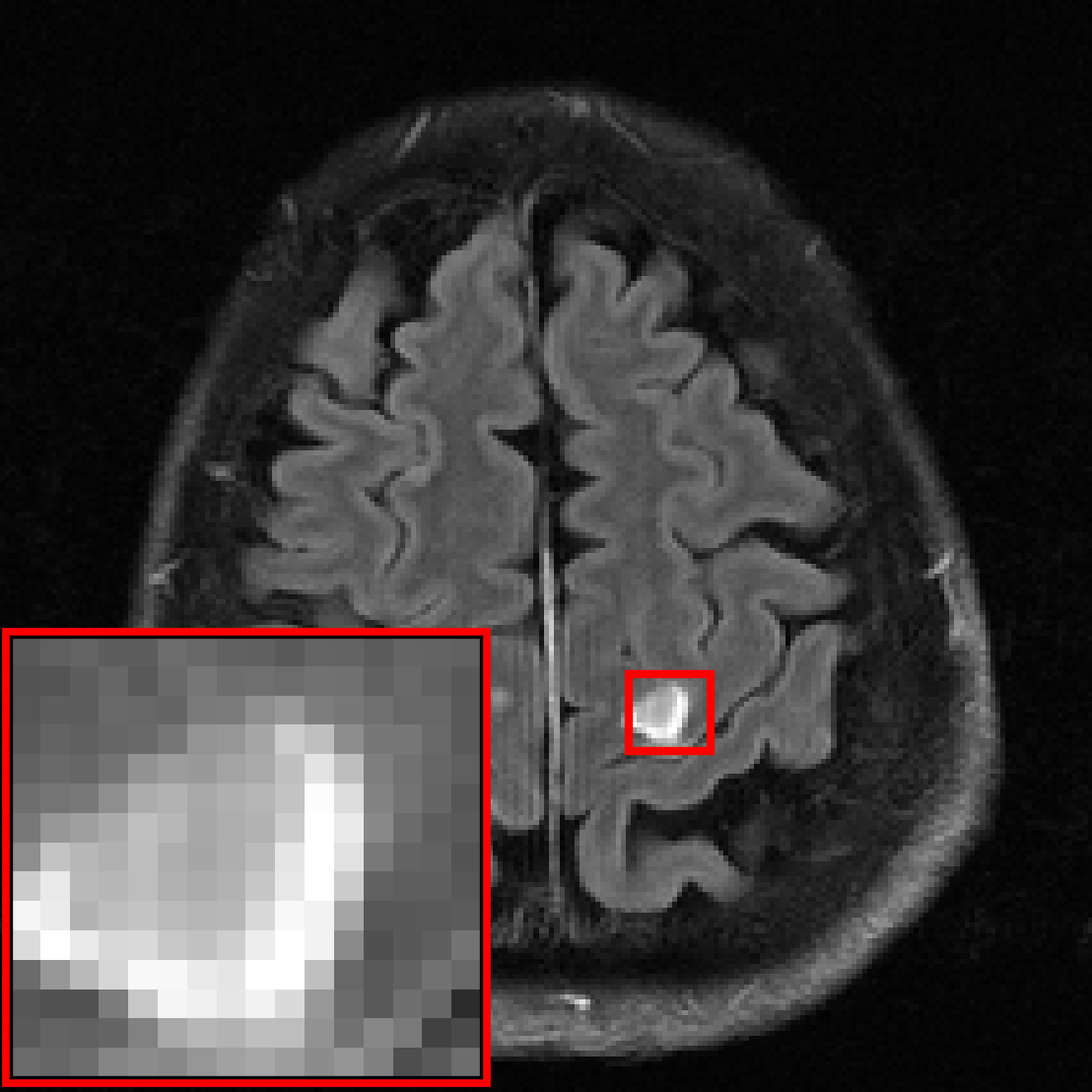}};
            \node (z1)[below = 0\linewidth of z0] {\includegraphics[width=\w]{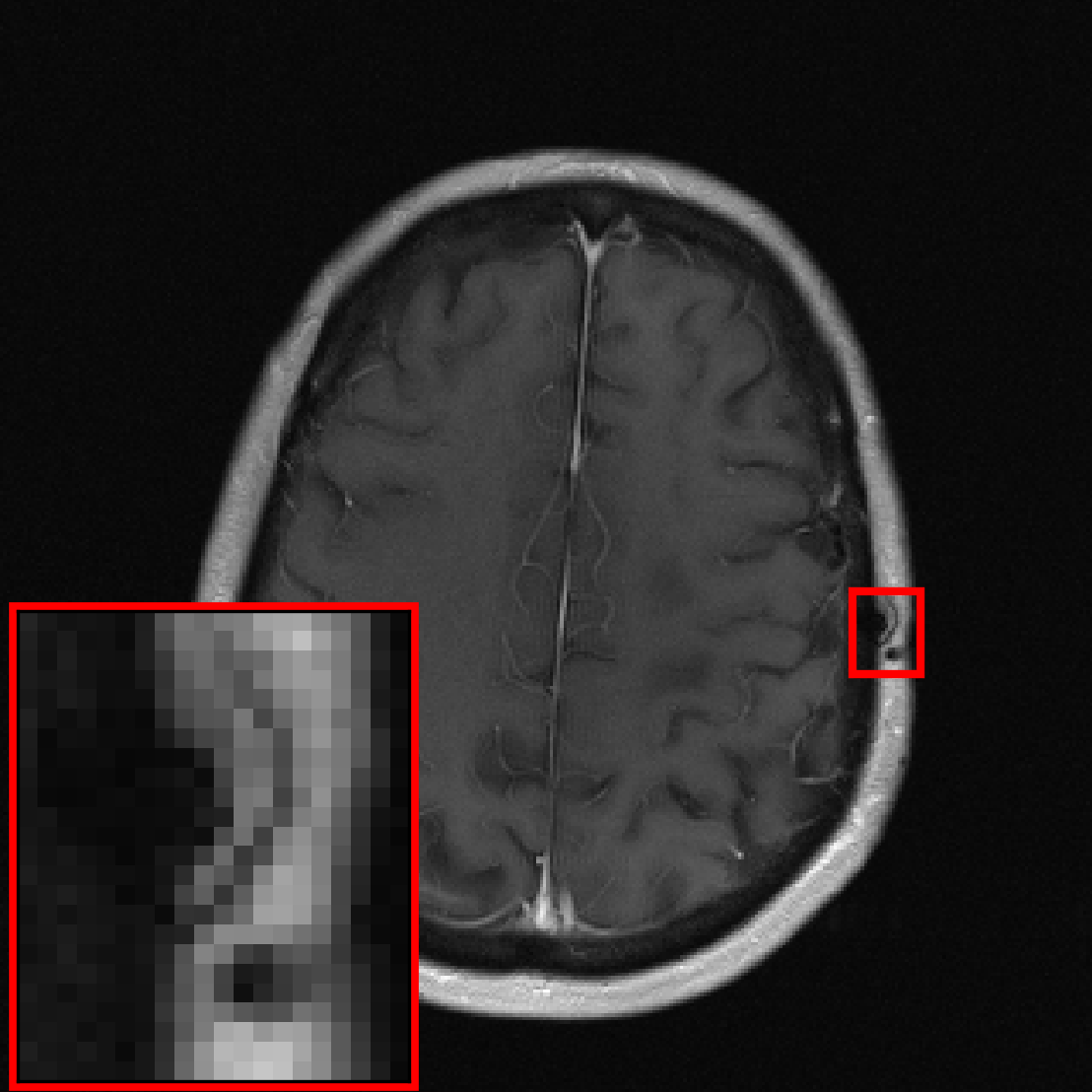}};
            \node (z2)[below = 0\linewidth of z1] {\includegraphics[width=\w]{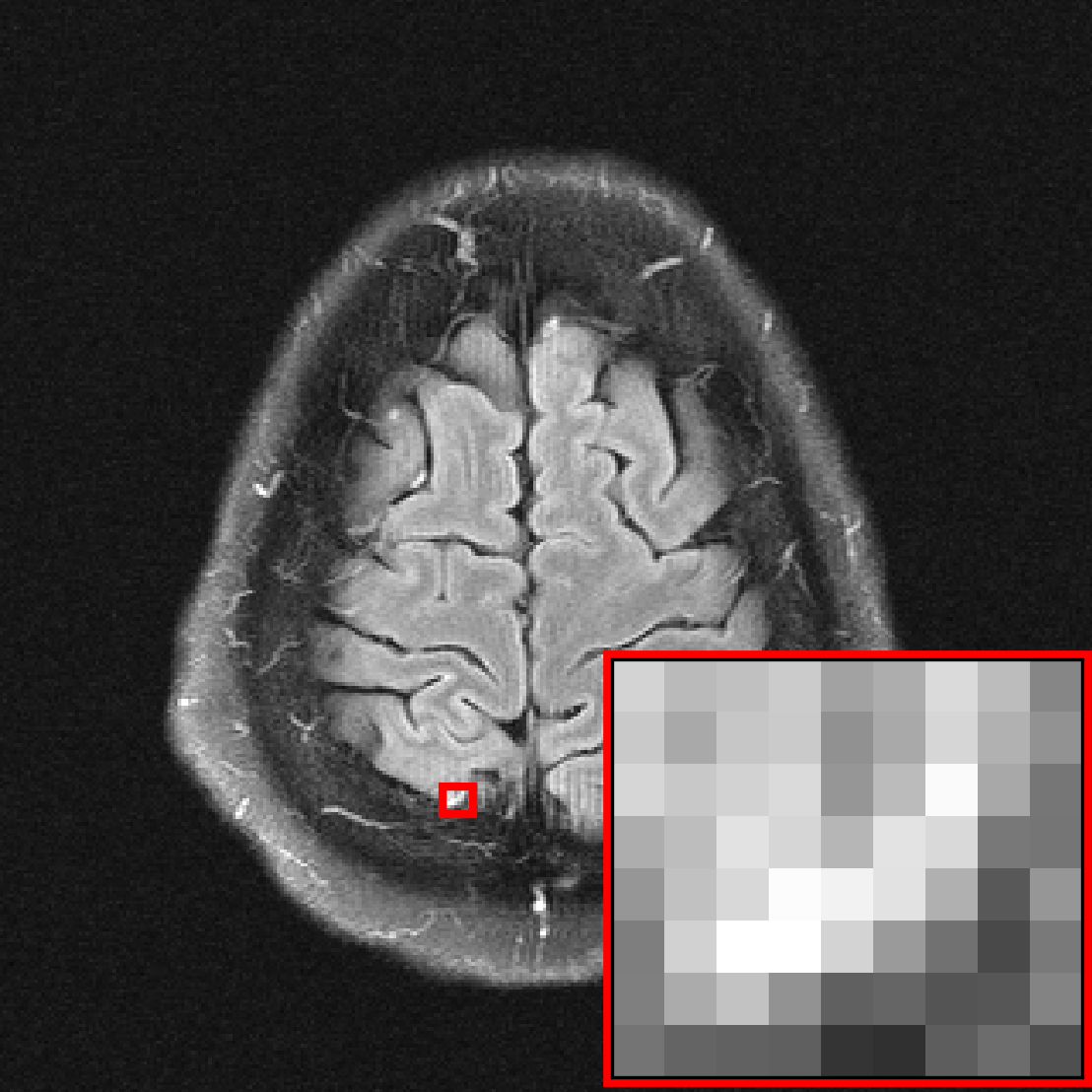}};
            
            \node[above = 0.0\linewidth of z0] {Ground truth};
            
        \end{tikzpicture}
    \end{minipage}
    \caption{Random selection of reconstructions of small pathologies, given by the VarNet when trained on images without pathologies ($P$), and on images without and with pathologies ($P+Q$).}
    \label{fig:pathology_reconstructions_small}
\end{figure}
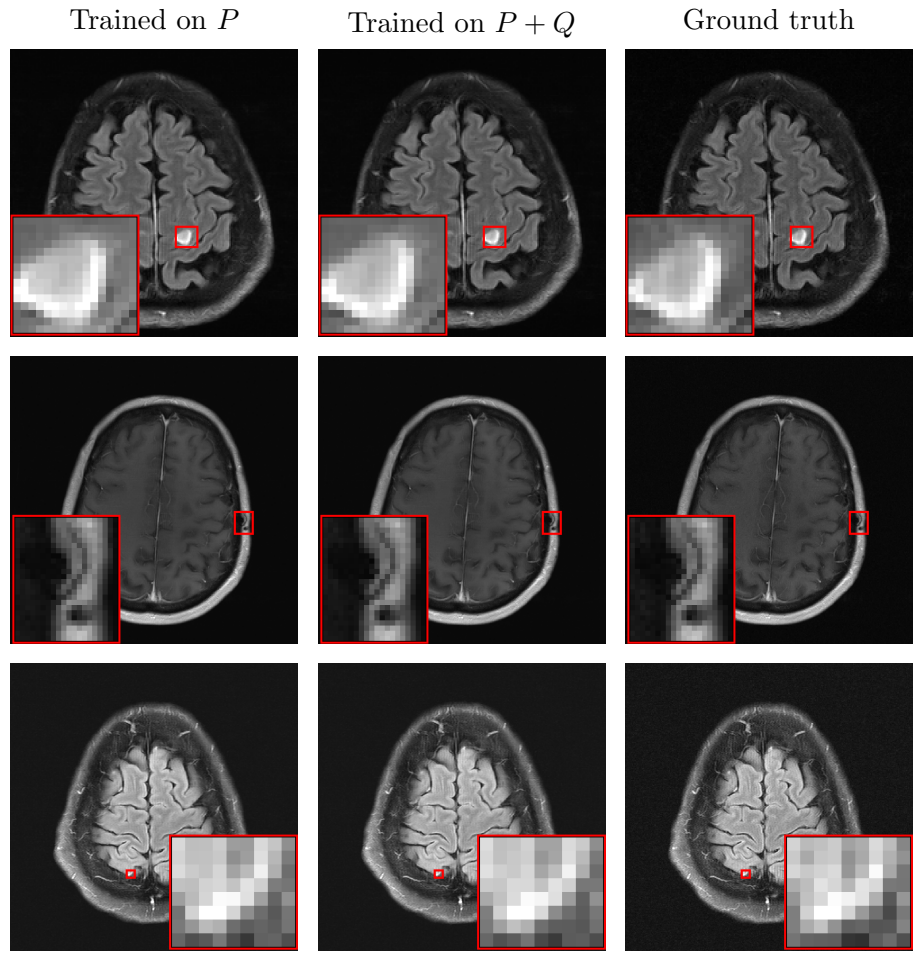
\section{Additional Results for Section~\ref{sec:pathology}}\label{app:pathology_eval}
Figure~\ref{fig:pathology_scatter} presents the reconstruction performance evaluated for individual images in the test set, focusing on small pathologies. The evaluation specifically targeted the pathology regions. Results are provided for VarNet trained solely on images without pathologies ($P$) and VarNet trained on images with and without pathologies ($P+Q$). Both models exhibit similar mean SSIM values for test images without pathology (approximately 0.957 SSIM) and also similar SSIM values for test images with small pathologies (approximately 0.948 SSIM).

Both models perform well for the majority of samples, indicated by high SSIM scores. In the low-SSIM regime where the SSIM is low for both models, some samples are better reconstructed by the model trained on $P+Q$ and some are better reconstructed by the model trained solely on $P$.

In Figure~\ref{fig:pathology_global}, we show reconstruction performance when SSIM is calculated across the entire image and not just for the pathology region as in the main body. It can be seen that even when evaluated globally, models trained on data without pathologies perform as well as models trained on data with and without pathologies.

In Figure~\ref{fig:pathology_reconstructions_small} we provide a selection of reconstruction examples for images with small pathologies, obtained by a VarNet trained on data without pathologies and a VarNet trained on data with and without pathologies. It can be seen that the reconstructions by the two models are essentially indistinguishable.

Our intuition on why models can generalize well for this particular distribution-shift is as follows: First, compared to the distributions shifts in Section~\ref{sec:robustness}, which include shifts between anatomies or image contrasts inducing strong differences in structure and content (see Figure~\ref{fig:sources}), the pathology distribution-shift is less drastic as many image characteristics stay the same, i.e, same anatomy, scanners, image modalities. Second, we believe that models for image reconstruction learn local priors rather than global ones. Pathologies consist of local patterns which are also present in other areas of an image without  pathologies. 


\begin{figure}[t]
    \centering
    \begin{minipage}[c]{0.4\linewidth}
        \begin{tikzpicture}
            \node (x) at (0.0\linewidth,0.0\linewidth) {\includegraphics[width=0.6\linewidth]{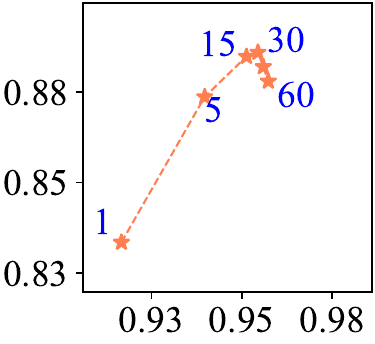}};
            \node[above = -0.03\linewidth of x, xshift=0.05\linewidth] {\small ViT};
            \node[left = 0.33\linewidth of x, rotate=90, font=\small, xshift=0.15\linewidth] at (0.\linewidth,0) {SSIM on $Q$};
            \node[below = -0.04\linewidth of x, xshift=0.05\linewidth] {\small SSIM on $P$};
            \node (a) [right = 0.04\linewidth of x, yshift=0.18\linewidth]  {\includegraphics[width=0.2\linewidth, ]{figs/imgs/AXT2-Prisma_fit.jpg}};   
            \node (b) [right = 0.01\linewidth of x, yshift=0.15\linewidth]  {\includegraphics[width=0.2\linewidth, ]{figs/imgs/AXT2-Aera.jpg}};
            \node[above = 0.0\linewidth of b, align=center, font=\small] {$P$: fm brain, T2 \phantom{1}};
            \node (c) [right = 0.01\linewidth of x, yshift=-0.15\linewidth] {\includegraphics[width=0.2\linewidth, ]{figs/imgs/CORPD_FBK-Skyra.jpg}};
            \node[above = -0.04\linewidth of c] {\small $Q$: fm knee};
        \end{tikzpicture}
    \end{minipage}
    \begin{minipage}[c]{0.4\linewidth}
    \begin{tikzpicture}
            \node (x) at (0.0\linewidth,0.0\linewidth) {\includegraphics[width=0.6\linewidth]{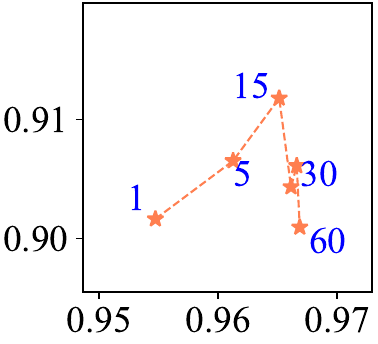}};
            \node[above = -0.03\linewidth of x, xshift=0.05\linewidth] {\small VarNet};
            \node[left = 0.33\linewidth of x, rotate=90, font=\small, xshift=0.15\linewidth] at (0.\linewidth,0) {SSIM on $Q$};
            \node[below = -0.04\linewidth of x, xshift=0.05\linewidth] {\small SSIM on $P$};
            \node (a) [right = 0.04\linewidth of x, yshift=0.18\linewidth]  {\includegraphics[width=0.2\linewidth, ]{figs/imgs/AXT2-Prisma_fit.jpg}};   
            \node (b) [right = 0.01\linewidth of x, yshift=0.15\linewidth]  {\includegraphics[width=0.2\linewidth, ]{figs/imgs/AXT2-Aera.jpg}};
            \node[above = 0.0\linewidth of b, align=center, font=\small] {$P$: fm brain, T2 \phantom{1}};
            \node (c) [right = 0.01\linewidth of x, yshift=-0.15\linewidth] {\includegraphics[width=0.2\linewidth, ]{figs/imgs/CORPD_FBK-Skyra.jpg}};
            \node[above = -0.04\linewidth of c] {\small $Q$: fm knee};
        \end{tikzpicture}
    \end{minipage}
    \caption{Distributional overfitting for ViT and VarNet. Models are trained on data from distribution $P$, and evaluated at different training epochs (1, 5, 15, 30, 45, 60) on $P$ and $Q$.}
    \label{fig:dist_overfit_vit_varnet}
\end{figure}
\begin{figure}
    \begin{tikzpicture}
            \node (x0) at (0.0\linewidth,0.0\linewidth) {\includegraphics[width=0.24\linewidth]{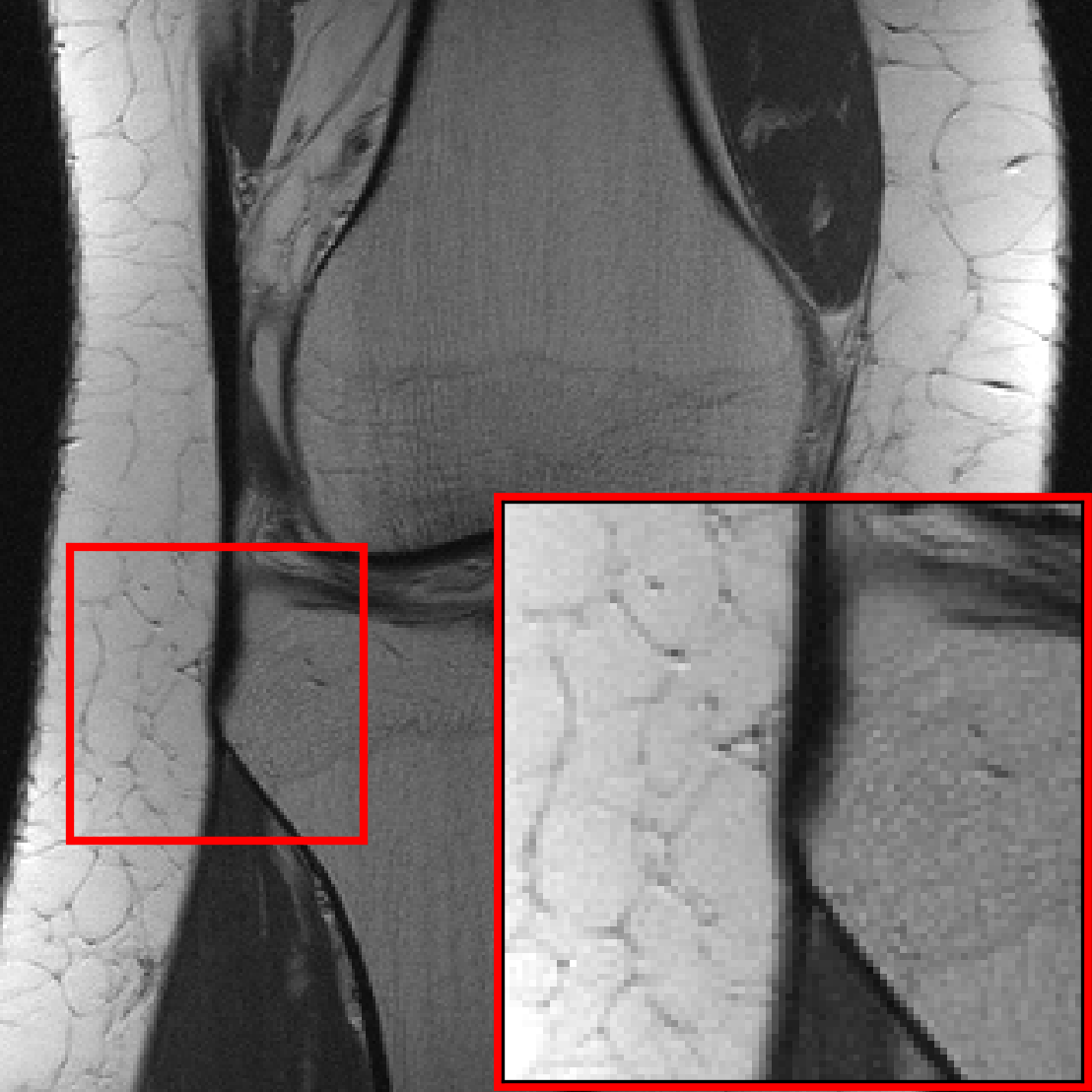}};
            \node (x1)[right = 0\linewidth of x0] {\includegraphics[width=0.24\linewidth]{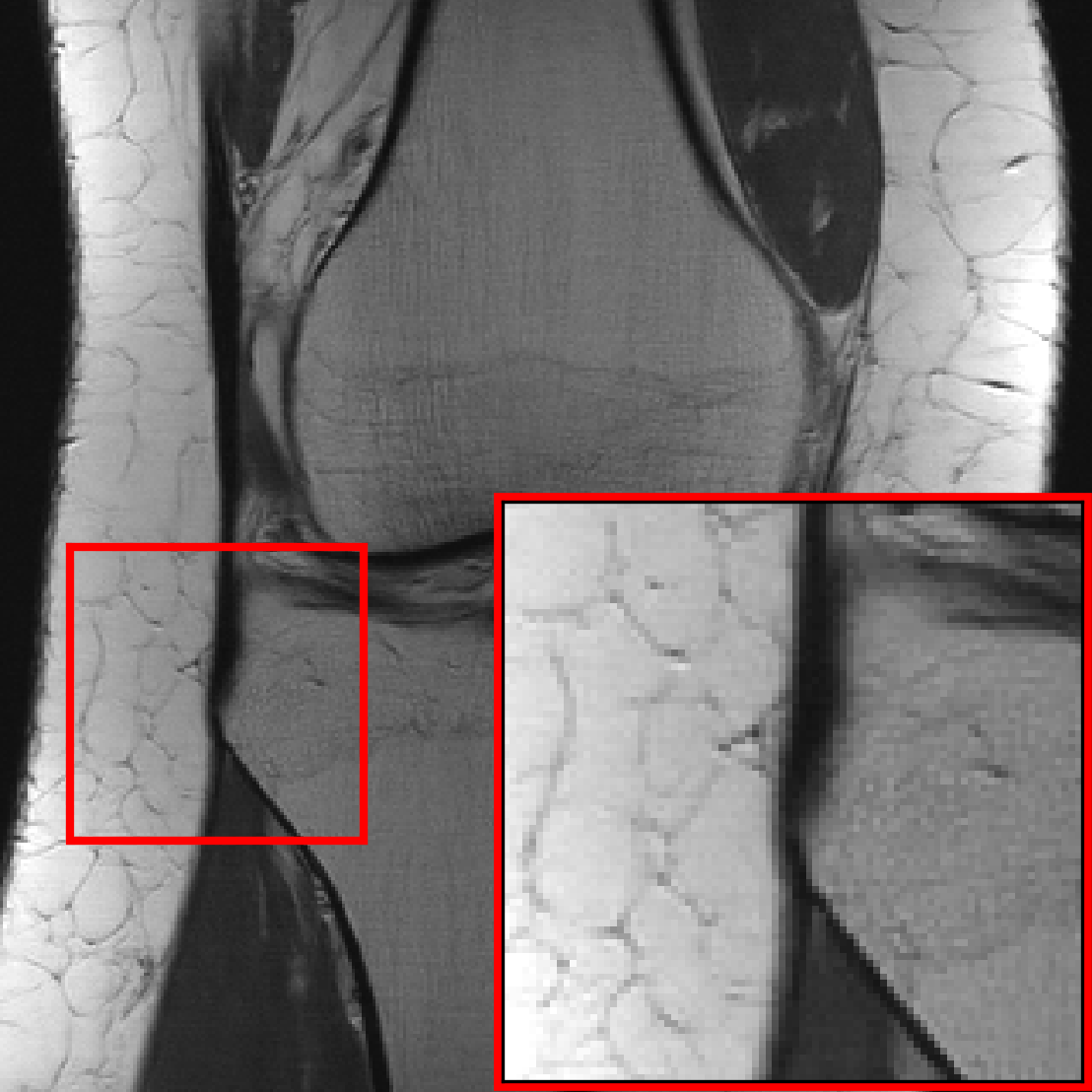}};
            \node (x2)[right = 0\linewidth of x1] {\includegraphics[width=0.24\linewidth]{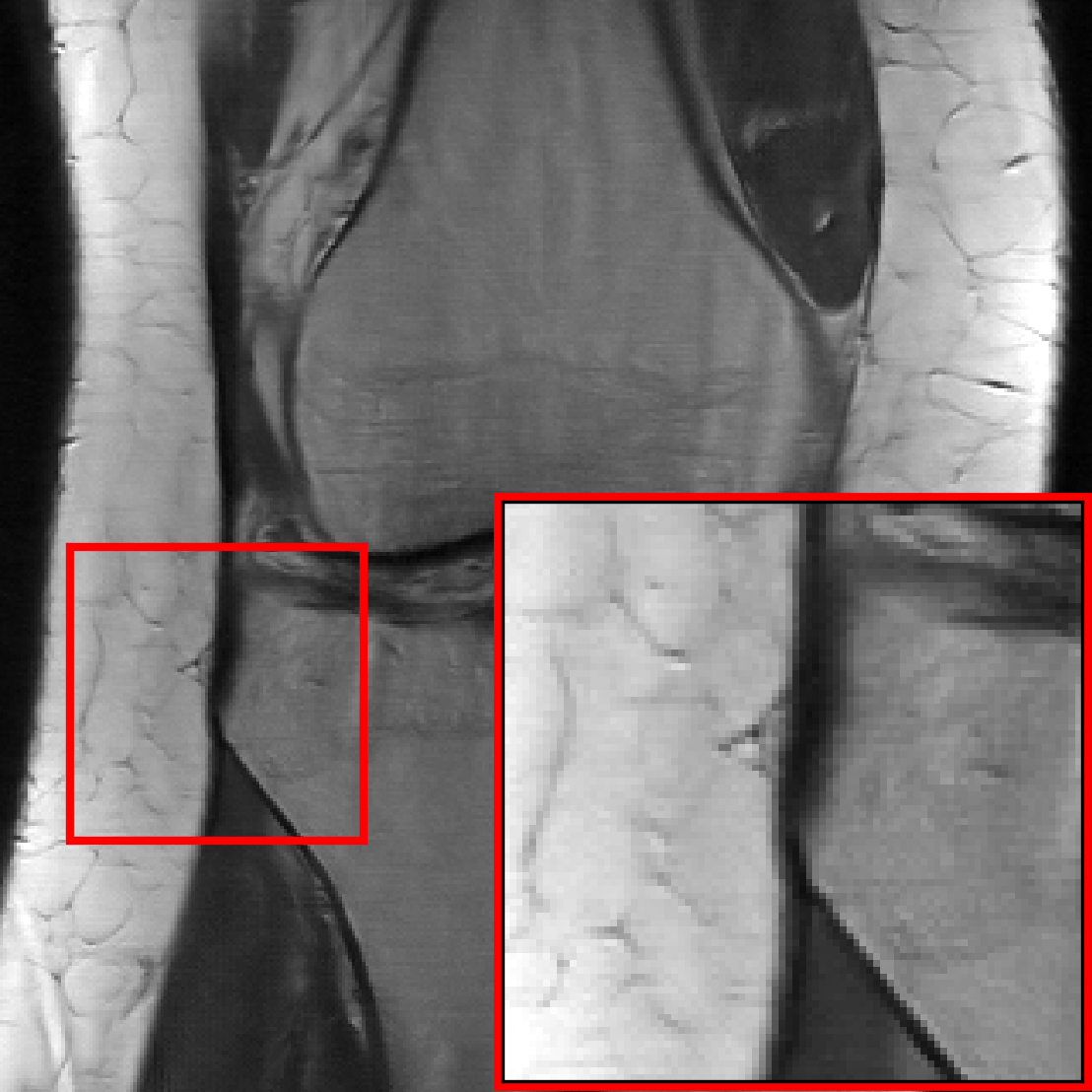}};
            \node (x3)[right = 0\linewidth of x2] {\includegraphics[width=0.24\linewidth]{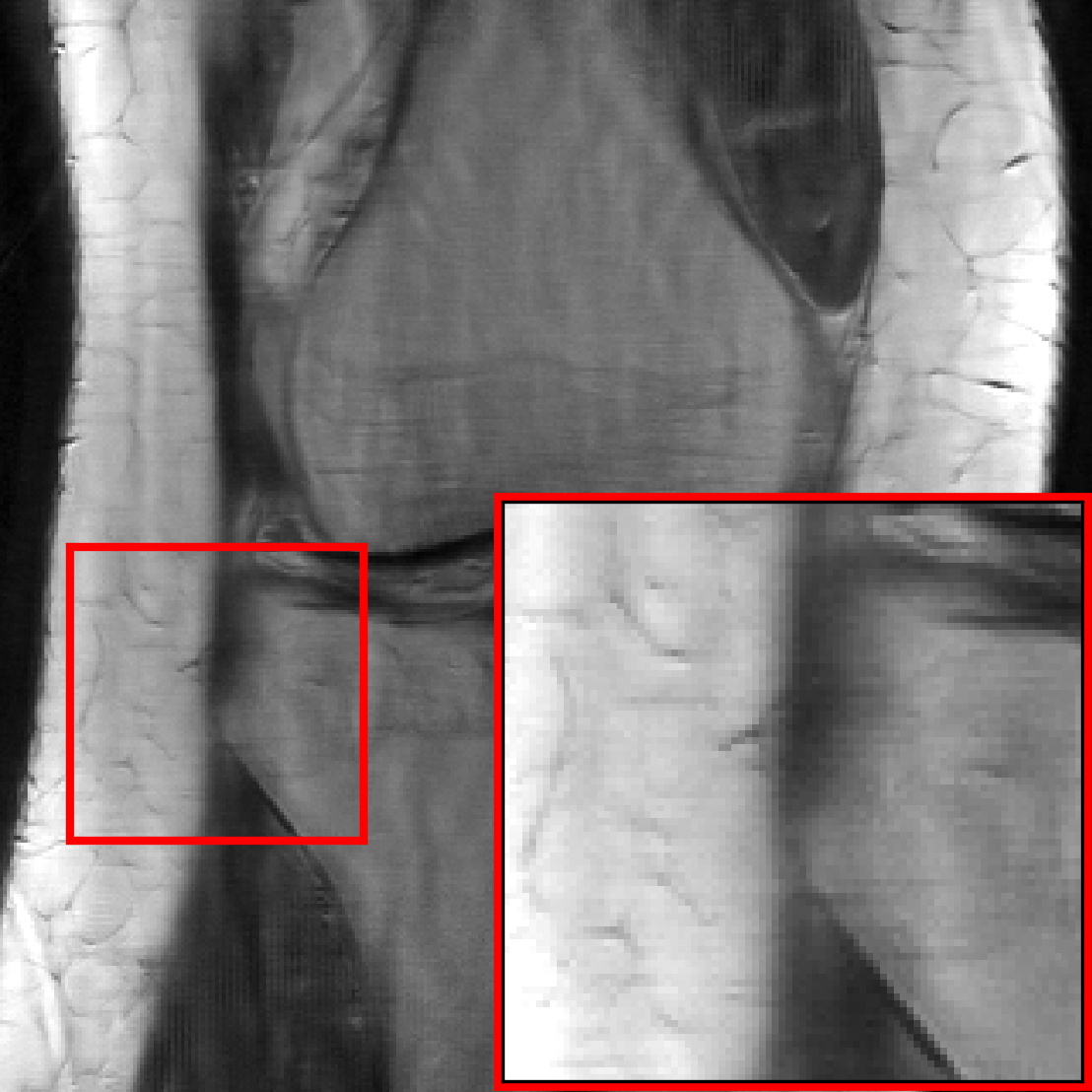}};

            \node[above = 0.0\linewidth of x0] {Ground truth, $Q$ (knee)};
            \node[above = 0.0\linewidth of x1, align=center] {Trained on $Q$ (knee)\\0.964 SSIM};
            \node[above = 0.0\linewidth of x2, align=center] {Trained on $P$ (T2 brain)\\\textbf{Epoch 15}, 0.918 SSIM};
            \node[above = 0.0\linewidth of x3, align=center] {Trained on $P$ (T2 brain)\\\textbf{Epoch 60}, 0.877 SSIM};
        \end{tikzpicture}
    \caption{Reconstruction example showing how distributional overfitting affects the VarNet from Figure~\ref{fig:dist_overfit_vit_varnet}. The VarNet trained for 60 epochs on distribution $P$ (fastMRI T2-weighted brain) produces more blurriness and artifacts in the reconstruction of a sample from distribution $Q$ (fastMRI knee) compared to the same VarNet trained for 15 epochs. This is despite the fact, that the VarNet trained for 60 epochs performs better in-distribution ($P$), as seen in Figure~\ref{fig:dist_overfit_vit_varnet}.}
    \label{fig:dist_overfit_varnet_example}
\end{figure}
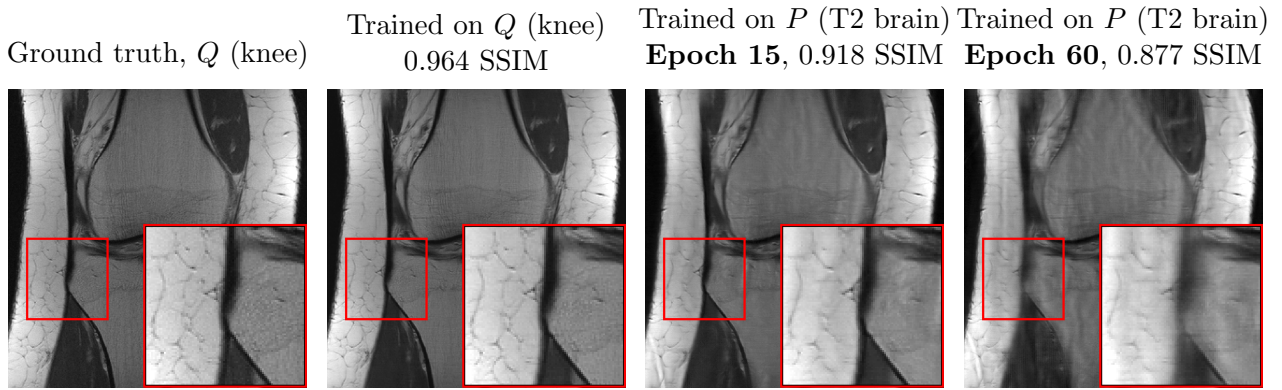
\section{Additional Results on Distributional Overfitting}\label{app:dist_overfit}

In Section~\ref{sec:dist_overfit}, we discussed distributional overfitting for the U-net; here we demonstrate that distributional overfitting happens equally for the ViT and VarNet. 
Figure~\ref{fig:dist_overfit_vit_varnet} demonstrates for a distribution-shift from fastMRI T2-weighted brain to fastMRI knee that ViT and VarNet also suffer from distributional overfitting. Figure~\ref{fig:dist_overfit_varnet_example} further illustrates distributional overfitting for the VarNet on a random reconstruction example.
 
For image classification, studies have shown that using SGD instead of Adam can improve generalizability of neural networks~\citep{zhouTheoreticallyUnderstandingWhy2020}. To see whether distributional overfitting could be mitigated by using SGD instead of Adam, we trained U-nets with SGD on fastMRI brain data, the same data as in Section~\ref{sec:robustness} for non-T2 weighted images. We trained the models for a maximum of 100 epochs, either with zero momentum or with 0.9 momentum, and maximal learning rates of 0.1, 0.01 and 0.001. Other hyperparameters are the same as in Appendix~\ref{app:models_training}. We observed that distributional overfitting was not mitigated by using SGD when evaluating on fastMRI knee data, and we also did not observe better in-distribution nor out-of-distribution performance (on T2-weighted brain and fastMRI knee) compared to using Adam.

\section{Experiment Details and Additional Results for Section~\ref{sec:robust_models}} 
\label{app:details_robust_models}
We now discuss the experimental details for the results in section~\ref{sec:robust_models} on training a robust model for accelerated MRI on diverse data and provide additional analysis. 

\subsection{Preparation of Datasets}\label{app:dataprep_non-fm}
We convert all the dataset listed in Table~\ref{tbl:datasets} to follow the convention of the fastMRI knee and brain datasets, where the anatomies in images are vertically flipped, targets are RSS reconstructions, and the k-space is oriented such that the horizontal axis corresponds to the phase-encoding direction and the vertical axis corresponds to the read-out direction.

If predefined train and test splits are not already provided with a dataset, we randomly select 85\% of the volumes as training set and the remaining volumes as test set. If a dataset has a designated validation set that is separate from the test set, then we include the validation set in the training set.
For 3D MRI volumes, we synthesize 2D k-spaces by taking the 1D IFFT in the 3D k-space along either x, y or z dimension to create 2D volumes of different anatomical views (axial, sagittal and coronal). However, for the SKM-TEA dataset, we only consider the sagittal view. Depending on the dataset, the first and last 15-70 slices of the synthesized 2D volumes are omitted as we mostly observe pure noise measurements:
\begin{itemize}
    \item CC-359, sagittal view: First 15 and last 15 slices are omitted.
    \item CC-359, axial view: First 50 slices are omitted.
    \item CC-359, coronal view: First 25 and 15 slices are omitted.
    \item Stanford 3D, axial view: First 5 and last 5 slices are omitted.
    \item Stanford 3D, coronal view: First 40 and last 40 slices are omitted.
    \item Stanford 3D, sagittal view: First 30 and last 30 slices are omitted.
    \item 7T database, axial view: First 70 and last 70 slices are omitted.
    \item 7T database, coronal view: First 30 and last 30 slices are omitted.
    \item 7T database, sagittal view: First 30 and last 30 slices are omitted.
\end{itemize}
For the other datasets that are not mentioned above, all slices are used. Moreover, each of the volumes of the SKM-TEA dataset contains originally two echos due to the use of the qDESS sequence. We separate the two echos and count them as separate volumes.

\paragraph{fastMRI prostate T2.} Originally, each volume of the fastMRI prostate T2 dataset contains three averages~\citep{tibrewalaFastMRIProstatePublicly2023}: two averages sampling the odd k-space lines and one average sampling the even k-space lines. Then, for each average the authors estimate the missing k-space lines with GRAPPA~\citep{griswoldGeneralizedAutocalibratingPartially2002a} and perform SENSE~\citep{pruessmannSENSESensitivityEncoding1999a} reconstruction. The final ground truth image is then obtained by taking the mean across the three averages (see \href{https://github.com/cai2r/fastMRI_prostate/tree/main}{\color{blue} code} in the paper's GitHub repository). However, we convert the data as follows: we take the raw k-space and average the two averages corresponding to the odd k-space lines and then fill the missing even k-space lines with the average corresponding to the even k-space lines. This k-space serves as our k-space data. We then take this k-space and apply a 2D-IFFT and finally perform a RSS reconstruction and use this image as ground truth.

\subsection{Models, Training, and Evaluation}\label{app:robust_models_training}
The U-net used has 124M parameter with 4 pooling layers and 128 channels in the first pooling layer. The maximal learning rate is set to 4e-4. The ViT has 127M parameters, where we use a patch-size of 10, an embedding dimension of 1024, 16 attention heads, and a depth of 10. The maximal learning rate is set to 2e-4. The VarNet contains 8 cascades, each containing an U-net
with 4 pooling layers and 12 channels in the first pooling layer. We use linear learning rate decay and gradients are clipped at a global $\ell_2$-norm of 1. For U-net and ViT, training is set for 40 epochs but we early stopped the models at epoch 24, and we use a mini-batch size of 8. The VarNet is trained for 40 epochs and we use a mini-batch size of 4. Since slice dimensions can vary across different volumes, the images within a mini-batch are chosen randomly from the same volume without replacement. We use the Adam optimizer with $\beta_1= 0.9$ and $\beta_2=0.999$ and the models are trained to maximize SSIM between model output and RSS target. Training was carried out on two NVIDIA RTX A6000 GPUs. Training the U-net took 384 GPU hours, the ViT took 480 GPU hours, and the VarNet took 960 GPU hours. 

\vsparagraph{Early stopping criterion.} Observing whether distributional overfitting occurs and its significance depends on both the training dataset and the test dataset as for some combinations we might not be able to observe distributional overfitting, e.g., when training on fastMRI knee data and evaluating on fastMRI brain data. However, our results suggest a common thread: when distributional overfitting occurs, in-distribution performance improvements are marginal. Therefore, we base our early stopping criterion on this observation.

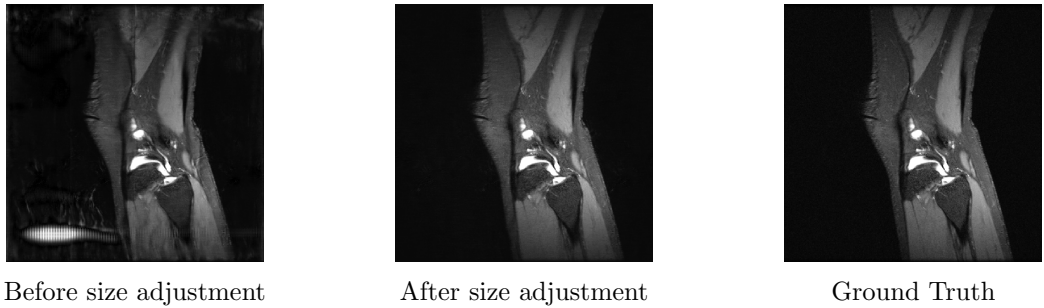
\begin{figure}[t!]
    \centering
    \begin{minipage}{0.9\linewidth}
    \centering
         \begin{tikzpicture}
            \node at (0.0\linewidth,0.0\linewidth) (a1) {\includegraphics[width=0.23\linewidth]{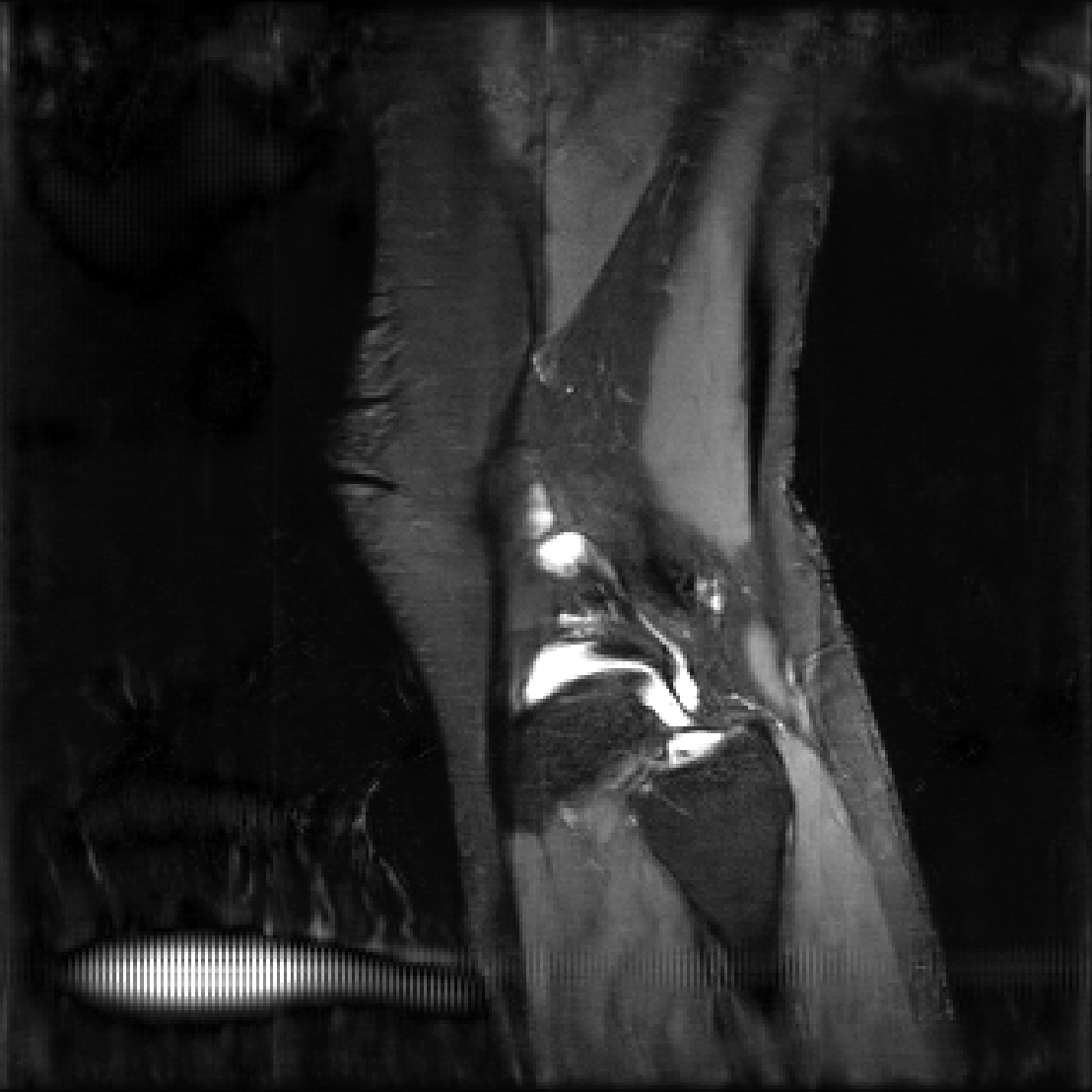}};
            \node[below = -0.0\linewidth of a1, font=\small]  {Before size adjustment};
            \node[right = 0.1\linewidth of a1] (a2) {\includegraphics[width=0.23\linewidth]{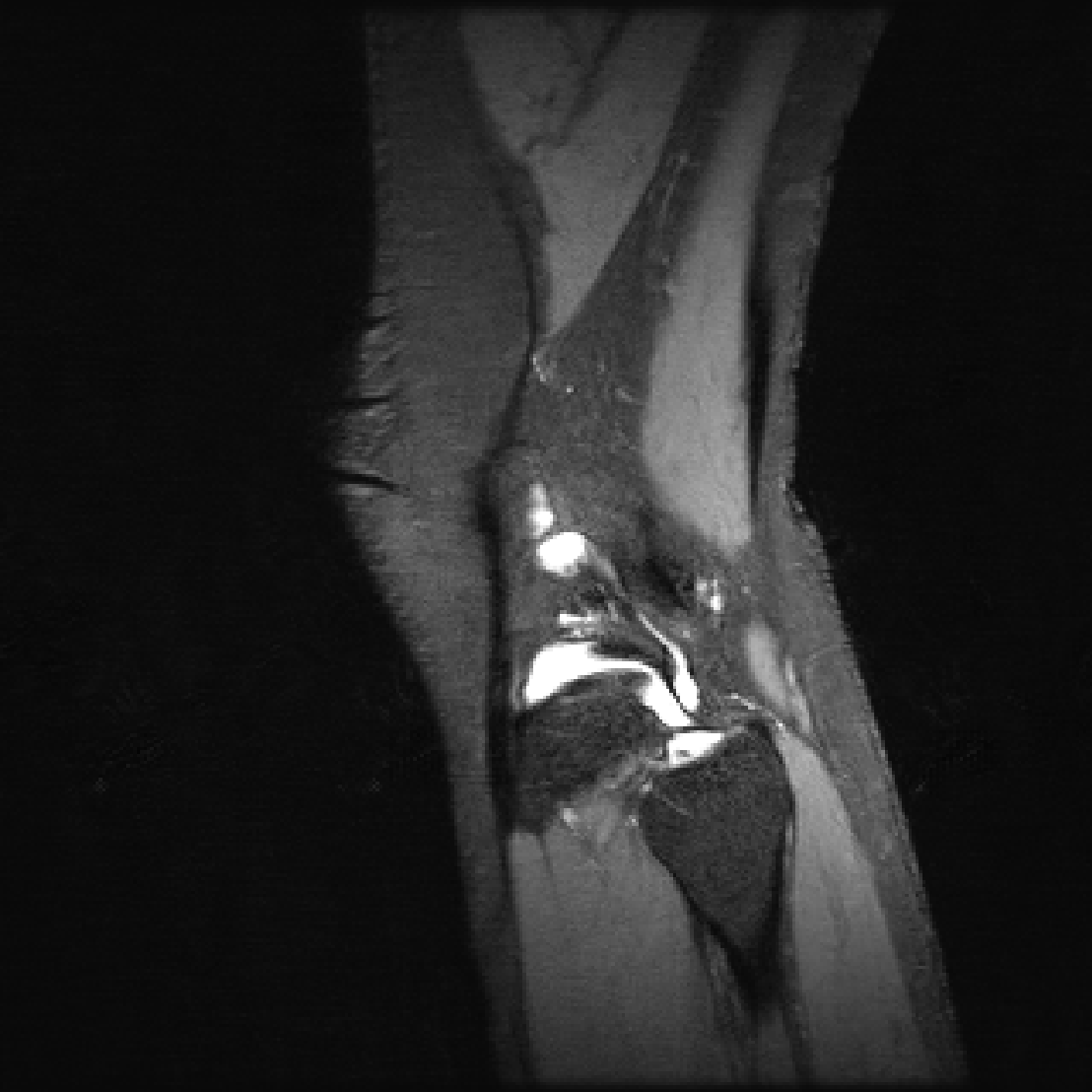}};
            \node[below = -0.0\linewidth of a2, font=\small]  {After size adjustment};
            \node[right = 0.1\linewidth of a2] (a3) {\includegraphics[width=0.23\linewidth]{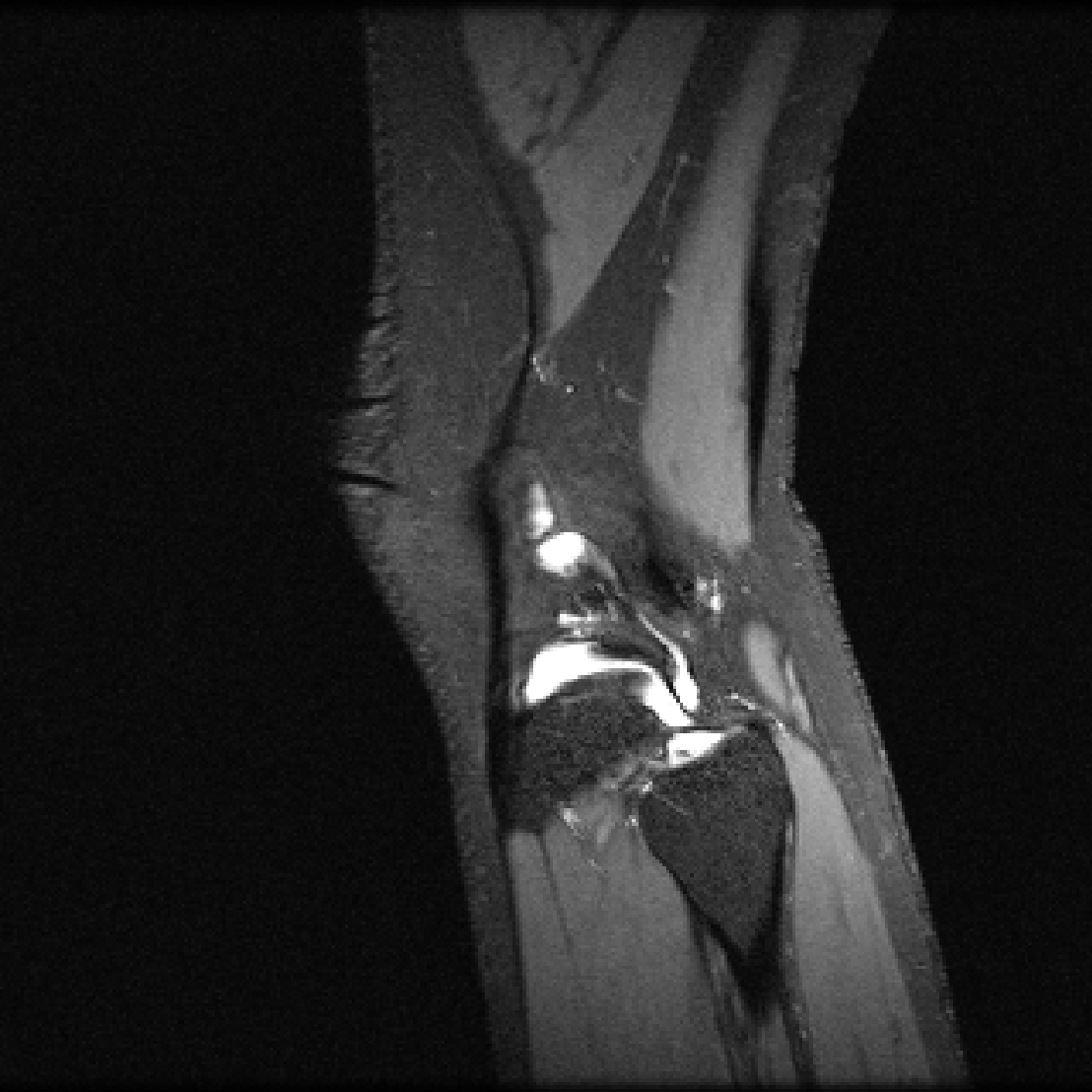}};
            \node[below = -0.0\linewidth of a3, font=\small]  {Ground Truth};
        \end{tikzpicture}
    \end{minipage}
    \caption{Shown are two reconstructions by a VarNet trained on fastMRI knee dataset and applied to a sample of the Stanford 3D sagittal view dataset. A mismatch in image size between training and test time can lead to artifacts (\textbf{left}). Adjusting the size of the input k-space mitigates these artifacts (\textbf{middle}).}
    \label{fig:size_adjustment}
\end{figure}

\vsparagraph{Resolution mismatch.} The U-net and ViT are trained on center-cropped zero-filled reconstructions, the VarNet is trained on the entire k-space and therefore on the full-sized image. For example, the average image size of fastMRI knee dataset is 640 $\times$ 360. Now, if we consider for example a distributions-shift from the fastMRI knee dataset to the Stanford 3D dataset which contains images of approximately half the size, we additionally introduce an artificial distribution-shift by having a mismatch between the image size from training to evaluation. 

To mitigate this artificial distribution-shift we implement the following steps during inference: given the undersampled k-space and mask, we first repeat the undersampled k-space one time in an interleaved fashion in horizontal direction and another time in vertical direction, and adjust the undersampling mask accordingly. The repeated k-space and mask serve as input to the VarNet and the output is center-cropped to the original image size. As can be seen in Figure~\ref{fig:size_adjustment}, these processing steps heavily reduces artifacts of the VarNets trained on the fastMRI datasets when evaluated on the Stanford 2D, CC-359 sagittal view, and M4Raw GRE dataset.

\begin{table}[t!]
    \centering
    \small
    \captionof{table}{
    Performance of models from Section~\ref{sec:robust_models} when evaluated on deep feature metrics LPIPS/DISTS (lower is better). For any architecture, the model trained on the diverse collection of datasets $\mathcal{D}_P$ performs better on out-of-distribution data than the models trained on fastMRI knee/brain while showing similar performance on the in-distribution fastMRI data.
    }
    \label{tbl:deep_metrics}
    \begin{adjustbox}{max width=\linewidth}
    \begin{tabular}{l l c c c c c c r}
    \toprule
    Model & \diagbox[width=0.12\linewidth]{Train}{Test} 
                            & fastMRI knee&fastMRI brain& CC-359 sag. & NYU         & Stanford 2D  & M4Raw GRE & Mean\\
    \midrule
    U-net & fastMRI knee    & 0.230/0.132 & 0.249/0.146 & 0.294/0.185 & 0.226/0.128 & 0.260/0.163 & 0.276/0.183 & 0.256/0.156\\
          & fastMRI brain   & 0.299/0.163 & 0.224/0.130 & 0.281/0.168 & 0.270/0.150 & 0.283/0.168 & 0.253/0.160 & 0.268/0.157\\
          & $\mathcal{D}_P$ & 0.230/0.132 & 0.224/0.132 & 0.247/0.149 & 0.221/0.128 & 0.257/0.154 & 0.249/0.160 & 0.238/0.142\\
          & \color{gray}Test distribution & \color{gray}0.230/0.132 & \color{gray}0.224/0.130 &\color{gray}0.215/0.136 & \color{gray}0.214/0.123 & \color{gray}0.246/0.143 & \color{gray}0.239/0.153 & \color{gray}0.228/0.136\\
    \midrule
    ViT   & fastMRI knee    & 0.225/0.132 & 0.246/0.146 & 0.295/0.185 & 0.219/0.128 & 0.258/0.163 & 0.265/0.183 & 0.251/0.156\\
          & fastMRI brain   & 0.280/0.163 & 0.220/0.130 & 0.264/0.168 & 0.254/0.150 & 0.258/0.168 & 0.253/0.160 & 0.255/0.157\\
          & $\mathcal{D}_P$ & 0.223/0.132 & 0.221/0.132 & 0.228/0.149 & 0.219/0.128 & 0.243/0.154 & 0.252/0.160 & 0.231/0.142\\
          & \color{gray}Test distribution & \color{gray}0.225/0.132 & \color{gray}0.220/0.130 &\color{gray}0.206/0.136 & \color{gray}0.207/0.123 & \color{gray}0.236/0.143 & \color{gray}0.237/0.153 & \color{gray}0.222/0.136\\
    \midrule
    VarNet& fastMRI knee    & 0.209/0.122 & 0.226/0.132 & 0.306/0.189 & 0.201/0.106 & 0.244/0.143 & 0.344/0.226 & 0.255/0.153\\
          & fastMRI brain   & 0.246/0.139 & 0.206/0.120 & 0.251/0.152 & 0.223/0.118 & 0.241/0.141 & 0.273/0.174 & 0.240/0.140\\
          & $\mathcal{D}_P$ & 0.210/0.122 & 0.206/0.119 & 0.173/0.117 & 0.198/0.095 & 0.228/0.131 & 0.269/0.179 & 0.214/0.127\\
          & \color{gray}Test distribution & \color{gray}0.209/0.122 & \color{gray}0.206/0.120 &\color{gray}0.161/0.105 & \color{gray}0.193/0.114 & \color{gray}0.226/0.131 & \color{gray}0.219/0.147 & \color{gray}0.202/0.123\\
    \bottomrule
    \end{tabular}
    \end{adjustbox}
\end{table}
\begin{table}[t!]
    \centering
    \small
    \captionof{table}{Quantification of artifacts in the reconstructions of the models from Section~\ref{sec:robust_models}. We quantify the prominence of artifacts by computing the variance of the Laplacian applied to the difference image, abs(target -- reconstruction), where lower values indicate fewer artifacts. Models trained on the diverse collection of datasets $\mathcal{D}_P$ tend to have less pronounced artifacts compared to models trained on the fastMRI dataset. 
    }
    \label{tbl:error_metrics}
    \begin{adjustbox}{max width=\linewidth}
    \begin{tabular}{l l c c c c c c r}
    \toprule
    Model & \diagbox[width=0.12\linewidth]{Train}{Test} 
                            & fastMRI knee&fastMRI brain& CC-359 sag. & NYU         & Stanford 2D  & M4Raw GRE & Mean\\
    \midrule
    U-net & fastMRI knee    & 257 & 184 & 389 & 153 & 357 & 246 & 264\\
          & fastMRI brain   & 294 & 157 & 348 & 167 & 334 & 197 & 245\\
          & $\mathcal{D}_P$ & 267 & 169 & 309 & 165 & 346 & 194 & 242\\
          & \color{gray}Test distribution & \color{gray}257 & \color{gray}157 &\color{gray}239 & \color{gray}147 & \color{gray}267 & \color{gray}156 & \color{gray}204\\
    \midrule
    ViT   & fastMRI knee    & 255 & 189 & 409 & 158 & 348 & 239 & 266\\
          & fastMRI brain   & 293 & 154 & 329 & 173 & 303 & 202 & 242\\
          & $\mathcal{D}_P$ & 255 & 157 & 288 & 158 & 265 & 194 & 220\\
          & \color{gray}Test distribution & \color{gray}255 & \color{gray}154 &\color{gray}227 & \color{gray}148 & \color{gray}249 & \color{gray}156 &\color{gray}198\\
          
    \midrule
    VarNet& fastMRI knee    & 245 & 174 & 434 & 183 & 318 & 331 & 281\\
          & fastMRI brain   & 271 & 149 & 292 & 214 & 323 & 247 & 249\\
          & $\mathcal{D}_P$ & 245 & 148 & 151 & 136 & 271 & 199 & 192\\
          & \color{gray}Test distribution & \color{gray}245 & \color{gray}149 &\color{gray}117 & \color{gray}138 & \color{gray}211 & \color{gray}139 & \color{gray}166\\

    \bottomrule
    \end{tabular}
    \end{adjustbox}
\end{table}

\vsparagraph{Output normalization.} Similar to Appendix~\ref{app:coil_shift}, we observed significant drops in SSIM on out-of-distribution evaluations due to hardly visible mismatches in terms of mean or variance between model output and target. Hence, we normalize the output of the models to have the same mean and variance as the target during evaluation. This normalization reduces the SSIM score's sensitivity to variations in brightness and contrast, enabling it to better reflect structural differences.

\subsection{Additional Metrics and Error Analysis}\label{app:additional_metrics}

Summarizing the reconstruction quality in a single number is difficult. 
While SSIM is a rather standard metric for evaluating the reconstruction quality, there are other metrics that agree better with radiologit's ratings, such as  such as LPIPS~\citep{zhangUnreasonableEffectivenessDeep2018} or DISTS~\citep{dingImageQualityAssessment2022}. LPIPS an DISTS are based on features of pretrained neural networks, and correlate relatively well with radiologist evaluations~\citep{adamsonUsingDeepFeature2023, kastryulinImageQualityAssessment2023}. Those metrics compute the distance between the ground-truth and the reconstruction in the feature space of pretrained neural networks. 

In Table~\ref{tbl:deep_metrics}, we evaluate the performance of models from Section~\ref{sec:robust_models} on LPIPS and DISTS. We find that, consistent with the SSIM results in Section~\ref{sec:robust_models}, the models trained on the diverse collection of datasets $\mathcal{D}_P$ perform better out-of-distribution than models trained on fastMRI knee or brain without sacrificing performance on the fastMRI datasets.

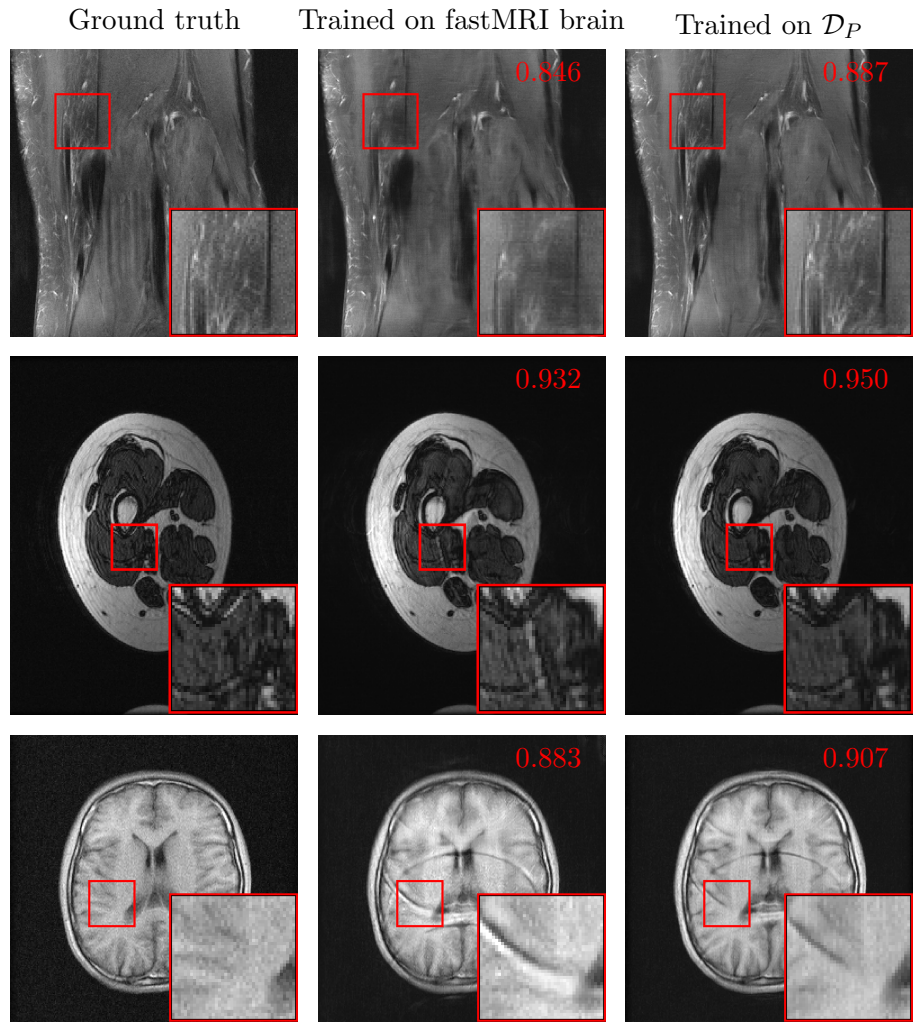
\begin{figure}[t!]
    \newcommand{\w}{0.33\linewidth}
    \centering
    \begin{minipage}{0.7\linewidth}
        \begin{tikzpicture}
            \node (x0) at (0.0\linewidth,0.0\linewidth) {\includegraphics[width=\w]{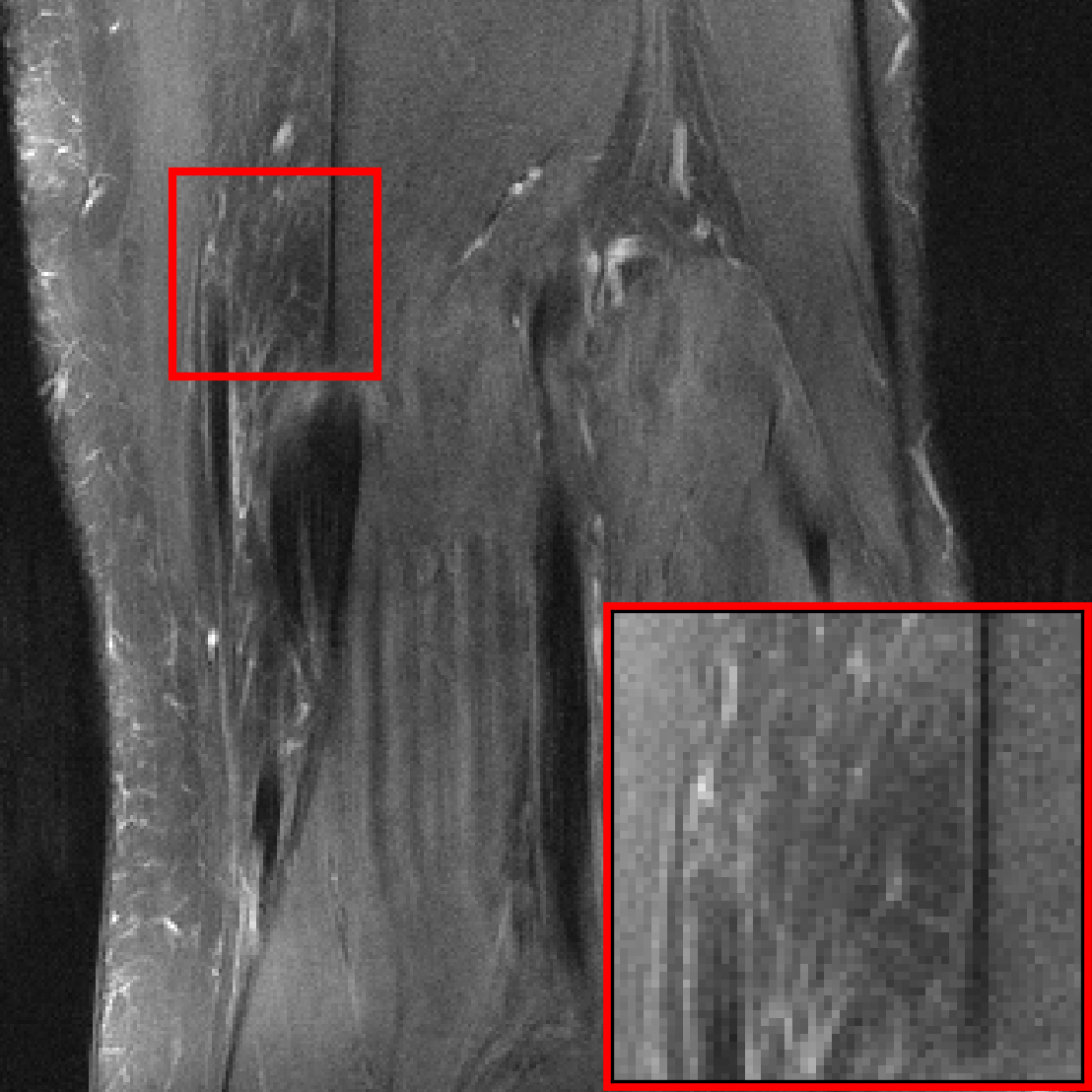}};
            \node (x1)[below = 0\linewidth of x0] {\includegraphics[width=\w]{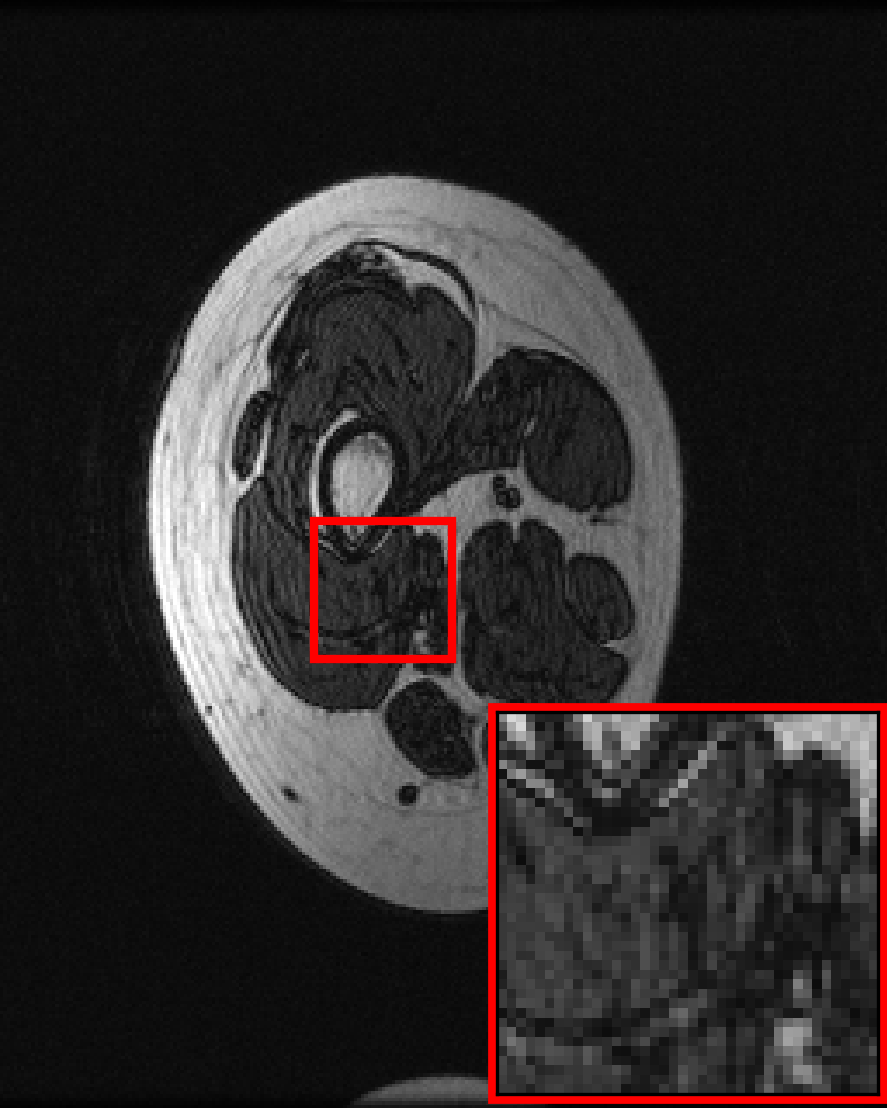}};
            \node (x2)[below = 0\linewidth of x1] {\includegraphics[width=\w]{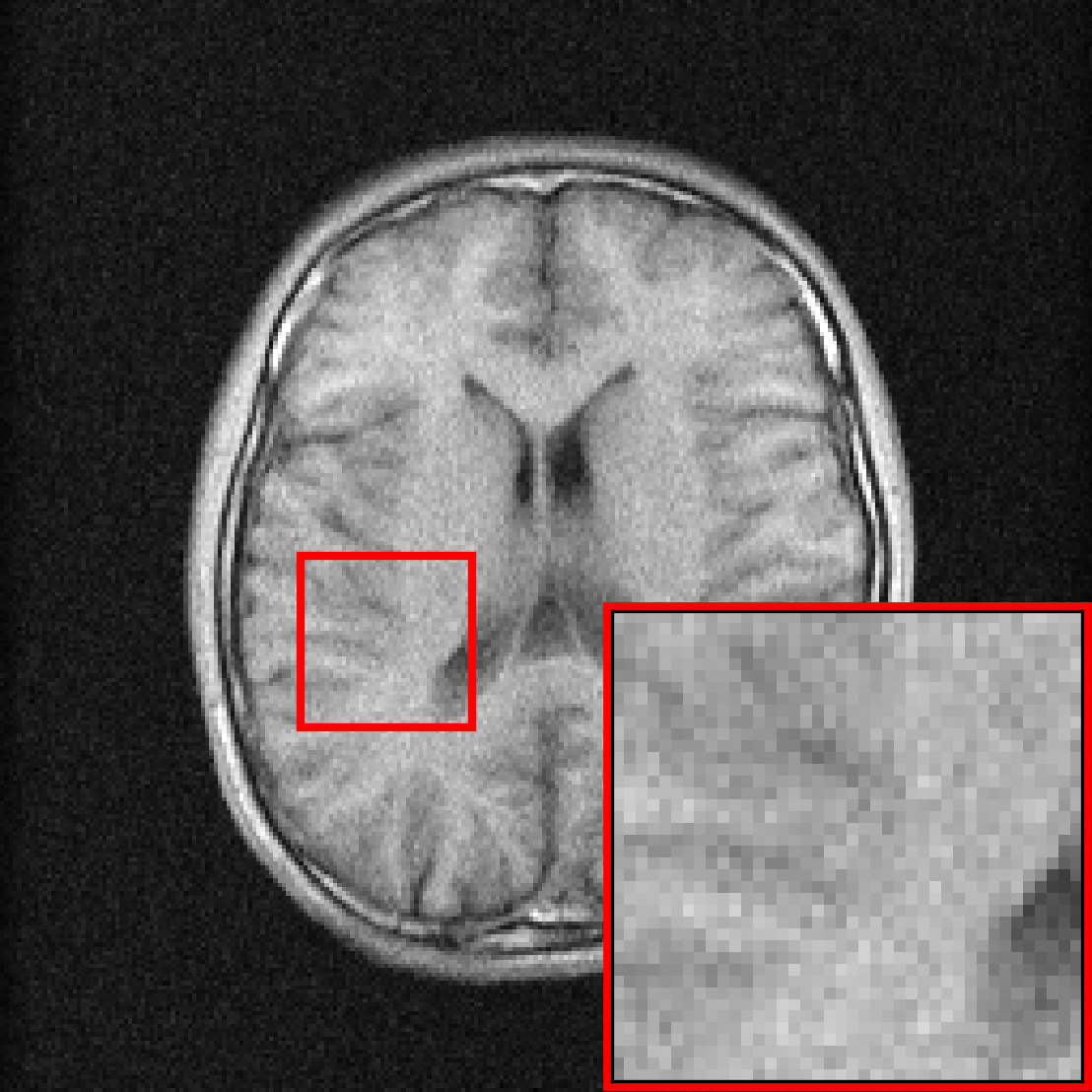}};
            
            \node[above = 0\linewidth of x0] {Ground truth};

            \node (y0) [right = 0.\linewidth of x0] {\includegraphics[width=\w]{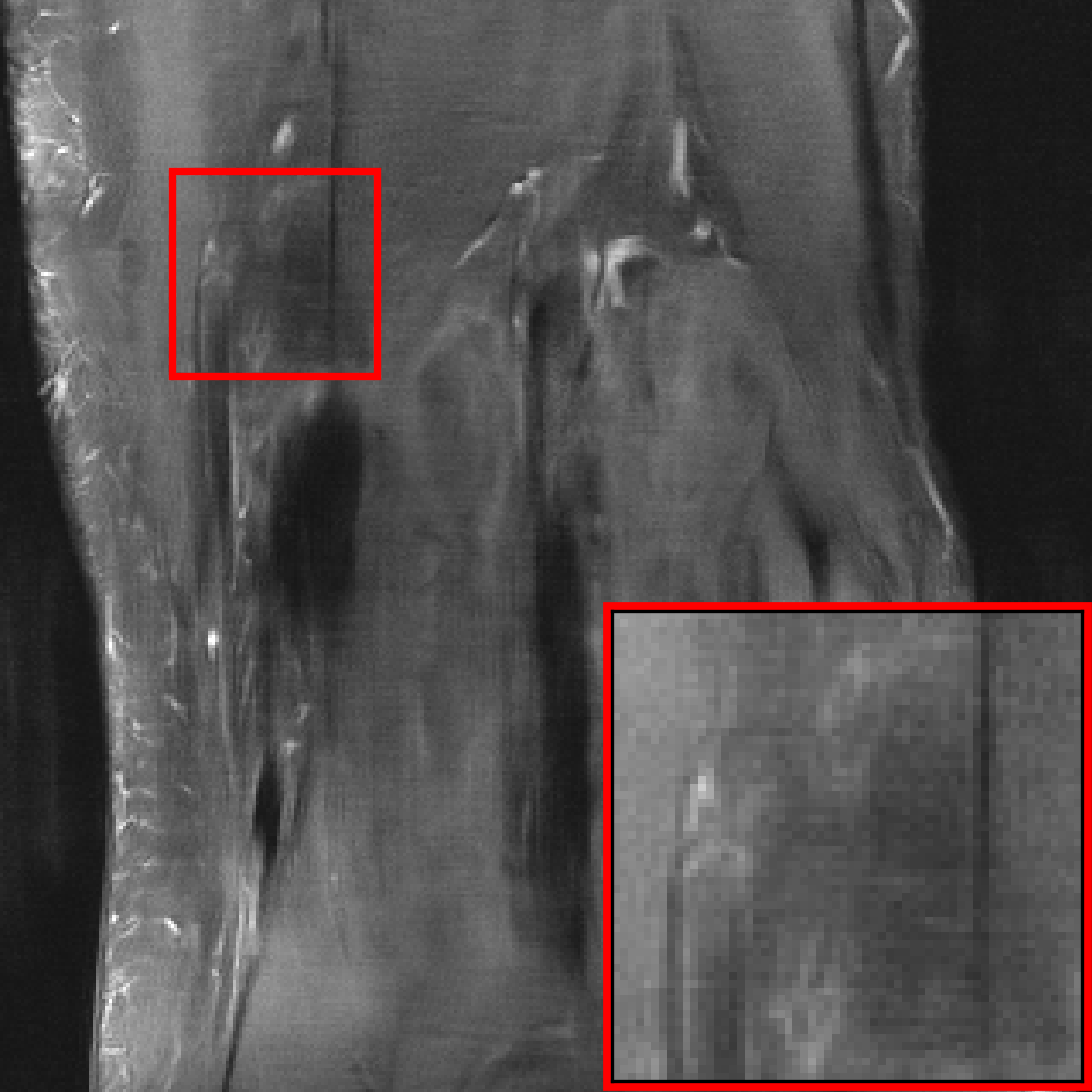}};
            \node[above = -0.06\linewidth of y0, xshift=0.1\linewidth] {\color{red}0.846};
            \node (y1)[below = 0\linewidth of y0] {\includegraphics[width=\w]{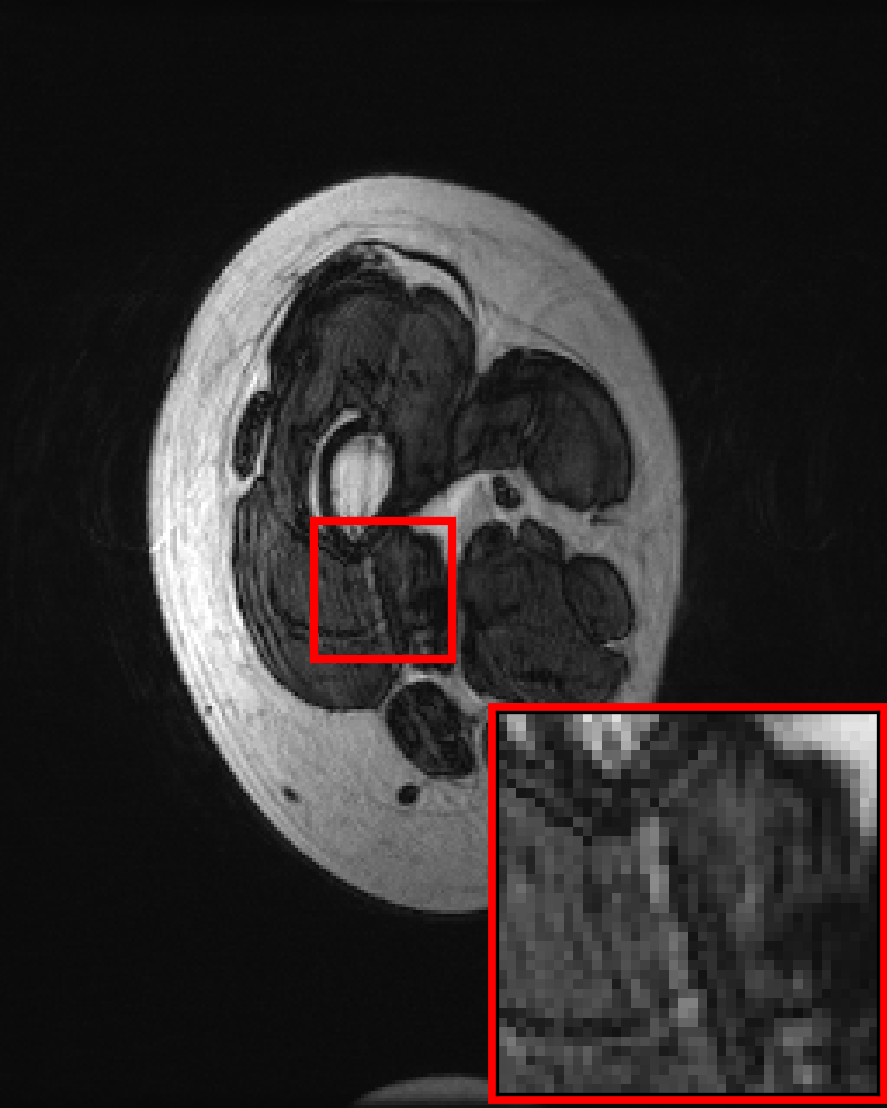}};
            \node[above = -0.06\linewidth of y1, xshift=0.1\linewidth] {\color{red}0.932};
            \node (y2)[below = 0\linewidth of y1] {\includegraphics[width=\w]{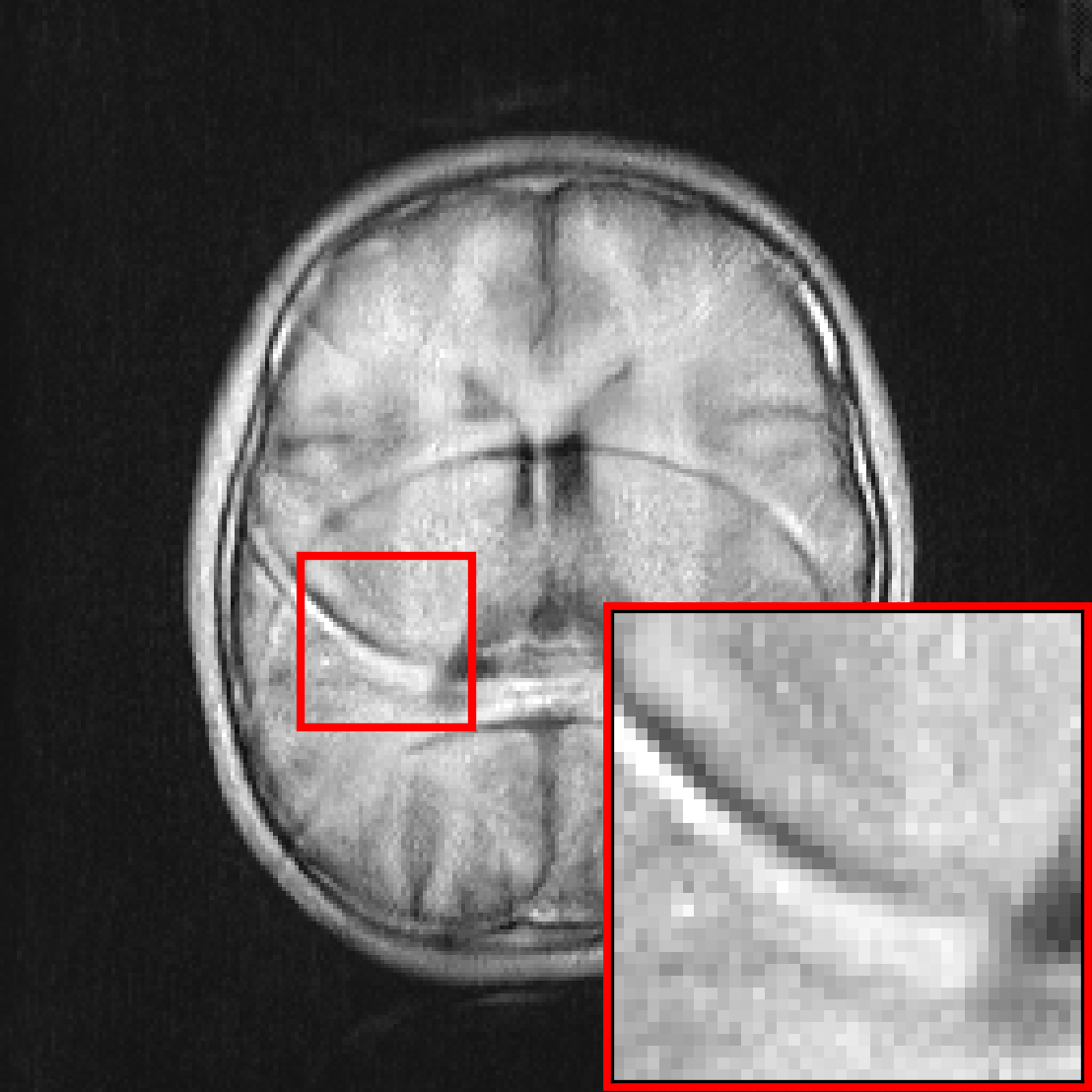}};
            \node[above = -0.06\linewidth of y2, xshift=0.1\linewidth] {\color{red}0.883};
            
            \node[above = 0.0\linewidth of y0] {Trained on fastMRI brain};
            \node (z0) [right = 0.\linewidth of y0] {\includegraphics[width=\w]{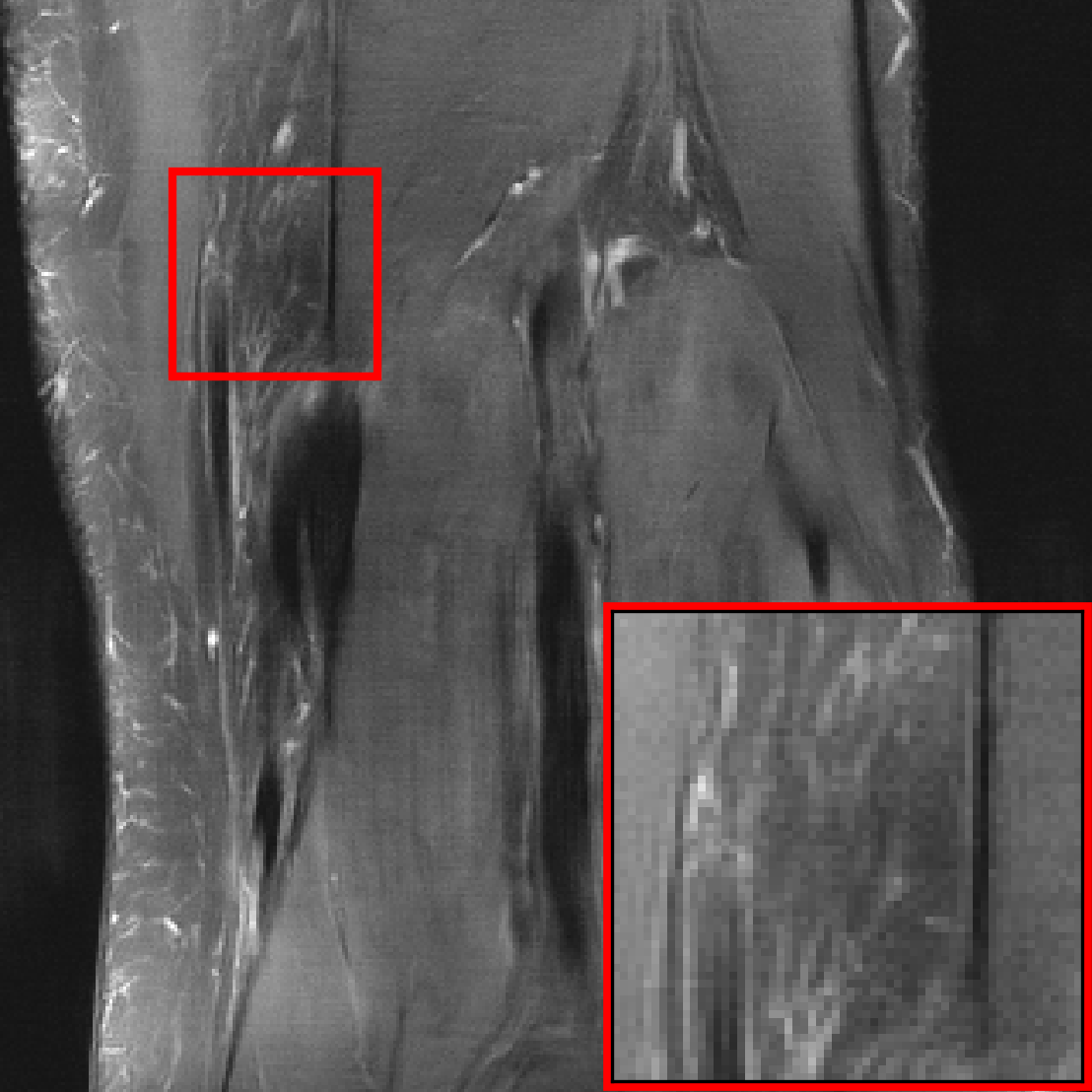}};
            \node[above = -0.06\linewidth of z0, xshift=0.1\linewidth] {\color{red}0.887};
            \node (z1)[below = 0\linewidth of z0] {\includegraphics[width=\w]{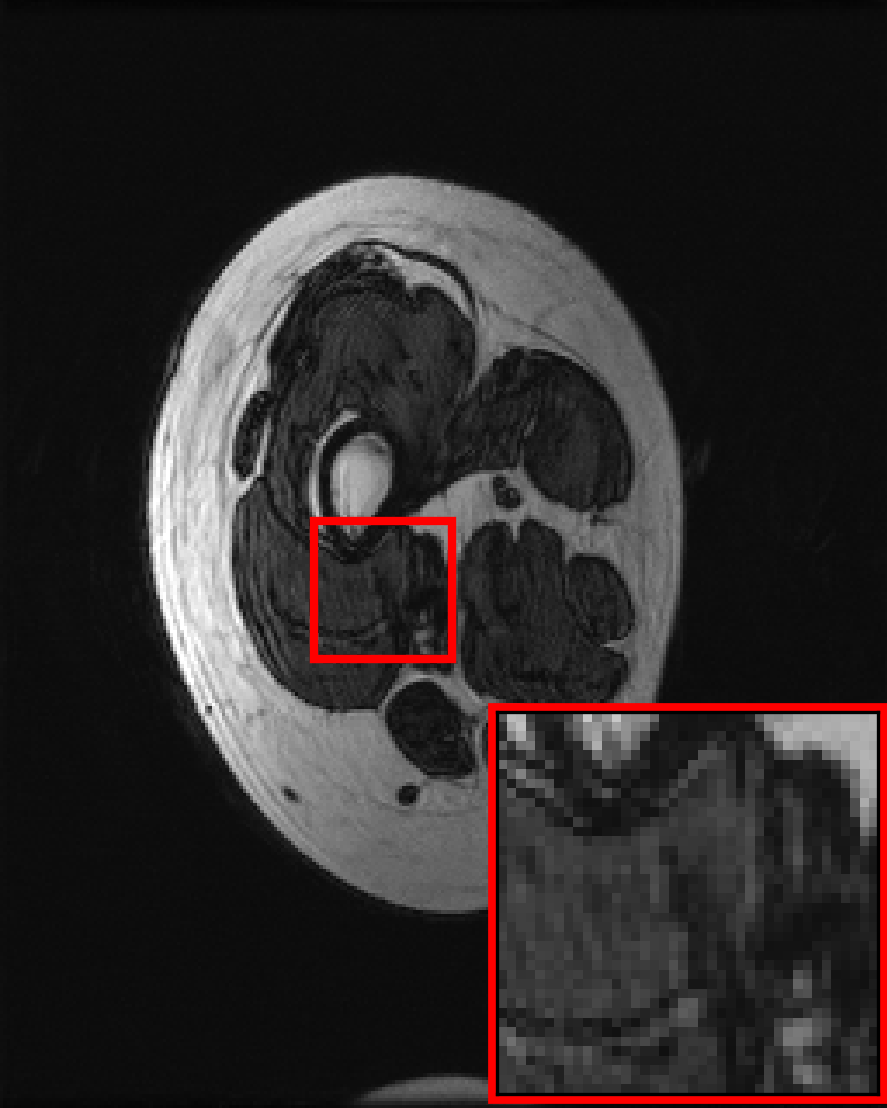}};
            \node[above = -0.06\linewidth of z1, xshift=0.1\linewidth] {\color{red}0.950};
            \node (z2)[below = 0\linewidth of z1] {\includegraphics[width=\w]{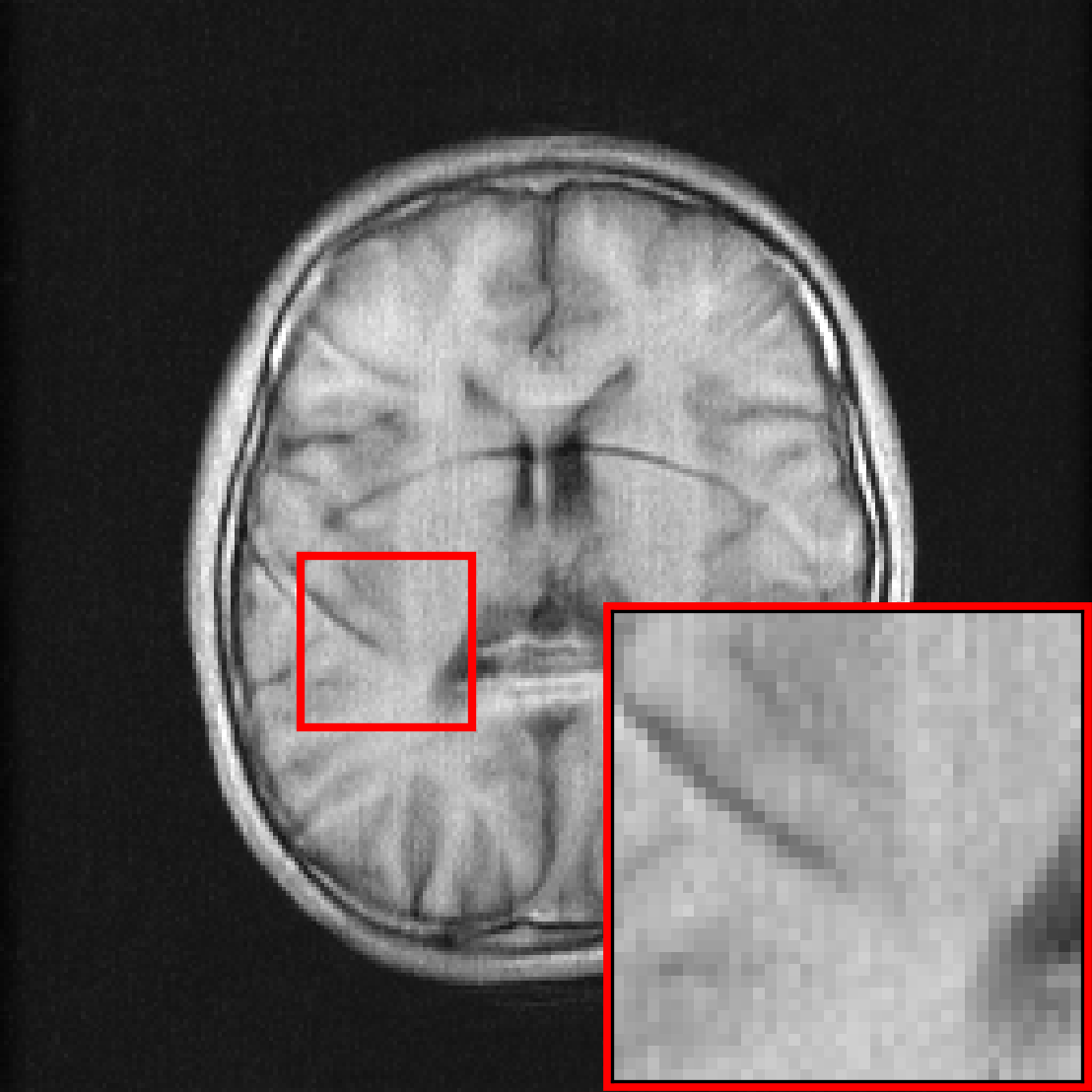}};
            \node[above = -0.06\linewidth of z2, xshift=0.1\linewidth] {\color{red}0.907};
            
            \node[above = 0.0\linewidth of z0, yshift=-0.01\linewidth] {Trained on $\mathcal{D}_P$};
            
        \end{tikzpicture}
    \end{minipage}
    \caption{Out-of-distribution reconstruction examples for the NYU, Stanford 2D, and M4Raw GRE datasets. Shown are reconstructions by the VarNet trained on fastMRI brain are compared to the VarNet trained on the collection of datasets ($\mathcal{D}_P$). The numbers are the SSIM between the reconstruction and the ground truth image.}
    \label{fig:reconstruction_examples}
\end{figure}

\vsparagraph{Quantification of artifacts.} When inspecting the reconstructions of our models in out-of-distribution setups, we observed significant errors caused by artifacts. For example, Figure~\ref{fig:P_recon} (middle) depicts a reconstruction severely affected by artifacts. To quantify the prominence of artifacts, we compute the variance of the Laplacian applied to the difference image, i.e., abs(target -- reconstruction), where lower values indicate fewer artifacts.

Table~\ref{tbl:error_metrics} shows the evaluation of the models from Section~\ref{sec:robust_models} on the prominence of artifacts. We observe that the models trained on the diverse data $\mathcal{D}_P$ tend to yield less pronounced artifacts.

In Figure~\ref{fig:reconstruction_examples}, we provide further out-of-distribution reconstruction examples by the VarNet trained on fastMRI brain and trained on the collection of datasets $\mathcal{D}_P$ for the NYU, Stanford 2D, and M4Raw GRE datasets. A reconstruction example, for CC-359 sagittal view is provided in Figure~\ref{fig:P_recon}.

\subsection{Finetuning Reduces Out-Of-Distribution Robustness}\label{app:finetuning_robustness}
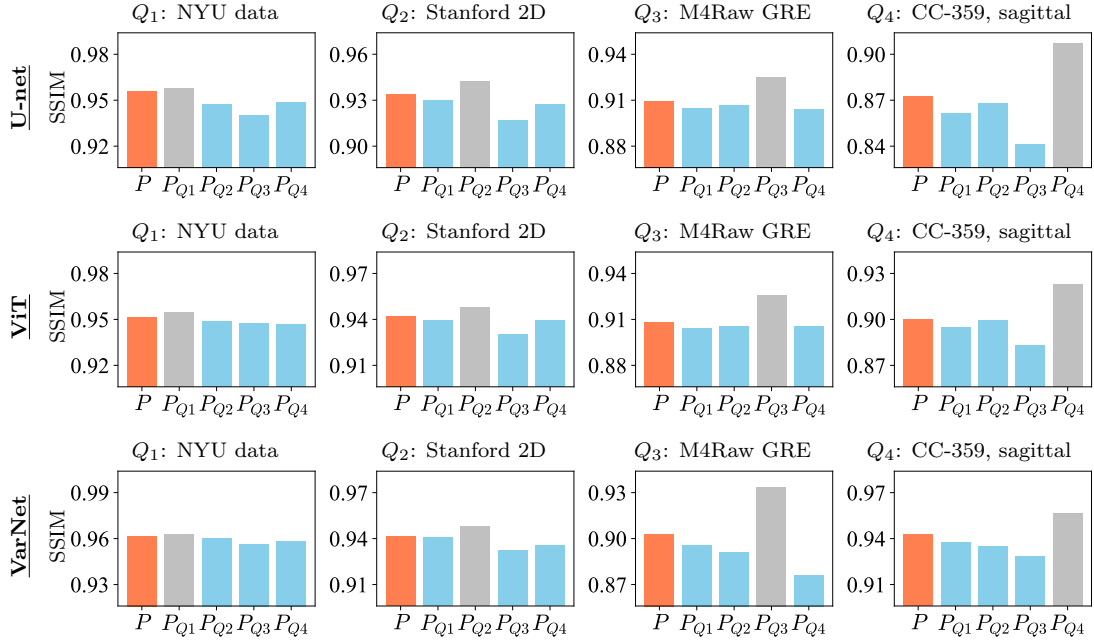
\begin{figure}[t!]
    \centering
    \begin{minipage}{1\linewidth} \hspace{0.05\linewidth}
    \begin{tikzpicture}
        \node (c) at (0.0\linewidth,0.0\linewidth) {\includegraphics[width=0.2\linewidth]{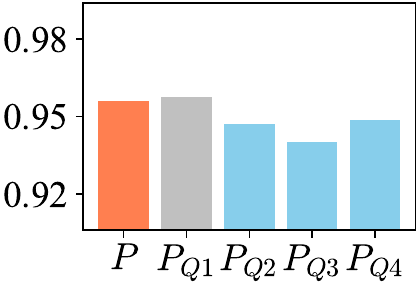}};
        \node[above = -0.01\linewidth of c, xshift=0.01\linewidth, font=\scriptsize] {$Q_1$: NYU data};
        \node[left = 0.01\linewidth of c, rotate=90, font=\scriptsize, xshift=0.045\linewidth, yshift=-0.009\linewidth] {SSIM};
        \node[left = 0.02\linewidth of c, rotate=90, font=\scriptsize, xshift=0.015\linewidth, yshift=0.01\linewidth, anchor=center] {\underline{\textbf{U-net}}};

        \node[right = -0.008\linewidth of c] (d) {\includegraphics[width=0.2\linewidth]{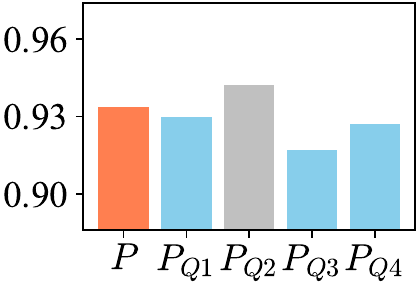}};
        \node[above = -0.01\linewidth of d, xshift=0.01\linewidth, font=\scriptsize] {$Q_2$: Stanford 2D};
        \node[right = -0.008\linewidth of d] (e) {\includegraphics[width=0.2\linewidth]{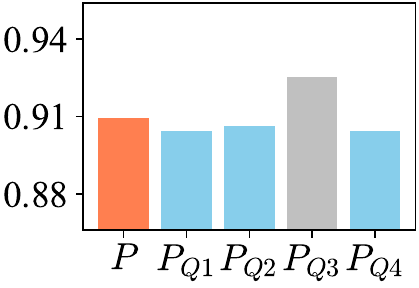}};
        \node[above = -0.01\linewidth of e, xshift=0.01\linewidth, font=\scriptsize] {$Q_3$: M4Raw GRE};
        \node[right = -0.008\linewidth of e] (f) {\includegraphics[width=0.2\linewidth]{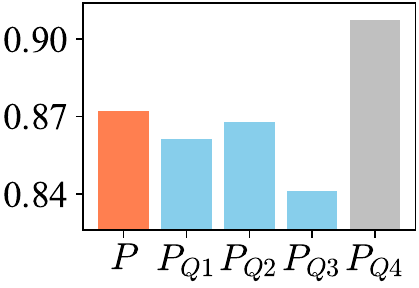}};
        \node[above = -0.01\linewidth of f, xshift=0.00\linewidth, font=\scriptsize] {$Q_4$: CC-359, sagittal};
    \end{tikzpicture}
    \end{minipage} \\ 
    \begin{minipage}{1\linewidth}  \hspace{0.05\linewidth}
    \begin{tikzpicture}
        \node (c) at (0.0\linewidth,0.0\linewidth) {\includegraphics[width=0.2\linewidth]{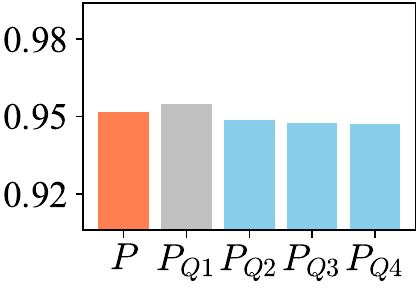}};
        \node[above = -0.01\linewidth of c, xshift=0.01\linewidth, font=\scriptsize] {$Q_1$: NYU data};
        \node[left = 0.01\linewidth of c, rotate=90, font=\scriptsize, xshift=0.045\linewidth, yshift=-0.009\linewidth] {SSIM};
        \node[left = 0.02\linewidth of c, rotate=90, font=\scriptsize, xshift=0.015\linewidth, yshift=0.01\linewidth, anchor=center] {\underline{\textbf{ViT}}};

        \node[right = -0.008\linewidth of c] (d) {\includegraphics[width=0.2\linewidth]{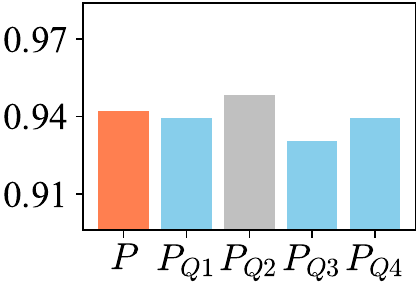}};
        \node[above = -0.01\linewidth of d, xshift=0.01\linewidth, font=\scriptsize] {$Q_2$: Stanford 2D};
        \node[right = -0.008\linewidth of d] (e) {\includegraphics[width=0.2\linewidth]{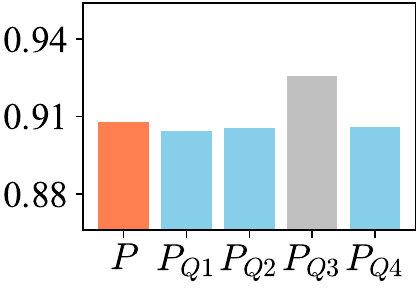}};
        \node[above = -0.01\linewidth of e, xshift=0.01\linewidth, font=\scriptsize] {$Q_3$: M4Raw GRE};
        \node[right = -0.008\linewidth of e] (f) {\includegraphics[width=0.2\linewidth]{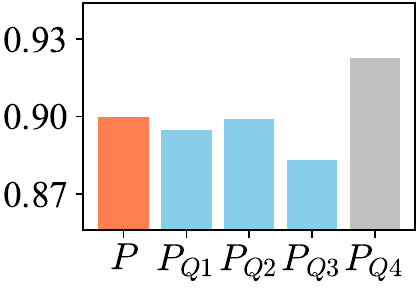}};
        \node[above = -0.01\linewidth of f, xshift=0.00\linewidth, font=\scriptsize] {$Q_4$: CC-359, sagittal};
    \end{tikzpicture}    
    \end{minipage}\\
    \begin{minipage}{1\linewidth}  \hspace{0.05\linewidth}
        \begin{tikzpicture}
            \node (c) at (0.0\linewidth,0.0\linewidth) {\includegraphics[width=0.2\linewidth]{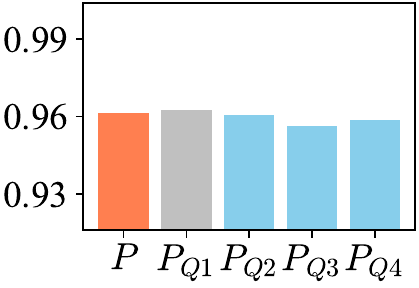}};
            \node[above = -0.01\linewidth of c, xshift=0.01\linewidth, font=\scriptsize] {$Q_1$: NYU data};
            \node[left = 0.01\linewidth of c, rotate=90, font=\scriptsize, xshift=0.045\linewidth, yshift=-0.009\linewidth] {SSIM};
            \node[left = 0.02\linewidth of c, rotate=90, font=\scriptsize, xshift=0.015\linewidth, yshift=0.01\linewidth, anchor=center] {\underline{\textbf{VarNet}}};
    
            \node[right = -0.008\linewidth of c] (d) {\includegraphics[width=0.2\linewidth]{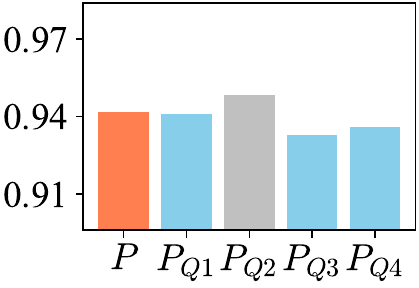}};
            \node[above = -0.01\linewidth of d, xshift=0.01\linewidth, font=\scriptsize] {$Q_2$: Stanford 2D};
            \node[right = -0.008\linewidth of d] (e) {\includegraphics[width=0.2\linewidth]{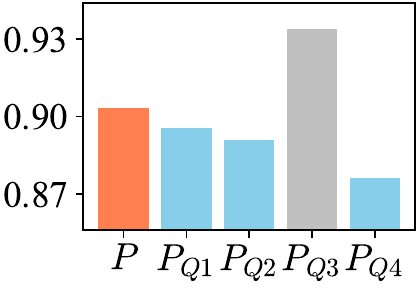}};
            \node[above = -0.01\linewidth of e, xshift=0.01\linewidth, font=\scriptsize] {$Q_3$: M4Raw GRE};
            \node[right = -0.008\linewidth of e] (f) {\includegraphics[width=0.2\linewidth]{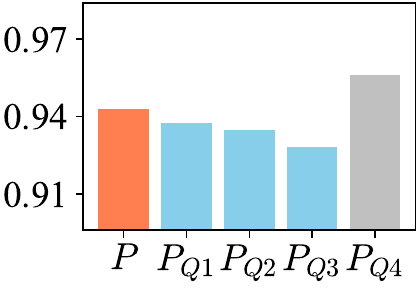}};
            \node[above = -0.01\linewidth of f, xshift=0.00\linewidth, font=\scriptsize] {$Q_4$: CC-359, sagittal};
        \end{tikzpicture}    
        \captionof{figure}{Fine-tuning reduces out-of-distribution robustness. 
        $P_{Qi}$ indicates the model obtained by fine-tuning the model trained on $\mathcal{D}_P$ (\textcolor{orange}{orange bar}) indicated by $P$, on one of the distributions $Q_1$, $Q_2$, $Q_3$ or $Q_4$. It can be seen that while the models perform better on the data they are fine-tuned on (\textcolor{gray}{gray bar}), the fine-tuned models perform worse on out-of-distribution data than the model $P$.}
        \label{fig:diverse_datasets_finetune}
        \end{minipage}
\end{figure}

Our results indicate that training a model on a diverse dataset enhances its robustness towards natural distribution-shifts. In this section we demonstrate that fine-tuning an already diversely trained model on a new dataset reduces its overall robustness. 

For this experiment, we take the models from Section~\ref{sec:robust_models} that were trained on $\mathcal{D}_P$ and fine-tune them on one of the four out-of-distribution datasets $\mathcal{D}_{Qi}$ (see last four rows in Table~\ref{tbl:datasets}). We denote the model fine-tuned on $Q_i$ by $P_{Qi}$. As depicted in Figure~\ref{fig:diverse_datasets_finetune}, the fine-tuned model $P_{Qi}$ exhibits improved performance on the specific distribution $Q_i$ it is fine-tuned on, as expected. However, the model under-performs on all other datasets in comparison to the model trained on $\mathcal{D}_P$ (i.e., prior to fine-tuning).

\end{document}